\documentclass[12pt]{article}
\usepackage[utf8]{inputenc}
\usepackage[export]{adjustbox}
\usepackage{authblk}
\usepackage{bbm}
\usepackage{pgfpages}
\usepackage{graphicx}
\usepackage{multicol}
\usepackage{csquotes}
\usepackage{multirow}
\usepackage{graphicx} 
\usepackage{booktabs} 
\usepackage{authblk}
\usepackage{caption}
\usepackage{subcaption}
\usepackage{stackengine}
\usepackage{amsmath}

\usepackage{comment}
\usepackage{lscape}
\usepackage{hyperref}
\hypersetup{
    colorlinks=true,
    linkcolor=blue,
    filecolor=blue,      
    urlcolor=blue,
    citecolor=blue,
}
\usepackage[capposition=top]{floatrow}

\usepackage[spanish,english]{babel}

\usepackage{amssymb}
\usepackage{stackrel}
\usepackage{natbib} 

\usepackage[a4paper, margin=0.85in]{geometry}

\title{\textbf{Spillovers of US Interest Rates} \\ Monetary Policy \& Information Effects\footnote{I would like to thank the Monetary and Fiscal History of Latin America project of the Becker Friedman Institute for their generous financial support. Particularly, I would like to thank Edward R. Allen, for supporting my research.}}

\author[1]{Santiago Camara}

\affil[1]{Northwestern University \& Red-NIE}
\date{\normalsize This Draft \today, First Draft: August 2nd 2021}

\begin{document}
    
\maketitle

\begin{abstract}
    This paper quantifies the international spillovers of US monetary policy by exploiting the high-frequency movement of multiple financial assets around FOMC announcements. I use the identification strategy introduced by \cite{jarocinski2020central} to identify two FOMC shocks: a pure US monetary policy and an information disclosure shock. These two FOMC shocks have intuitive and very different international spillovers. On the one hand, a US tightening caused by a pure US monetary policy shock leads to an economic recession, an exchange rate depreciation and tighter financial conditions. On the other hand, a tightening of US monetary policy caused by the FOMC disclosing positive information about the state of the US economy leads to an economic expansion, an exchange rate appreciation and looser financial conditions. Ignoring the disclosure of information by the FOMC biases the impact of a US monetary policy tightening and may explain recent atypical findings.
    
    
\end{abstract}

\newpage
\section{Introduction} \label{sec:introduction}

The international spillovers of US monetary policy is a classic question in international macroeconomics, going back to \cite{fleming1962domestic}, \cite{mundell1963capital}, \cite{dornbusch1976expectations} and \cite{frenkel1983monetary}. The status of the US dollar as the global reserve currency, unit of account, invoice of international trade and its dominant role in financial markets implies that the Federal Open Market Committee's (FOMC) policy decisions have spillover effects on the rest of the world. In recent years, the conventional view that a US monetary tightening leads to negative international spillovers such as a recession, exchange rate depreciation and financial distress (\cite{svensson1989excess}, \cite{obstfeld1995exchange}, \cite{betts2000exchange}), has been challenged. In fact, a recent literature has found opposite empirical results, an increase in the US monetary policy rate has been associated with a depreciation of the US dollar and an economic boom and looser financial conditions in the rest of the world (\cite{stavrakeva2019dollar}, \cite{ilzetzki2021puzzling}). In this paper, I show that the atypical dynamics documented in this recent literature are can be explained by the disclosure of information about the US economy that takes place around FOMC meetings. This ``information disclosure'' effect contaminates the identification of monetary policy shocks and biases the estimates of the international spillovers of US monetary policy shocks. Controlling for this information disclosure effect re-establishes the conventional view that a US monetary tightening leads to a recession, exchange rate depreciation and tighter financial conditions. 

The recent literature estimating atypical effects of US monetary policy shocks has identified them using the standard high frequency identification strategy, i.e. as unexpected movements in interest rates around FOMC announcements, as in \cite{gertler2015monetary} and \cite{nakamura2018high}. However, FOMC announcements both convey decisions about policy rates and disclose information about the present and future state of the US economy. Both Advanced (AE) and Emerging Market (EM) economies depend heavily on the US business cycle (for instance, because of its international trade with the US or because of the impact of the US economy on commodity goods' prices) or on the conditions in US financial markets (for example, due to the appetite for AEs and EMs' sovereign and corporate bonds and/or equity markets). As a result, these economies are affected by both the FOMC's policy decisions and the new information disclosed by the FOMC. Therefore, separately identifying US monetary policy shocks from information disclosure shocks is essential to study the impact of US monetary policy on EM economies. 

In line with this argument, I estimate a panel SVAR model using an identification scheme that allows me to separate two FOMC shocks: a pure US monetary policy shock (MP shock) and an information-disclosure shock (ID shock). I tell them apart using the methodology introduced by  \cite{jarocinski2020deconstructing} which imposes sign restrictions on the co-movement of the high-frequency surprises of interest rates and the S\&P 500 around FOMC meetings. This co-movement is informative as standard theory unambiguously predicts that a monetary policy tightening shock should lead to lower stock market valuation. This is because a monetary policy tightening decreases the present value of future dividends by increasing the discount rate and by deteriorating present and future firm's profits and dividends. Thus, MP shocks are identified as those innovations that produce a negative co-movement between these high-frequency financial variables. On the contrary, innovations generating a positive co-movement between the interest rates and the S\&P 500 correspond to ID shocks. This is a shock that occurs systematically at the time of the central bank policy announcements, but that is different from the standard monetary policy shock. By separately identifying these two structural shocks into a panel SVAR, I find that MP shocks produce conventional international spillovers, i.e. a recession, exchange rate depreciation and financial distress. On the contrary, an ID shock produces an economic expansion, an exchange rate appreciation and looser financial conditions.

Then, I argue that the recently found atypical dynamics can be attributed by not controlling for the systematic disclosure of information around FOMC meetings. I show that following the standard high frequency identification scheme to identify US interest rate shocks leads to dynamics which are an average of those arising from a MP and ID shock. In particular, this leads to a US interest rate tightening associated with a significant expansion of industrial production and equity indexes for both AE and EM economies. Overall, I argue that not controlling for the information disclosure around FOMC meetings biases the quantifying of international spillovers of US interest rates. 

\noindent 
\textbf{Related literature.} This paper relates to three main strands of literature. First, this paper contributes to a long strand of literature which has focused on identifying and quantifying the international spillovers of US monetary policy shocks and their transmission channels. A significant share of this literature has found that a US tightening is associated with negative international spillovers such as an economic recession, an exchange rate depreciation or fall of the value of the country's currency and tighter overall financial conditions. Examples of this literature are \cite{eichenbaum1995some} and \cite{uribe2006country} using data from the 1980s, 1990s, and more recently \cite{dedola2017if}, \cite{vicondoa2019monetary} using data up to the late 2000s. During the rest of the paper I will refer to these results as the conventional view or impact of a US tightening. The contribution to this literature is twofold. First, this paper innovates  by introducing an identification scheme that clearly purges any information content included in US monetary policy decisions and finds that conventional results still hold for a time sample of the 2000s and 2010s. Second, this paper contributes to the literature by empirically studying the international spillovers of the disclosure of information by the FOMC as a transmission channel of US interest rates. I show that this transmission channel is quantitatively sizable and has not been considered by the previous literature as a meaningful transmission channel of US interest rates.

This paper also relates to a more recent literature in international economics which have found an atypical association between the US interest rates and AE and EM dynamics. Particularly, \cite{ilzetzki2021puzzling} argue that there has been a significant change over time in the transmission of US monetary policy shocks on the rest of the world. The authors find that while in the 1980s and early 1990s a US tightening lead to the conventional results described in the previous paragraph, in the last two decades there has been a shift whereby increases in US interest rates depreciate the US dollar but stimulate the rest of the world economy. The authors label this shift as a puzzling change in the transmission of US interest rates. Consequently, in the rest of the paper I will refer to these responses of AEs and EMs to a US tightening as atypical dynamics. Another example of these atypical dynamics is \cite{canova2005transmission} which finds that after a US tightening, Latin American currencies appreciate while the conventional view would expect a currency depreciation.\footnote{Evidence of atypical dynamics can also be found in an influential paper of the conventional view as \cite{uribe2006country}. In the working paper, the authors explore an identification scheme different from the one presented in the actual paper, which allows for real domestic variables to react contemporaneously to innovations in the US interest rate. Under this alternative identification strategy, the point estimate of the impact of a US-interest-rate shock on output and investment is slightly positive. This lead to adoption of a different identification scheme.} I contribute to this literature by introducing an identification scheme that deconstruct shocks around FOMC announcements following the methodology introduced by \cite{jarocinski2020deconstructing}. This identification scheme allows me to identify an information disclosure shock which entirely explains the atypical dynamics found by this recent literature. Additionally, this identification scheme allows me to identify a pure US monetary policy shock, which re-establishes the results presented by the conventional view. Consequently, by deconstructing monetary policy shocks I am able to match both the conventional and atypical results. 

Third, this paper relates to a recent literature which has studied the spillovers of US interest rates over the rest of the world by using identification strategies that control for possible informational effects around FOMC meetings.  For instance, \cite{jarocinski2022central} uses the identification strategy introduced by \cite{jarocinski2020deconstructing} to demonstrate that central bank information effects are an important channel of the transatlantic spillover of monetary policy, as they account for a significant share of the co-movement of German and US government bond yields around ECB and FOMC policy announcements. Another example is \cite{degasperi2020global} which uses the identification strategy constructed by \cite{miranda2021transmission} which controls for potential ``signalling information'' effects around FOMC meetings.    Lastly, \cite{camararamirez2022} uses the identification strategy constructed by \cite{bauer2022reassessment} which controls for the Federal Reserve's ``responding to news'' informational effect. This paper's key contribution to the literature is that it actively seeks to identify the spillovers of the US interest rates through the systematic disclosure of information around FOMC meetings. While the main focus of \cite{degasperi2020global} and \cite{camararamirez2022} is to study the different transmission channels of US interest rates, they disregard the transmission of US interest rates through informational effects. In this paper, I show that this is sizeable transmission channel, leading to quantitatively large spillovers on the rest of the world.\footnote{While this paper innovates by presenting an empirical analysis of the international spillovers of the FOMC's disclosure of information, the theoretical literature has already studied its potential impacts, see \cite{ahmed2021us}.} Furthermore, this paper shows that this systematic information disclosure around FOMC meetings biases estimates of US monetary policy shocks using the standard high-frequency identification strategy, leading to the recent expansionary effect of US interest rate hikes on the rest of the world. While \cite{degasperi2020global} and \cite{camararamirez2022}, suggest that informational effects may explain these recent puzzling dynamics, they do not seek to answer this question. Additionally, I show that alternative identification strategies that seek to purge for any ``informational effects'' around FOMC meetings yield remarkably similar results to this paper's benchmark results. I take this result as supporting evidence that recent atypical dynamics can be attributed to the systematic disclosure of information around FOMC meetings.

\section{Data, Methodology \& Identification} \label{sec:data_methodology_identification}

In this section, I describe the construction of my dataset, describe the panel SVAR methodology used and delineate the identification strategy used across the paper.

\noindent
\textbf{Data.} First, I describe the sample of AE and EM countries, the different datasets used across the paper to construct out sample of macroeconomic and financial variables, and the source of the high-frequency surprises and FOMC shocks. The benchmark specification analyzes the international transmission of US monetary spillovers on 18 countries, 9 Emerging Markets and 9 Advanced Economies at the monthly frequency for the period January 2004 to December 2016. Table \ref{tab:country_list} presents the different countries in the analysis. 
\begin{table}[ht]
    \centering
    \caption{Country List}
    \label{tab:country_list}
    \begin{tabular}{c | c}
       Emerging Markets & Advanced Economies  \\ \hline \hline
    Brazil	&	Australia	\\
    Chile	&	Canada	\\
    Colombia	&	France	\\
    Hungary	&	Iceland	\\
    Indonesia	&	Italy	\\
    Mexico	&	Japan	\\
    Peru	&	South Korea	\\
    Philippines	&	The Netherlands	\\
    South Africa	&	Sweden	\\
    \end{tabular}
\end{table}
The reasoning behind this choice of time sample is twofold. First, a main motivation of this paper is to be able to explain and deconstruct the atypical dynamics found after a US tightening in recent times. Consequently, I estimate the empirical models during a time period where the abnormal dynamics are found in previous papers.\footnote{For instance, \cite{ilzetzki2017country} founds that the puzzling dynamics arise during the 1990s and remain during the 2000s.} The second reason is the lack of data availability for Emerging Market economies in the late 1990s and early 2000s. Additionally, during the late 1990s and early 2000s several EM economies experienced significant monetary and fiscal policy changes (for instance the implementation of inflation targetting regimes and fiscal policy rules).\footnote{An example of these policy changes is that around a third of emerging and developing countries shifted from pro-cyclical to counter-cyclical fiscal policies between the late 1990s to the early 2000s.} Thus, to construct a rich but balanced panel, I start the sample in January 2004. My benchmark model specification includes five macroeconomic and financial variables: (i) nominal exchange rate with respect to the US dollar, (ii) industrial production index, (iii) CPI index, (iv) domestic lending rates, (v) equity index.\footnote{In order to construct a harmonized dataset of macroeconomic and financial variables for both AE and EM economies, I source all of the datasets from the IMF, which guarantees that the variables in the dataset are constructed following closely related methodologies. Variables ``industrial production'', ``consumer price index'', ``nominal exchange rate with respect to the US dollar'', ''equity index'' and ``lending rates'' are sourced from the IMF's International Financial Statistics dataset. For additinal details on the construction of the dataset see Appendix \ref{sec:appendix_data_details}.}

Lastly, I describe the source of the high frequency surprises used to construct the two FOMC shocks. I follow \cite{jarocinski2022central} and define the high frequency surprise of the interest rate as the first principal component of the 30 minute window surprises in interest rate derivatives with maturities up to 1 year. In particular, I use the first principal component of the surprises in the current month and 3-month Fed Funds Futures and the 2, 3, and 4 quarters ahead 3-month eurodollar futures.\footnote{Other papers exploiting the first principal component of surprises in different interest rates are, among others, \cite{gurkaynak2004actions} and \cite{nakamura2018high}.} For the stock market surprise, I use the 30 minute window surprise in the S\&P 500 index. This data is sourced directly from the dataset constructed by \cite{jarocinski2022central}.\footnote{Both the time series and the the structural shocks and the replication codes to compute shocks can be directly downloaded from the authors' website. See \url{https://marekjarocinski.github.io/}.}

\noindent
\textbf{Methodology.} Next, I describe the panel SVAR model methodology estimated across the paper. The model specification is a pooled panel SVAR as presented by \cite{canova2013panel}. This type of model considers the dynamics of several countries simultaneously, but assuming that the dynamic coefficients are homogeneous across units, and coefficients are time-invariant. In this framework, this implies that country $i$'s variables only depend on structural shocks  and the lagged values of country $i$'s variables. Although the possible interactions and inter-dependencies across countries is an interesting topic on itself, I abstract from this possibility in this paper. In Section \ref{sec:robustness_checks_additional_results} I discuss heterogeneous responses across different countries by partitioning the benchmark sample across different dimensions. Additionally, as a robustness check, I show that results are robust when estimating the mean-group estimator proposed by \cite{pesaran1995estimating} and using local projection techniques.

\noindent
\textbf{Identification strategy.} Lastly, I describe the identification strategy that allows me to recover two distinct FOMC shocks and estimate their impact on both AE and EM economies. The identification strategy combines the two structural FOMC shocks recovered by using the identification methodology developed by \cite{jarocinski2020deconstructing} and \cite{jarocinski2022central} with a standard Choleski ordering identification strategy. The identification strategy introduced by \cite{jarocinski2020deconstructing} exploit the high-frequency surprises of multiple financial instruments to recover two distinct FOMC shocks: a pure monetary policy (MP) shock and an information disclosure (ID) shock. The authors impose sign restrictions conditions on the co-movement of the high-frequency surprises of interest rates and the S\&P 500 around FOMC meetings. This co-movement is informative as standard theory unambiguously predicts that a monetary policy tightening shock should lead to lower stock market valuation.\footnote{This is because a monetary policy tightening decreases the present value of future dividends by increasing the discount rate and by deteriorating present and future firm's profits and dividends.} MP shocks are identified as those innovations that produce a negative co-movement between these high-frequency financial variables. On the contrary, innovations generating a positive co-movement between interest rates and the S\&P 500 correspond to ID shocks. 

Note that imposing a sign restriction over the co-movement of the high-frequency surprises of the interest rates and the S\&P 500 does not uniquely identify the underlying structural shocks. In terms of \cite{jarocinski2020deconstructing} and \cite{jarocinski2022central}, there are ``different rotations'' of the decomposition matrix that satisfy the sign restriction condition. Previous papers have chosen different approaches to deal with this non-uniqueness.\footnote{For instance, \cite{jarocinski2020deconstructing} include the high frequency surprises of the interest rates and the S\&P 500 in their VAR and specify an agnostic flat prior over the space of admissible rotations. \cite{andrade2021delphic} use the average admissible rotation angle for a similar decomposition. \cite{jarocinski2022central} recovers the structural shocks by choosing a rotational angle that imposes a relationship between the relative variances of the MP shock and that of the high-frequency surprises of the interest rate.} For my benchmark specification, I follow the approach introduced by \cite{jarocinski2022central} to deal with this non-uniqueness problem and use the median rotation, which here boils down to using the median of the admissible rotation angles. As robustness checks, I also consider the rotation angle which pins down the decomposition imposing a 0.88 ratio between the variance of the MP shock and the variance of the high-frequency interest rate surprise and imposing a uniform prior on the space of admissible rotations.\footnote{In Appendix \ref{subsec:appendix_model_details_identification} I present the different steps of this procedure by following the methodology introduced by \cite{jarocinski2022central}.}$^{,}$\footnote{ Following this procedure allows me to recover the time series of the two FOMC shocks and introduce them into the panel SVAR model in Section \ref{sec:main_results} and into a Local Projection regression as shown in Section \ref{sec:robustness_checks_additional_results} as a robustness check.} I show that results remain qualitatively and quantitatively similar when using different approaches to deal with this non-uniqueness problem. 

In order to identify and quantify the international spillovers of the two FOMC shocks over the rest of world I combine the recovered FOMC shocks with a standard Choleski identification strategy. I order the vector of identified structural shocks $m_t$ first with the vector of country $i$ specific macroeconomic and financial variables second. Within the vector $m_t$ is order the two FOMC shocks with $i^{\text{MP}}$ first and $i^{\text{ID}}$ second. \textit{A priori}, this implies that the first ordered FOMC shock is allowed to contemporaneously impact the second ordered FOMC shock but the latter does not impact the former on impact. Given that by construction these shocks are mutually orthogonal these ordering decisions should only lead to efficiency losses and do not introduce any systematic bias. In Section \ref{sec:robustness_checks_additional_results}, I show that results hold and are quantitatively similar when re-ordering the shocks in vector $m_t$ and by estimating the impact of each FOMC shock separately by defining vector $m_t$ as containing only one of the two FOMC shocks at a time.

\section{Spillovers of Monetary Policy \& Information Effects} \label{sec:main_results}

This section presents the main results of this paper. I estimate and quantify the impact of the two FOMC shocks: a pure monetary policy (MP) shock and an information disclosure (ID) shock. I compare the resulting impulse response functions with those arising from following the standard identification strategy (``Standard HFI'') of only using the high-frequency surprise of the policy interest rate. First, I show that the two FOMC shocks have completely opposite spillovers over the rest of the world. Second, I argue that the presence of information disclosure shocks biases the results arising from the standard identification strategy and may explain recently found atypical dynamics.

\begin{figure}[ht]
         \centering
         \caption{Impulse Response to One-Standard-Deviation Shock \\ \footnotesize Benchmark Specification}
         \includegraphics[scale=0.325]{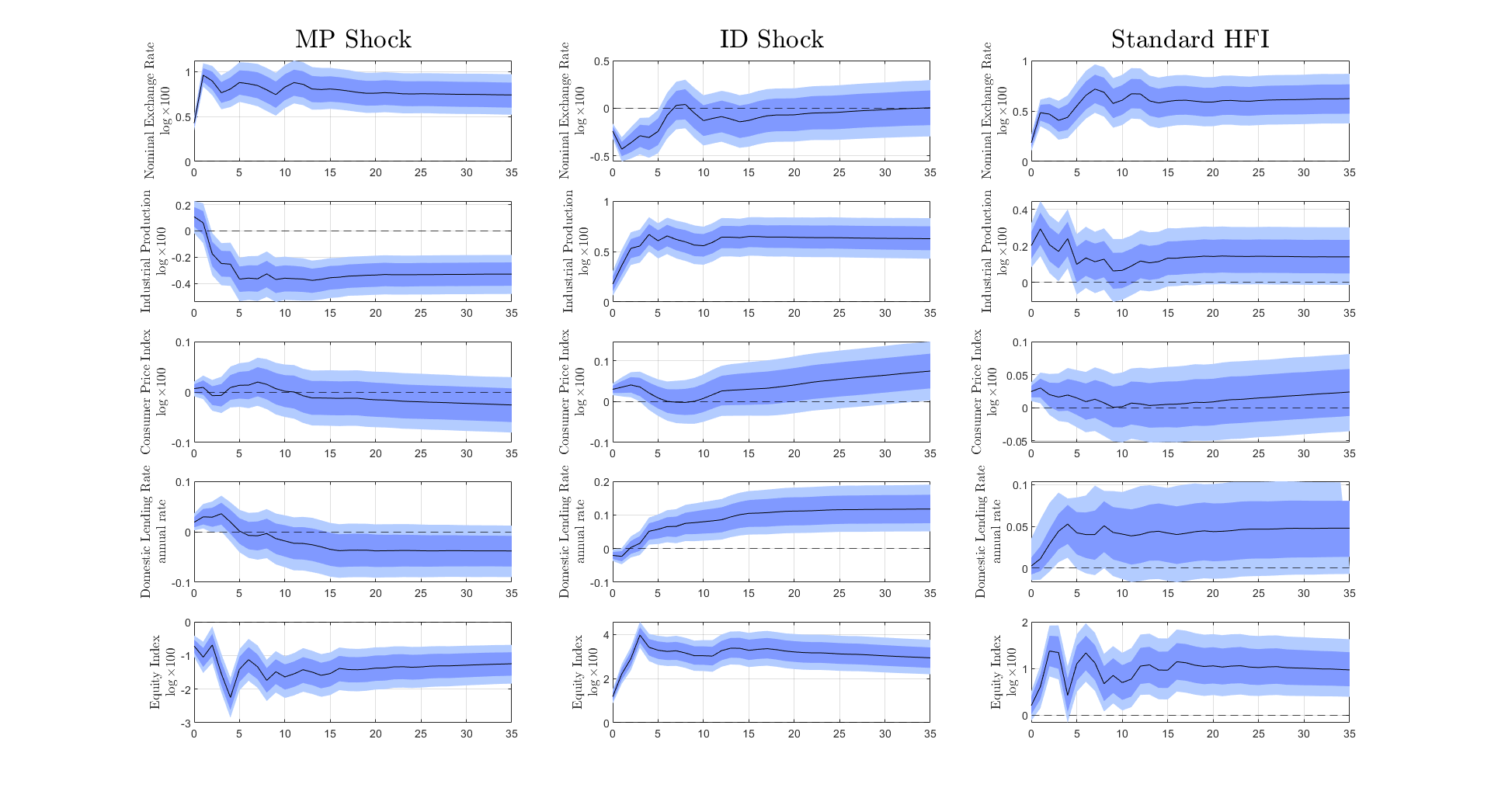}
         \label{fig:Benchmark}
         \floatfoot{\scriptsize \textbf{Note:} The black solid line represents the median impulse response function. The dark shaded area represents the 16 and 84 percentiles. The light shaded are represents the 5 and 95 percentiles. The figure is comprised of 15 sub-figures ordered in five rows and three columns. Every row represents a different variable: (i) nominal exchange rate, (ii) industrial production index, (iii) consumer price index, (iv) lending rate, (v) equity index. The first column presents the results for the MP or ``Pure US Monetary Policy'' shock, the middle column presents the results for the ID or ``Information Disclosure'' shock, and the last column presents the results for the interest rate composite high frequency surprise or ``Standard HFI''. In the text, when referring to Panel $(i,j)$, $i$ refers to the row and $j$ to the column of the figure.}
\end{figure}

First, I start by testing whether the deconstruction of US interest rate movements into the two distinct FOMC shocks matter for quantifying the spillovers of US monetary policy. The first and second columns of Figure \ref{fig:Benchmark} present the impulse response functions of the macroeconomic and financial variables to a MP and ID shock, respectively.
Comparing the figures across these two columns leads to a first important conclusion. This is that the identification scheme based on sign restriction of the US high-frequency financial surprises separately identifies two distinct economic shocks. If the co-movement between the high frequency of the policy interest rate composite and the S\&P 500 was uninformative, the impulse response functions presented in the first and second columns of Figure \ref{fig:Benchmark} should exhibit the same results. Comparing the results presented in these figures, it is straightforward to conclude that this is not the case. For instance, the behavior of the industrial production index is completely opposite across figures, with a MP shock leading to a persistent decline in industrial output and a ID shock leading to a persistent increase in it. Hence, correctly identifying the different FOMC shocks is crucial to accurately quantify the spillovers of US monetary policy shocks.

Next, I describe in greater detail the impulse response functions. The first column of Figure \ref{fig:Benchmark} shows the responses of domestic macro and financial variables to a MP shock. First, Panel 1.1 shows that a one-standard-deviation MP shock leads to a 50 basis point depreciation on impact of the nominal exchange rate. The nominal exchange rate further depreciates during the first 6 months after the shock, reaching a level 100 basis points greater to the pre-shock levels. This depreciation continues to be significantly different from zero even 3 years after the initial shock. Panel 2.1 in the second row shows that after a brief two month expansion, the industrial production index shows a hump shaped, persistent decrease, reaching a level 40 basis point below its pre-shock levels 10 months after the initial shock. Furthermore, this decrease in industrial production persists even 3 years after the initial shock. Panel 3.1 shows that a MP shock does not lead to a significant response of the consumer price index. \textit{A priori}, a MP shock affects consumer prices through two opposite channels. On the one hand, the nominal exchange rate depreciation increases the domestic price of imported goods. On the other hand, the economic recession (shown by the drop in industrial production) may reduce inflationary pressures. Panel 4.1 on the fourth row shows that a MP shock leads to a short-lived increase in domestic lending rates to the private sector. This increase peaks between 3 and 5 months after the initial shock at 10 basis points above pre-shock levels, quickly returning to this level 10 months after the initial shock. Finally, Panel 5.1 on the last row shows that a MP shock leads to a significant and persistent drop in the equity index. This drop is between 100 and 200 basis points during the first 6 months after the initial shock. Moreover, the drop in the equity index is persistent, remaining below its pre-shock levels 3 years after the initial shock.\footnote{A possible concern arising from the dynamics of the equity index on Panel 5.1 of Figure \ref{fig:Benchmark} is that the drop in the equity index is driven by the depreciation of the nominal exchange rate, shown in Panel 1.1. Note that, as described in Appendix \ref{sec:appendix_data_details} the equity index is defined in domestic currency. Furthermore, the drop in the equity index is between 50\% and 100\% greater in magnitude than the increase in the nominal exchange rate. Thus, Panel 5.1 shows that the equity index drops in value both in domestic currency and in US dollars.}

The spillovers of a ID shock are completely opposite to those of a MP shock. The second column of Figure \ref{fig:Benchmark} presents the impulse response functions of a one-standard-deviation ID shock. Panel 1.2 shows that a ID shock leads to a 25 basis points appreciation of the exchange rate on impact. The exchange rate continues appreciating reaching a level 40 basis points below its pre-shock level on the following month. The exchange rate returns to its pre-shock level between 6 and 10 months after the initial shock. Panel 2.2 shows that the industrial production index shows a persistent hump shaped expansion. After a 10 basis point increase on impact, industrial output increases to a level 60 basis point above pre-shock levels. Additionally, industrial output remains 20 basis points above pre-shock levels 3 years after the shock. Panel 3.2 shows that the consumer price index increases on impact between 25 and 50 basis points. This index exhibits a moderate increase even 3 years after the initial shock. Panel 4.2 shows that domestic lending rates decrease for the first 5 months after the initial shocks. This initial decrease in domestic lending rates accompanied with a higher inflation (see Panel 3.2) suggests that real rates exhibit a decrease after a FOMC interest rate hike caused by an ID shock. Lastly, Panel 5.2 shows that a ID shock leads to a persistent increase in the equity index. The index jumps close to 200 basis points on impact, reaching a peak of almost 400 basis points above pre-shock levels 4 months after the initial shock. The significant and persistent expansion in the equity index accompanied with lower lending rates provide evidence of an ID shock leading to looser financial conditions. Overall, the first and second columns of Figure \ref{fig:Benchmark} show that the two FOMC shocks lead to almost completely opposite impulse response functions for the full sample of countries. 

Next, I argue that following the ``Standard HFI'' strategy, using only the high-frequency surprise of the policy interest rate composite, may lead to biased impulse response functions. In particular, I replace the benchmark specification of vector $m_t$ which contains the two FOMC shocks with the high-frequency surprise of the policy interest rate composite and of the S\&P 500 index. The third column of Figure \ref{fig:Benchmark} exhibits the impulse response functions under this identification strategy. Across the different variables, the impulse responses are an average of the responses presented for the MP and ID shocks in the first and middle columns. Most shockingly, the ``Standard HFI'' strategy leads to a qualitatively different response of the industrial production and equity indexes from that arising after a MP shock. Under the ``Standard HFI'' strategy, the industrial production exhibits a one-year increase above pre-level shocks. On impact, industrial production increases by 20 basis points, peaking close to 25 basis points in the following two months. Similarly, the equity index increases 100 basis points above its pre-shock level persistently after a ``Standard HFI'' interest rate shock. In addition to this qualitative differences, there are notable quantitative differences in the impulse response functions of other variables. Panel 1.3 shows that under the ``Standard HFI'' strategy, the nominal depreciation is 50\% smaller than that implied by a MP shock (25 versus 50 basis points). This quantitative difference continues during the first year after the shock. Consequently, following the ``Standard HFI'' strategy may lead to underestimating the depreciation of the exchange rate after a pure US monetary policy shock.

\begin{landscape} 
\begin{figure}[ht]
    \centering
    \caption{Impulse Response to One-Standard-Deviation Shock \\ \footnotesize Separate Samples for Adv. \& Emerging Economies}
    \label{fig:Benchmark_Adv_and_EME}
     \centering
     \begin{subfigure}[b]{0.495\textwidth}
         \centering
         \includegraphics[width=\textwidth,height=9.5cm]{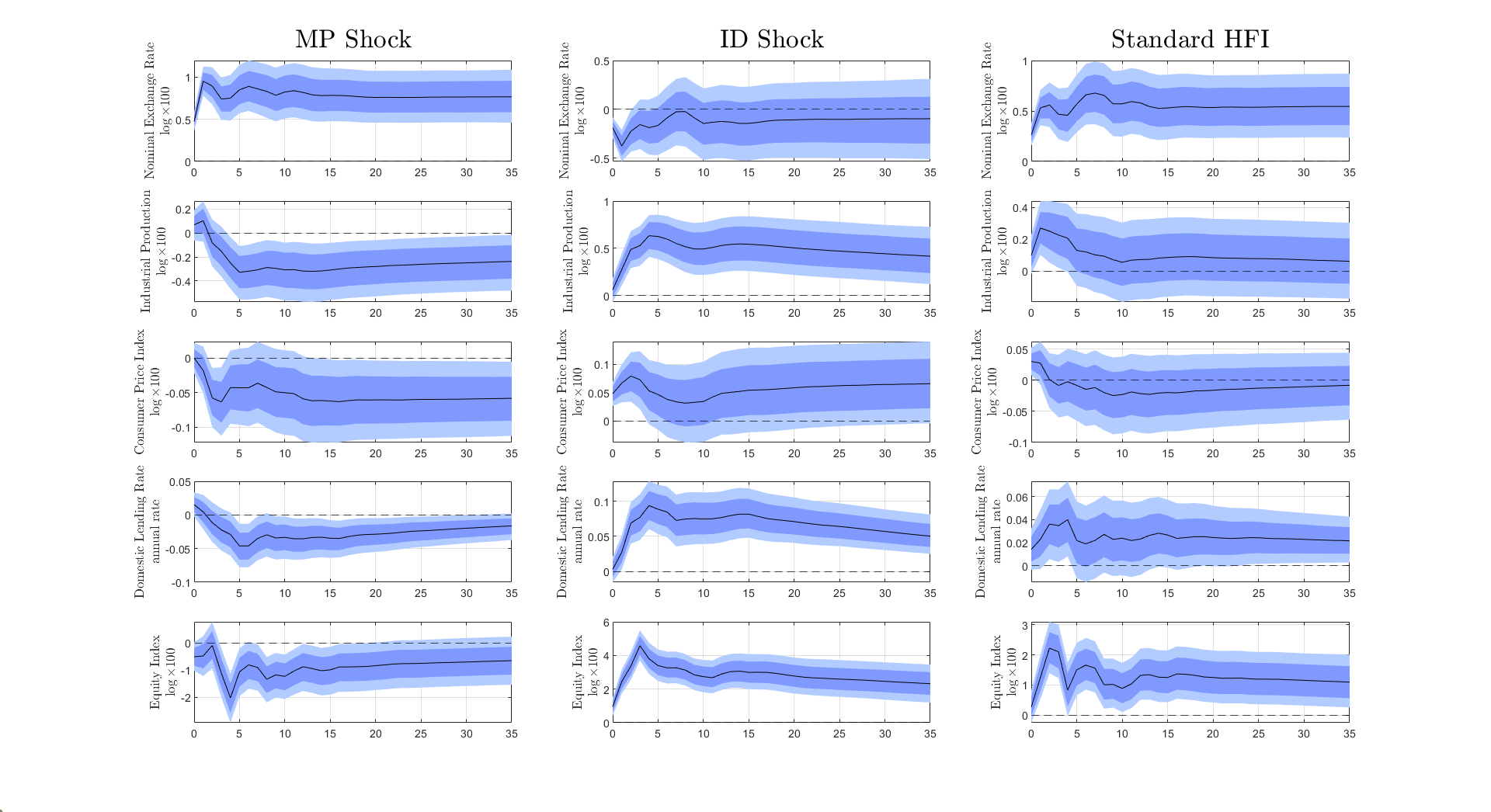}
         \caption{Advanced Economies}
         \label{fig:Benchmark_Adv}
     \end{subfigure}
     \hfill
     \begin{subfigure}[b]{0.495\textwidth}
         \centering
         \includegraphics[width=\textwidth,height=9.5cm]{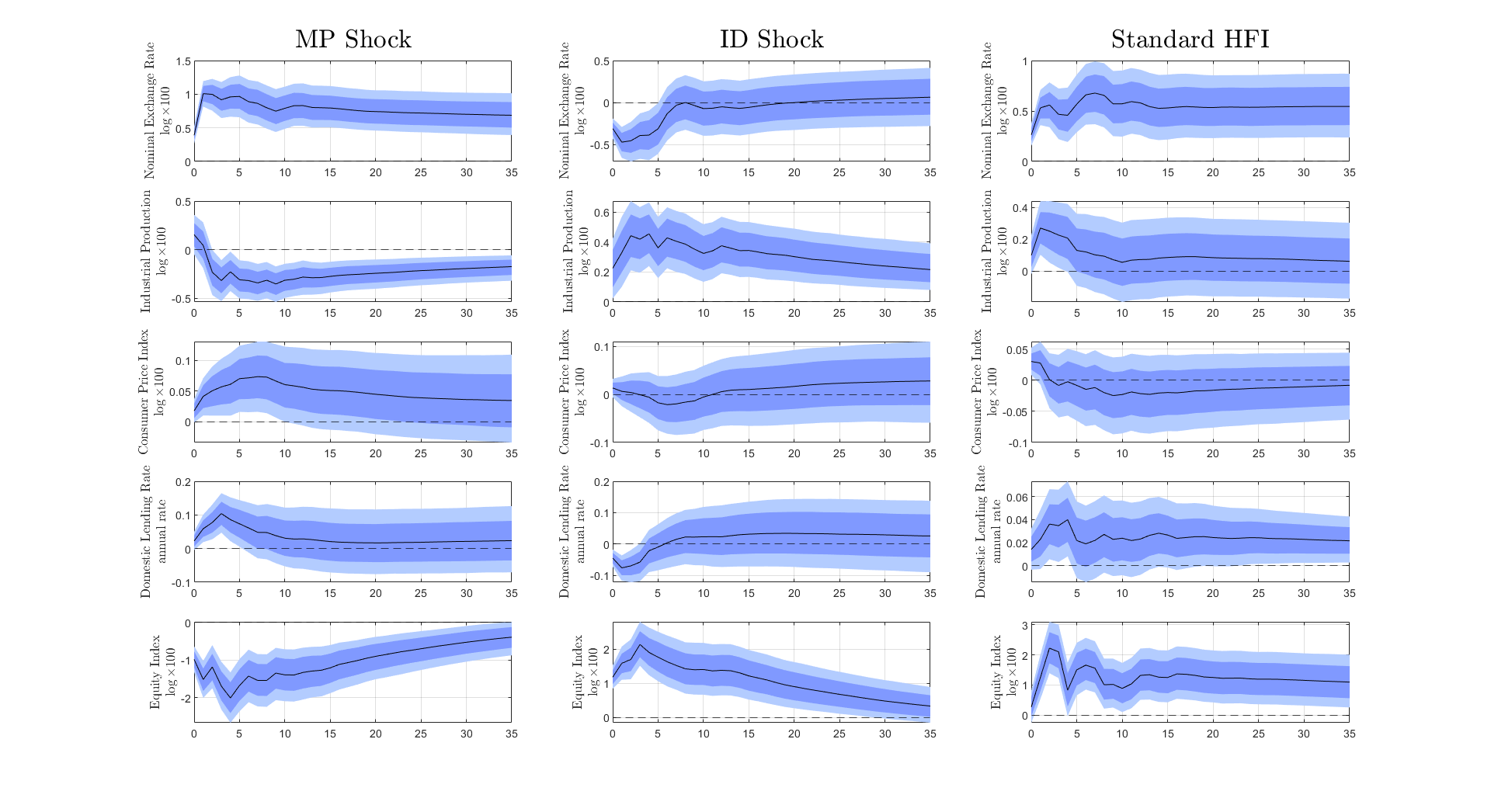}
         \caption{Emerging Market Economies}
         \label{fig:Benchmark_EM}
     \end{subfigure} 
     \floatfoot{\scriptsize \textbf{Note:} The black solid line represents the median impulse response function. The dark shaded area represents the 16 and 84 percentiles. The light shaded are represents the 5 and 95 percentiles. Figure \ref{fig:Benchmark_Adv} on the left presents the results for the panel of ``Advanced Economies'' comprised of: Australia, Canada, France, Iceland, Italy, Japan, South Korea, The Netherlands and Sweden. Figure \ref{fig:Benchmark_EM} presents the results for the panel of ``Emerging Market Economies'' comprised of: Brazil, Chile, Colombia, Hungary, Indonesia, Mexico, Peru, Philippines and South Africa. Each figure is comprised of 15 sub-figures ordered in five rows and three columns. Every row represents a different variable: (i) nominal exchange rate, (ii) industrial production index, (iii) consumer price index, (iv) lending rate, (v) equity index. The first column presents the results for the MP or ``Pure US Monetary Policy'' shock, the middle column presents the results for the ID or ``Information Disclosure'' shock, and the last column presents the results for the interest rate composite high frequency surprise or ``Standard HFI''. In the text, when referring to Panel $(i,j)$, $i$ refers to the row and $j$ to the column of the figure.}
\end{figure}
\end{landscape}

Finally, I test whether the results are driven by the composition of countries in the sample. Figure \ref{fig:Benchmark_Adv_and_EME} shows the resulting impulse response functions for a sample of Advanced Economies (on the left panel in Figure \ref{fig:Benchmark_Adv}) and Emerging Market economies {on the right panel in Figure \ref{fig:Benchmark_EM}} separately. Across the two sub-samples, the main results presented in Figure \ref{fig:Benchmark} hold, with the two FOMC shocks leading to completely opposite dynamics and the ``Standard HFI'' leading to a weighted average of them. Still, there are some noticeable differences. First, after an MP shock, Emerging Market economies exhibit a persistent increase in the consumer price index while Advanced Economies exhibit a significant drop. As argued by \cite{garcia2020revisiting} and \cite{auclert2021exchange}, Emerging Market economies depend relatively more in imported goods for both consumer and intermediate input goods. Thus, one would expect that greater exchange rate pass through in EMs relative to AE, all else equal. Similarly, the impact of a MP shock on domestic lending rates is different across the two sub-samples. While lending rates in AE moderately decrease by 5 basis points, they sharply increase in EMs, peaking above 10 basis points. This may be driven by Emerging Market economies having relatively less sophisticated and smaller domestic financial markets and, thus, exhibiting a relatively greater dependence on international financial markets than AE (see \cite{dages2000foreign,broner2013emerging,cortina2018corporate,abraham2020growth}). While lending rates decrease for Advanced Economies, the fact that industrial production and equity indexes decrease for both country sub-samples suggests that a MP shock leads to overall tighter financial conditions.

\noindent

To summarize, by introducing an identification scheme which deconstruct US interest rates movements into two different FOMC shocks, I am able to show that an increase in US interest rates has completely different international spillovers depending on the underlying economic shock. While a pure monetary policy shock leads to conventional results, an information disclosure shock leads to an economic expansion, an exchange rate appreciation and looser financial conditions. Moreover, I argued that following the ``Standard HFI'' strategy underestimates the spillovers of a US monetary policy shock and leads to atypical dynamics, such as an expansion of industrial production and equity indexes. Lastly, I showed that these results are present for both the sub-samples of Advanced and Emerging Market economies with only minor caveats.

\section{Additional Results \& Robustness Checks} \label{sec:robustness_checks_additional_results}

This section presents additional results and robustness checks that complement and support the findings presented in Section \ref{sec:main_results}. First, I show that the main results are also present using local projection techniques a la \cite{jorda2005estimation} to estimate impulse response functions. Second, I show that the main results are robust when using different ordering identification assumptions between the two FOMC shocks. Third, I show that the main results are robust to using different approaches to deal with the non-uniqueness problem inherent to the sign-restriction identification strategy. Lastly, I briefly describe potential sources of quantitatively heterogeneity in the spillovers of the two FOMC shocks. 

\noindent
\textbf{Alternative model specifications.} I start by showing that the main results presented in Section \ref{sec:main_results} are robust to estimating the impulse response functions to the two FOMC shocks using local projection techniques a la \cite{jorda2005estimation}. We estimate the following two empirical specifications.
\begin{align} 
    y_{i,t+h} = \beta^{MP}_{h} i^{\text{MP}}_t + \beta^{ID}_{h} i^{\text{ID}}_t + \sum^{J_y}_{j=1} \delta^{j}_i y_{i,t-j} + \sum^{J_x}_{j=1} \alpha^{j}_i x_{i,t-j} + \sum^{J_i}_{j=1} \left( \phi^{j}_i i^{\text{MP}}_{t-j} + \varphi^{j}_i i^{\text{ID}}_{t-j} \right) + \epsilon_{i,t} \label{eq:LP_pooled} \\
    y_{i,t+h} = \beta^{MP}_{h} i^{\text{MP}}_t + \beta^{ID}_{h} i^{\text{ID}}_t + \sum^{J_y}_{j=1} \delta^{j}_i y_{i,t-j} + \sum^{J_x}_{j=1} \alpha^{j}_i x_{i,t-j} + \sum^{J_i}_{j=1} \left( \phi^{j}_i i^{\text{MP}}_{t-j} + \varphi^{j}_i i^{\text{ID}}_{t-j} \right) +\Gamma_i+\Gamma_{i,t} + \epsilon_{i,t}  \label{eq:LP_fixed_effects_date}
\end{align}
The first empirical specification in Equation \ref{eq:LP_pooled}, which I label ``Pooled Specification'', measures outcome variable $y_{i,t+h}$ from country $i$'s at time horizon of $h$ months from date $t$. Coefficients $\beta^{MP}_{h}$ and $\beta^{ID}_{h}$ give the impulse response of outcome variable $y$ at horizon $h$ of a pure monetary policy (MP) and of an information disclosure (ID) shock, respectively. The specification includes $J_y$ lags of the outcome variable, $y_{i}$, $J_x$ lags of the other ``country specific'' variables $x_{i}$ and $J_i$ lags of the FOMC shocks. I select the number of lags by computing the Schwarz's Bayesian information criterion (SBIC) selection statistic for each country separately. I set $J_y = J_x $ equal to 1 as it is the optimal number of lags according to the SBIC statistic for all but one country, and I choose $J_i = 2$.\footnote{The exception is Indonesia, for which the SBIC statistic suggest the choice of 2 lags. As a robustness check, I computed the Hannan and Quinn information criterion statistic which also suggest using one lag for the vast majority of countries in my sample.} I also estimate the ``Pooled Specification'' using the ``Standard HFI'' by replacing $i^{MP}_t$ and $i^{ID}_t$ with the high-frequency surprises of the policy interest rate and the S\&P 500.\footnote{Note that this specification of the ``Standard HFI'' is in line with the analogous specification for the SVAR model in Section \ref{sec:main_results}.} The second specification, presented in Equation \ref{eq:LP_fixed_effects_date} includes country specific fixed effects given by $\Gamma_c$ and includes a linear time trend given by $\Gamma_{i,t}$. The final specification, in Equation \ref{eq:LP_fixed_effects_date} is consistent with the specification estimated by \cite{ilzetzki2021puzzling}. The standard errors are clustered at the country and time level, which control for the fact that all countries are hit by the FOMC shocks simultaneously.

Figures \ref{fig:LP_Pooled}, \ref{fig:LP_Panel} 
and \ref{fig:LP_Panel_Trend} in Appendix \ref{sec:appendix_additional_results} show that the findings in Section \ref{sec:main_results} are robust to estimating the impulse response functions using local projection techniques under the ``Pooled Specification'', with country fixed effects and with country fixed effects plus a time trend specification. First, by comparing the estimated dynamics on the first and second columns of these figures, it is clear that under a local projection methodology the two FOMC shocks lead to opposite spillovers on the rest of the world. On the one hand, a MP shock lead to the conventional results of a drop in industrial output, a persistent nominal exchange rate depreciation and tighter financial conditions shown by higher lending rates and a drop in the equity index. On the other hand, an ID shock leads to a nominal exchange rate appreciation, a persistent increase in the industrial production index and an increase in the equity index. Results are also quantitatively close to those presented in Section \ref{sec:main_results}. The last column of these figures shows that the estimated impulse response functions under the ``Standard HFI'' strategy lead to an average of the dynamics presented in the first two columns which yield atypical dynamics. These atypical dynamics, particularly the expansion of industrial output and the increase in the equity index, provide further evidence that the systematic disclosure of information around FOMC meetings may bias the identification and quantification of the impact of international spillovers of US monetary policy shocks. 

Additionally, I show that the main results are robust to relaxing the assumption of homogeneous dynamic coefficients across countries. An alternative  way to estimate panel VAR models is to use the ``mean group estimator'' presented in \cite{pesaran1995estimating}.\footnote{The authors show that in a standard maximum likelihood framework, this estimation technique yields consistent estimates.} Under this framework it is assumed that the $N$ countries of the model are characterised by heterogeneous dynamic coefficients, but that these coefficients are random processes sharing a common mean. Therefore, the parameters of interest are the average, or mean effects of the group. Carrying out the same assumption for the residual variance-covariance matrix, namely that it is heterogeneous across countries but is characterised by a common mean, then a single and homogeneous model is estimated for all the countries.\footnote{For greater detail see \cite{pesaran1995estimating}.}$^{,}$\footnote{Given that this model is estimated using non-Bayesian techniques, I change the number of lags of the endogenous variables to 1 to take into account the exponentially increasing number of coefficients and the limited number of observations.} Figure \ref{fig:MG_Benchmark} in Appendix \ref{sec:appendix_additional_results} shows the impulse response functions of the mean group estimator and compares them with the main results. Overall, the impulse response functions of both methods are quantitatively and qualitatively inline: MP and ID shocks lead to qualitatively opposite dynamics and the ``Standard HFI'' approach leads to the atypical dynamics an expansion in the industrial production and equity indexes. I interpret this result as supporting evidence of the main results presented in Section \ref{sec:main_results}.\footnote{Figure \ref{fig:MG} in Appendix \ref{sec:appendix_additional_results} presents the impulse response functions and the credibility intervals of the Mean Group estimator of the ``mean'' impact of the MP, ID and Standard HFI shocks. This figure provides additional evidence in support of the main results presented in Section \ref{sec:main_results}.}

\noindent
\textbf{Alternative shock ordering.} The benchmark specification sets a specific order between the two FOMC shocks inside the vector of variables $m_t$, with the MP shock ordered first and the ID shock ordered second. While by construction the two FOMC shocks are orthogonal to each other, introducing them into the SVAR model may lead to spurious correlations due to the finite sample and the imposition of ordering strategies. To test the robustness of the benchmark results I re-estimate the impulse response functions using two alternative specifications of vector $m_t$: (i) ordering the ID shock first and the MP shock second, and by estimating the impulse response functions introducing one FOMC shock at a time.\footnote{An additional test of the potential biases introduced by this identification strategy is to study the impulse response functions of the FOMC shock on each other. Figure \ref{fig:Benchmark_All_Variables} in Appendix \ref{sec:appendix_additional_results} presents the impulse response functions of the benchmark specification for the two FOMC shocks and the set of macroeconomic and financial variables. Under the benchmark ordering strategy, the impact of the FOMC structural shocks on the other shock is small (an order of magnitude smaller than the driving shock) and with uncertainty bounds containing the non-significant response.} Figures \ref{fig:MP_Reordered} and \ref{fig:CBI_Reordered} in Appendix \ref{sec:appendix_additional_results} show that the main results are robust both qualitatively and quantitatively to re-ordering the two FOMC shocks, i.e. $m_t = [i^{ID}_t,  i^{MP}_t]$; and to estimating the model by introducing these shocks one at a time, i.e. estimating the impulse response functions with only one shock at a time with $m_t = [i^{MP}_t]$ and $m_t = [i^{ID}_t]$, separately, one at a time.

\noindent
\textbf{Alternative approach to solving non-uniqueness in sign-restrictions.} The benchmark specification identifies the two FOMC shocks by choosing the median rotational angle that satisfies the sign restriction conditions. I show that the main results are robust to using different approaches to deal with the non-uniqueness problem of sign restrictions. 

First, I pin down the recovery of the two FOMC shocks by using an external moment condition as in \cite{jarocinski2022central}. I compute the ratio of the variance of the MP shock to the variance of the total high-frequency surprise of the interest rate around FOMC meetings when following the ``poor man's sign restriction'' approach. As presented in \cite{jarocinski2020deconstructing}, in the ``poor man's" approach each central bank announcement is classified as conveying a MP or ID shock depending on the sign of the co-movement of the high-frequency surprises.\footnote{For instance, if the co-movement between the interest rate and the S\&P 500 is negative, the announcement is classified as a MP shock and the ID shock is imputed a 0. If the co-movement between the interest rate and the S\&P 500 is positive, the announcement is classified as a ID shock and the MP shock is imputed a 0.} Following this approach the ratio of the variance MP shock to the variance of the high-frequency surprise of the interest rate is 0.88.\footnote{In \cite{jarocinski2022central} the author argues that following the ``poor man's'' monetary policy shock decomposition, a MP shock accounts for 88\% of the variance of the high frequency interest rate surprises. This external moment condition allows the author to pin down the decomposition of FOMC shocks. For further details see Appendix \ref{subsec:appendix_model_details_identification} or \cite{jarocinski2022central}.} Figure \ref{fig:Benchmark_j} in Appendix \ref{sec:appendix_additional_results} presents the impulse response functions using this alternative decomposition of MP and ID shocks. Overall, results are qualitatively in line with the main results presented in Section \ref{sec:main_results} with only minor quantitative differences.\footnote{On the one hand, under this alternative identification strategy the impact of an MP shock on industrial production is slightly greater in magnitude than the one using the median rotational angle. On the other hand, under this alternative identification strategy the impact of an ID shock on industrial production is smaller in magnitude than the one using the median rotational angle.}$^{,}$\footnote{Additionally, Figures \ref{fig:Benchmark_Adv_j} and \ref{fig:Benchmark_EM_j} show that the main results are also robust for the sub-samples of Advanced Economies and Emerging Markets, respectively.}

Second, I show that results are robust to using a uniform prior over the space of rotational angles that satisfy the sign restriction conditions (see \cite{rubio2010structural} and \cite{jarocinski2020deconstructing}). As argued in Section \ref{sec:data_methodology_identification}, the sign restriction conditions only provide set identification, i.e., conditionally on each draw of the VAR parameters there are multiple values of shocks and impulse responses that are consistent with the restrictions. When computing uncertainty bounds we take all these values into account weighting them according to the uniform prior on the rotation angles.\footnote{To compute the posterior draws of the impulse response functions we proceed as follows. First, as described by \cite{jarocinski2022central}, I normalize the continuum of rotational angles that yield acceptable decompositions of the FOMC shocks $\alpha \in (0,1)$. Next, I discretize the acceptable rotational angles to 99 percentiles, $p=0.01,0.02,\ldots,0.98,0.99$. I randomly choose percentile $p$ through a uniform prior and decompose the high-frequency surprises into MP and ID shocks. With the recovered FOMC shocks I estimate the impulse response functions following the same approach as described in Section \ref{sec:data_methodology_identification}, using 5,000 draws from the posterior distribution and discarding the first 500. This approach takes into account the set identification inherent to the sign restriction condition and builds on the SVAR methodology used in Section \ref{sec:main_results} to obtain the main results. For efficiency purposes, I estimate the impulse response functions over the 99 discretized acceptable rotational angles using 5,000 draws and discarding the first 500, leaving me with $4500 \times 99$ estimated impulse response functions. I put a uniform prior over all potential impulse response functions and subsequently take 10,000 random draws from it. Taking 10,000 random draws or plotting all $4500 \times 99$ draws yield similar results.} Figure \ref{fig:Sign_Restriction_90} in Appendix \ref{sec:appendix_additional_results} shows the resulting impulse response functions which are quantitatively and qualitatively in line with the results presented in Section \ref{sec:main_results}. Imposing a uniform prior over the space of acceptable rotations smooths out the median response and slightly increases the confidence intervals. Nevertheless, a MP and ID shock lead to opposite dynamics which are significantly different from each other and from zero. To provide intuition behind this result Figure \ref{fig:Sign_Restriction_Medians} in Appendix \ref{sec:appendix_additional_results} plots the median impulse response functions across the 99 different percentiles of acceptable rotational angles, $p=0.1,\ldots,0.99$. On the one hand, across all rotational angles, a MP shock causes an exchange rate depreciation and a drop in industrial production. On the other hand, across all rotational angles, a ID shock causes an exchange rate appreciation and an expansion of industrial production. However, the distribution of impulse response functions seem to be skewed. The top decile angles lead to dynamics close to the ones arising from the median angle than the bottom decile angles. All in all, considering alternative rotational angles which satisfy the sign restriction conditions yield results quantitatively and qualitatively in line with those presented in Section \ref{sec:main_results}.

\noindent
\textbf{Additional evidence from alternative identification strategies.} I provide supporting evidence of informational effects biasing the standard high-frequency approach by considering alternative identification strategies proposed by the literature. We consider the identification strategies proposed by \cite{miranda2021transmission} and \cite{bauer2022reassessment}. These alternative identification strategies purge the high-frequency surprises of interest rates around FOMC of information effects by orthogonalizing from variables in the information set of the FOMC and/or from the public.\footnote{I describe in detail this alternative identification strategies in Appendix \ref{sec:appendix_additional_results_strategies}.}

I begin by showing that estimating the international spillovers of US monetary policy using un-cleansed or un-orthogonalized shocks leads to the atypical dynamics presented in Section \ref{sec:main_results}. To do this I estimate the impulse response functions of the un-orthogonalized and orthogonalized shocks constructed by \cite{bauer2022reassessment}.\footnote{I compare the uncleansed and cleansed proposed by \cite{bauer2022reassessment} as it expands the benchmark standard high-frequency sample by considering the movements of interest rates around FOMC meetings and other events in which the chairman of the Federal Reserve presented speeches. The un-cleansed shocks used by \cite{miranda2021transmission} are analogous to those used in the main results of this paper in Section \ref{sec:main_results}.} Figure \ref{fig:BS_vs_MPS} in Appendix \ref{sec:appendix_additional_results_strategies} presents this comparison. On the left column, the un-orthogonalized shock leads to a atypical dynamics, such as a non-depreciation on impact of the exchange rate and a mild expansion of the industrial production and equity indexes. On the right column, the orthogonalized shocks lead to a significant depreciation on impact which is persistent across time, and a significant drop in both the industrial production and equity index. Thus, the un-cleansed shocks lead to dynamics similar to those arising from following the ``Standard HFI'' strategy used in Section \ref{sec:main_results}. Additionally, using a monetary policy shock purged of information effects leads to results similar to those estimated for the MP shock in Section \ref{sec:main_results}.

Lastly, I compare the impulse response functions of the MP shock in this paper, with those arising from monetary policy shocks proposed by these alternative strategies. Figure \ref{fig:Other_Information} in Appendix \ref{sec:appendix_additional_results_strategies} presents the result of this exercise. The first conclusion arising from this figure is that the atypical dynamics found in the literature are not present when controlling for informational effects around FOMC meetings. The second conclusion arising from this figure is that identification strategies that control for informational effects yield both quantitatively and qualitatively similar impulse response functions. I interpret these results as supporting evidence that pure monetary policy shocks still yield to the conventional results found by a previous literature and that not controlling for informational effects lead to biased estimates and atypical dynamics.\footnote{The key advantage of following the approach introduced by \cite{jarocinski2020deconstructing} is that exploiting the high-frequency movement of multiple financial assets around FOMC meetings allows to identify two FOMC shocks with distinct interpretations. As shown in Section \ref{sec:main_results}, changes in the Federal Reserve's interest rates driven by the disclosure of information can lead to quantitatively sizeable international spillovers. Additionally, \cite{jarocinski2022central} shows that this information disclosure shock is robust to the critiques presented by \cite{bauer2022reassessment}. }

\noindent
\textbf{Potential sources of quantitative heterogeneity.} In Appendix \ref{sec:appendix_heterogeneity} I describe two potential sources of quantitatively heterogeneity in the international impact of US interest rates: (i) countries' exchange rate regimes and (ii) countries' reliance on the export of commodity goods. I also briefly comment how previous literature has theorized that these country characteristics may influence the quantitative impact of the spillovers of US interest rates and describe my results. The results of Section \ref{sec:main_results} are present across all the different sub-samples. In terms of exchange rate regimes, countries with managed exchange rate regimes exhibit relatively smaller impact of FOMC shocks (see Figures \ref{fig:ERR_MP_Peru} to \ref{fig:ERR_CBI_PyI} in Appendix \ref{sec:appendix_heterogeneity}). In terms of the relative importance of commodity goods in export bundles, countries with a higher dependence of commodity goods exhibit smaller quantitative impacts of FOMC shocks (see \ref{fig:EMC_MP} and \ref{fig:EMC_CBI} in Appendix \ref{sec:appendix_heterogeneity}).

\section{Conclusion} \label{sec:conclusion}

In this paper I have argued that the international spillovers of a US interest rates hike depend critically on the underlying structural shock that causes the tightening. Using the identification strategy proposed by \cite{jarocinski2020deconstructing}, I deconstruct FOMC shocks into a pure US monetary policy (MP) shock and an information disclosure (ID) shock. Introducing these FOMC shocks into a SVAR model with both Advanced and Emerging Market economies I find that the two shocks lead to qualitatively opposite spillovers. On the one hand, a MP shock leads to a nominal exchange rate depreciation, a persistent drop in industrial output and overall tighter financial conditions. On the other hand, a ID shock leads to a short lived exchange rate appreciation, a persistent expansion in industrial output and overall looser financial conditions. 

Furthermore, I argue that following the standard high-frequency identification of US monetary policy leads to atypical spillover dynamics as it does not control for the systematic disclosure of information about the state of the US economy around FOMC announcements. I show that this identification strategy leads to impulse response functions which are an average of those resulting from the MP and ID shocks. In particular, under this identification strategy, a US interest rate tightening leads to a significant expansion of industrial production and equity indexes. Moreover, by not controlling for the disclosure of information, this identification strategy introduces an attenuation bias in terms of the quantitative impact US monetary policy shocks on the variables which do not show atypical dynamics.

Lastly, I show that the main results are robust to a battery of different model and sample specifications. Results hold when considering Advanced and Emerging Market economies separately, when estimating impulse response functions using local projection methodologies, and when estimating the model in sub-samples of countries with different exchange rate regimes and different exposures to commodity markets.

\newpage

\large
\noindent
\textbf{Acknowledgements:} \normalsize

\noindent
I owe an unsustainable debt of gratitude to Giorgio Primiceri for his guidance and advice. I would also like to thank the comments by two anonymous referees, Martin Eichenbaum, Matt Rognlie, Guido Lorenzoni and Larry Christiano. I would like to thank the Monetary and Fiscal History of Latin America project of the Becker Friedman Institute for their generous financial support. Particularly, I would like to thank Edward R. Allen, for supporting my research.  I would like to thank the attendees of Northwestern University's macro lunch seminar, Javier Garcia Cicco, Husnu Dalgic and Yong Cai. Susan Belles provided helpful comments.

\newpage
\bibliography{main.bib}

\newpage
\appendix

\section{Data Details} \label{sec:appendix_data_details}

In this section of the appendix I provide additional details on the construction of the sample used across the paper. The source of the macroeconomic and financial data used for the construction of the variables in the benchmark variable specification is the IMF's ``International Financial Statistics''.\footnote{To access the IMF's IFS datasets go to \url{https://data.imf.org/?sk=4c514d48-b6ba-49ed-8ab9-52b0c1a0179b}.}

First, the benchmark specification is comprised of five variables:
\begin{enumerate}
    \item Nominal Exchange Rate
    
    \item Industrial Production index
    
    \item Consumer Price Index
    
    \item Lending Rate

    \item Equity Index
\end{enumerate}
Next, I present additional details for the construction of each of the variables
\begin{itemize}
    \item \underline{Nominal Exchange Rate:} The variable's full name at the IMF IFS data set is ``Exchange Rates, National Currency Per U.S. Dollar, Period Average, Rate''. 
    
    \item \underline{Industrial Production Index:} In order to construct countries' ``Industrial Production Index'' I rely on three variables of the IMF IFS' dataset:
    
    \begin{itemize}
        \item Economic Activity, Industrial Production, Index
        
        \item Economic Activity, Industrial Production, Seasonally Adjusted, Index
        
        \item Economic Activity, Industrial Production, Manufacturing, Index
    \end{itemize}
    Ideally, I would  construct the variable ``Industrial Production Index'' by choosing only one of the variables mentioned above. However, this is impossible as countries do not report to the IMF all three of these variables for our time sample, rey 2004 to December 2016. For instance, Peru provides neither the ``Economic Activity, Industrial Production, Index'' nor the ``Economic Activity, Industrial Production, Seasonally Adjusted, Index'', but does provide the ``Economic Activity, Industrial Production, Manufacturing, Index''. Visiting Peru's Central Bank statistics website, there is no ``Industrial Production Index'', but there is an ``Industrial Production, Manufacturing Index'', which coincides with the variable reported as ``Economic Activity, Industrial Production, Manufacturing, Index'' to the IMF.
    
    In order to deal with this, I establish the following priority between the three IMF IFS variables: (i) ``Economic Activity, Industrial Production, Seasonally Adjusted, Index'' (ii) ``Economic Activity, Industrial Production, Index'', (iii) ``Economic Activity, Industrial Production, Manufacturing, Index''. Table \ref{tab:data_details_industrial} below presents the IMF IFS variable used for each country.
    \begin{table}[ht]
        \centering
        \caption{Construction of Industrial Production Index}
        \footnotesize
        \label{tab:data_details_industrial}
        \begin{tabular}{l l}
\multicolumn{2}{c}{Emerging Markets} \\ \hline \hline
Brazil 	&	Economic Activity, Industrial Production, Seasonally Adjusted, Index	\\
Chile	&	Economic Activity, Industrial Production, Seasonally Adjusted, Index	\\
Colombia	&	Economic Activity, Industrial Production, Seasonally Adjusted, Index	\\
Hungary	&	Economic Activity, Industrial Production, Seasonally Adjusted, Index	\\
Indonesia	&	Economic Activity, Industrial Production, Manufacturing, Index	\\
Mexico	&	Economic Activity, Industrial Production, Seasonally Adjusted, Index	\\
Peru	&	Economic Activity, Industrial Production, Manufacturing, Index	\\
Philippines	&	Economic Activity, Industrial Production, Manufacturing, Index	\\
South Africa	&	Economic Activity, Industrial Production, Manufacturing, Index	\\
			\\
\multicolumn{2}{c}{Advanced Economies} \\ \hline \hline
Australia	&	Economic Activity, Industrial Production, Seasonally Adjusted, Index	\\
Canada	&	Economic Activity, Industrial Production, Seasonally Adjusted, Index	\\
France	&	Economic Activity, Industrial Production, Seasonally Adjusted, Index	\\
Iceland	&	Economic Activity, Industrial Production, Seasonally Adjusted, Index	\\
Italy	&	Economic Activity, Industrial Production, Seasonally Adjusted, Index	\\
Japan	&	Economic Activity, Industrial Production, Seasonally Adjusted, Index	\\
South Korea	&	Economic Activity, Industrial Production, Seasonally Adjusted, Index	\\
The Netherlands	&	Economic Activity, Industrial Production, Seasonally Adjusted, Index	\\
Sweden	&	Economic Activity, Industrial Production, Seasonally Adjusted, Index	\\
        \end{tabular}
    \end{table}
I believe that every one of the three variables considered reflects the actual industrial production index. From Table \ref{tab:data_details_industrial}, it is clear that when ``Economic Activity, Industrial Production, Seasonally Adjusted, Index'' is not available, the non-seasonally adjusted is also not available. Also, the fact that results are significant for Emerging Market Economies and Advanced Economies separately, see Figures \ref{fig:Benchmark_EM} and \ref{fig:Benchmark_Adv} respectively, suggest that the paper's main results are not driven by the choice of variable. This conclusion also follows from the fact that the paper's main results are robust to the several sub-sample exercises carried out in Section \ref{sec:robustness_checks_additional_results}.

\item \underline{Consumer Price Index:} Data for all countries except Australia is constructed using the variable ``Prices, Consumer Price Index, All items, Index'' from IMF IFS data set. Australia does not report a monthly CPI series to the IMF-IFS data set. Furthermore, the Australian Bureau of Statistics provides only quarterly data on their consumer price index.\footnote{See \url{https://www.abs.gov.au/statistics/economy/price-indexes-and-inflation/consumer-price-index-australia/jun-2022}.} Thus, for the case of Australia I proxy the monthly consumer price index by using the ``Prices, Producer Price Index, All Commodities, Index''. Once again, given that the paper's main results are robust to the different exercises that partition the sample, I believe that using this proxy variable does not guide any of the of the results presented in the paper. 

\item \underline{Lending Rate:} For all countries except France, the Netherlands and Sweden I use the IMF-IFS' ``Monetary and Financial Accounts, Interest Rates, Other Depository Corporations Rates, Lending Rates, Lending Rate, Percent per Annum'' variable. For France, the Netherlands and Sweden I use the OECD Stats, ``Monthly Monetary and Financial Statistics (MEI)'' dataset variable ``Short-term interest rates, Per cent per annum''.

\item \underline{Equity Index:} In order to construct countries' ``Equity Index'' I rely on two variables of the IMF IFS' dataset:
    
    \begin{itemize}
        \item Monetary and Financial Accounts, Financial Market Prices, Equities, Index
        
        \item Monetary and Financial Accounts, Financial Market Prices, Equities, End of Period, Index
    \end{itemize}
    
    I establish a priority: (i) ``Monetary and Financial Accounts, Financial Market Prices, Equities, Index'' (ii) ``Monetary and Financial Accounts, Financial Market Prices, Equities, End of Period, Index''. Again, the data coverage is not complete for all countries for the full sample period of January 2004 to December 2016. Table \ref{tab:data_details_equity} presents index used for every country.
    \begin{table}[ht]
        \centering
        \caption{Construction of Equity Index}
        \footnotesize
        \label{tab:data_details_equity}
        \begin{tabular}{l l}
\multicolumn{2}{c}{Emerging Markets} \\ \hline \hline
Brazil 	&	Equities, End of Period, Index	\\
Chile	&	Equities, Index	\\
Colombia	&	Equities, End of Period, Index	\\
Indonesia	&	Equities, Index	\\
Hungary 	&	Equities, Index	\\
Mexico	&	Equities, End of Period, Index	\\
Peru	&	Equities, End of Period, Index	\\
Philippines	&	Equities, Index	\\
South Africa	&	Equities, Index	\\
			\\
\multicolumn{2}{c}{Advanced Economies} \\ \hline \hline
Australia	&	Equities, End of Period, Index	\\
Canada	&	Equities, Index	\\
France &	Equities, Index	\\
Iceland &	Equities, Index	\\
Italy &	Equities, Index	\\
Japan	&	Equities, Index	\\
South Korea	&	Equities, Index	\\
the Netherlands	&	Equities, Index	\\
Sweden	&	Equities, Index	\\
	\\
        \end{tabular}
    \end{table}
\end{itemize}

\newpage
\section{Model Details} \label{sec:appendix_model_details}

In this section of the appendix I provide additional details on the panel SVAR model and the identification strategy described in Section \ref{sec:data_methodology_identification}. Section \ref{subsec:appendix_model_details_panel_svar_model} presents details on the panel SVAR model while Section \ref{subsec:appendix_model_details_identification} presents additional technical details on the recovery of structural shocks. 

\subsection{Panel SVAR Model} \label{subsec:appendix_model_details_panel_svar_model}

In this section of the appendix I provide additional details on the estimation of the Structural VAR model presented in Section \ref{sec:data_methodology_identification}. In its most general form, a panel SVAR model comprises of N countries or units, $n$ endogenous variables, $p$ lagged values and $T$ time periods. The pooled panel SVAR model can be written as
\begin{align} \label{eq:pooled_estimator}
\begin{pmatrix}
y_{1,t} \\
y_{2,t} \\
\vdots  \\
y_{N,t}
\end{pmatrix}
&=C+
\begin{pmatrix}
A^1 \quad 0 \quad \cdots \quad 0 \\
0 \quad  A^1 \quad \cdots \quad 0 \\
\vdots \quad \vdots \quad \ddots \quad \vdots \\
0 \quad 0 \quad \cdots \quad A^1 
\end{pmatrix}
\begin{pmatrix}
y_{1,t-1} \\
y_{2,t-1} \\
\vdots  \\
y_{N,t-1}
\end{pmatrix}
+ \cdots \nonumber \\
\\
&+ \nonumber
\begin{pmatrix}
A^p \quad 0 \quad \cdots \quad 0 \\
0 \quad  A^p \quad \cdots \quad 0 \\
\vdots \quad \vdots \quad \ddots \quad \vdots \\
0 \quad 0 \quad \cdots \quad A^p 
\end{pmatrix}
\begin{pmatrix}
y_{1,t-p} \\
y_{2,t-p} \\
\vdots  \\
y_{N,t-p}
\end{pmatrix}
+
\begin{pmatrix}
\epsilon_{1,t} \\
\epsilon_{2,t} \\
\vdots \\
\epsilon_{N,t}
\end{pmatrix}
\end{align}
where $y_{i,t}$ denotes an $n \times 1$ vector of $n$ endogenous variables of country $i$ at time $t$ and $A^{j}$ is an $n \times n$ matrix of coefficients providing the response of country $i$ to the $j^{th}$ lag at period $t$. Note that by assuming that $A^j_1 = A^j_n = A^j$ for $j=1,\ldots,n$ implies the assumption that the estimated coefficients are common across countries. $C$ is a $Nn\times1$ vector of constant terms which are also assumed to be common across countries. Lastly, $\epsilon_{i,t}$ is an $n \times 1$ vector of residuals for the variables of country $i$, such that
\begin{align*}
    \epsilon_{i,t} \sim \mathcal{N}\left(0,\Sigma_{ii,t}\right)
\end{align*}
with 
\begin{align*}
    \epsilon_{ii,t} &= \mathbb{E} \left(\epsilon_{i,t} \epsilon_{i,t}' \right) = \Sigma_c \quad \forall i \\
    \epsilon_{ij,t} &= \mathbb{E} \left(\epsilon_{i,t} \epsilon_{j,t}' \right) = 0 \quad \text{for } i \neq j
\end{align*}
The last two equations imply that, as for the model's auto-regressive coefficients, the innovation's variance is equal across countries. 

The model described by Equation \ref{eq:pooled_estimator} is estimated using Bayesian methods. In order to carry out the estimation of this model I first re-write the model. In particular, the model can be reformulated in compact form as

\begin{align} \label{eq:model_compact_form}
\underbrace{\begin{pmatrix}
y_{1,t}' \\
y_{2,t}' \\
\vdots  \\
y_{N,t}'
\end{pmatrix}}_{Y_t, \quad N \times n}
&=
\underbrace{\begin{pmatrix}
y_{1,t-1}' \ldots y_{1,t-p}' \\
y_{2,t-1}' \ldots y_{2,t-p}' \\
\vdots  \ddots \vdots \\
y_{N,t-1}' \ldots y_{N,t-p}'
\end{pmatrix}}_{\mathcal{B}, \quad N \times np}
\underbrace{\begin{pmatrix}
\left(A^{1}\right)' \\
\left(A^{2}\right)' \\
\vdots \\
\left(A^{N}\right)'
\end{pmatrix}}_{X_t, \quad np \times n}
+
\underbrace{\begin{pmatrix}
\epsilon_{1,t}' \\
\epsilon_{2,t}' \\
\vdots \\
\epsilon_{N,t}'
\end{pmatrix}}_{\mathcal{E}_t, \quad N \times n}
\end{align}
or
\begin{align}
    Y_t = X_t \mathcal{B} + \mathcal{E}_t
\end{align}
Even more, the model can be written in vectorised form by stacking over the $T$ time periods 
\begin{align}
    \underbrace{vec\left(Y\right)}_{NnT \times 1} = \underbrace{\left(I_n \otimes X \right)}_{NnT \times n np} \quad \underbrace{vec\left(\mathcal{B}\right)}_{n np \times 1} \quad + \quad \underbrace{vec\left(\mathcal{E}\right)}_{NnT \times 1}
\end{align}
or
\begin{align}
    y = \Bar{X} \beta + \epsilon
\end{align}
where $\epsilon \sim \mathcal{N}\left(0, \Bar{\Sigma}\right)$, with $\Bar{\Sigma} = \Sigma_c \otimes I_{NT}$.

The model described above is just a conventional VAR model. Thus, the traditional Normal-Wishart identification strategy is carried out to estimate it. The likelihood function is given by
\begin{align}
    f\left(y | \Bar{X} \right) \propto |\Bar{\Sigma}|^{-\frac{1}{2}} \exp \left(-\frac{1}{2} \left(y - \Bar{X}\beta\right)' \Bar{\Sigma}^{-1} \left(y - \Bar{X}\beta\right) \right)
\end{align}
As for the Normal-Wishart, the prior of $\beta$ is assumed to be multivariate normal and the prior for $\Sigma_c$ is inverse Wishart. For further details, see \cite{dieppe2016bear}. All of the panel SVAR model computations are carried out using the BEAR Toolbox version 5.1.

\subsection{Identification Strategy \& Shock Recovery} \label{subsec:appendix_model_details_identification}

In this section of the appendix, I present additional details on the sign-restriction identification strategy presented in Section \ref{sec:data_methodology_identification} which recovers the two FOMC structural shocks. As stated in Section \ref{sec:data_methodology_identification}, this identification strategy follows \cite{jarocinski2020deconstructing} and \cite{jarocinski2022central}.

The identification strategy introduced by \cite{jarocinski2022central} exploit the high-frequency surprises of multiple financial instruments to recover two distinct FOMC shocks: a pure monetary policy (MP) shock and information disclosure (ID) shock. In particular, the authors impose sign restrictions conditions on the co-movement of the high-frequency surprises of interest rates and the S\&P 500 around FOMC meetings. This co-movement is informative as standard theory unambiguously predicts that a monetary policy tightening shock should lead to lower stock market valuation. This is because a monetary policy tightening decreases the present value of future dividends by increasing the discount rate and by deteriorating present and future firm's profits and dividends. Thus, MP shocks are identified as those innovations that produce a negative co-movement between these high-frequency financial variables. On the contrary, innovations generating a positive co-movement between interest rates and the S\&P 500 correspond to ID shocks. 

\begin{figure}[ht]
    \centering
    \caption{Scatter Plot of Interest Rate \& S\&P 500 Surprises around FOMC Meetings}
    \label{fig:HF_total}
     \centering
     \begin{subfigure}[b]{0.495\textwidth}
         \centering
         \includegraphics[width=\textwidth]{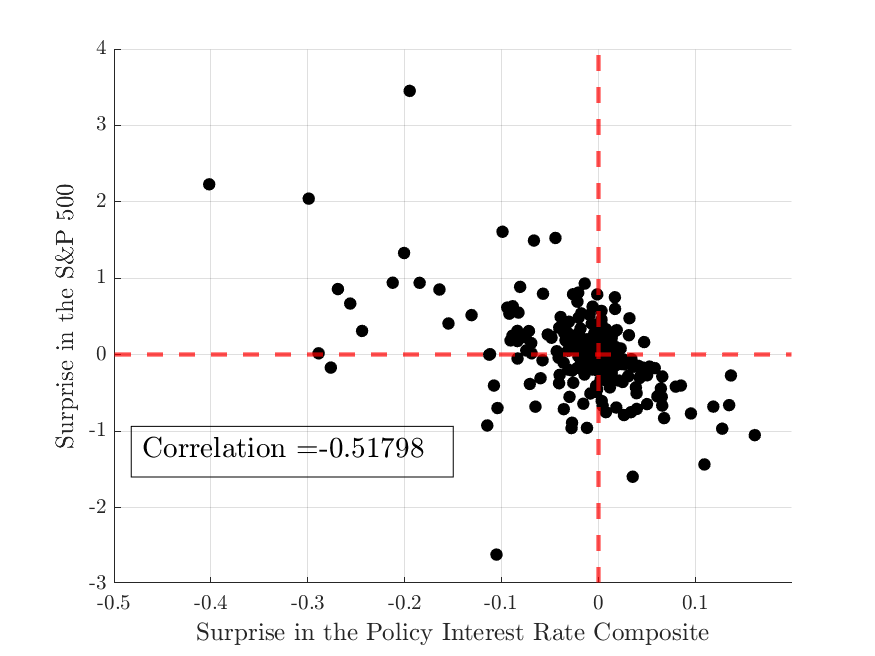}
         \caption{1990 - 2018}
         \label{fig:scatter_full_sample}
     \end{subfigure}
     \hfill
     \begin{subfigure}[b]{0.495\textwidth}
         \centering
         \includegraphics[width=\textwidth]{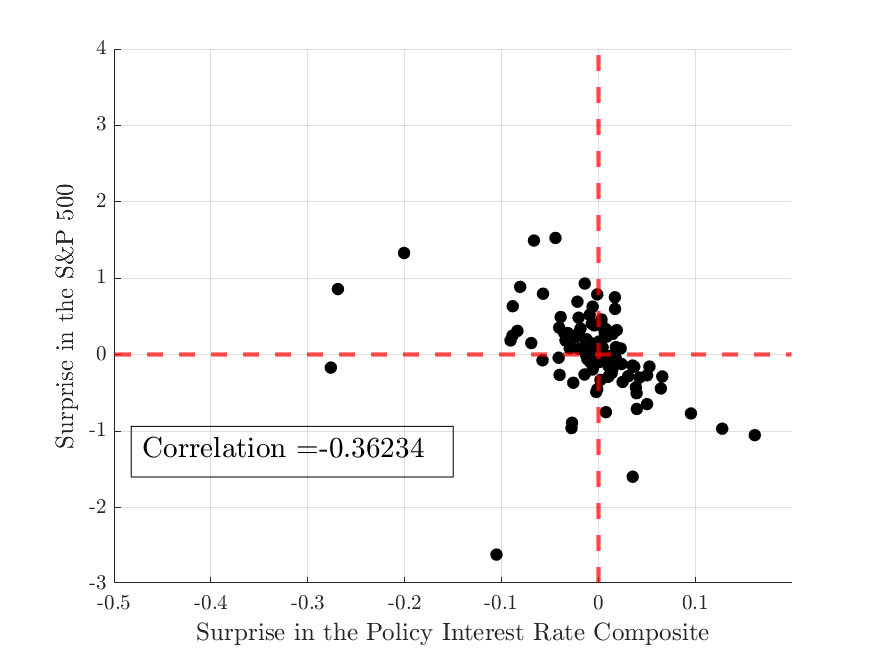}
         \caption{2004 - 2016}
         \label{fig:scatter_my_sample}
     \end{subfigure} 
     \floatfoot{\textbf{Note:} The left panel presents data for the period 1990-2018, in line with the sample constructed by \cite{jarocinski2022central}. The right panel presents data for the period 2004-2016, the sample for the empirical exercises in Section \ref{sec:main_results}. The high-frequency changes of the Policy Interest Rate Composite and the S\&P 500 are computed using 30-minute windows around FOMC meetings. Each black filled circle represents a different FOMC meeting.}
\end{figure}

Figure \ref{fig:HF_total} presents a scatter plot of the high-frequency surprises of interest rates and the S\&P 500 around FOMC meetings. 
The co-movement of these high-frequency surprises is remarkably different across FOMC meetings. Although a majority of FOMC meetings exhibit a negative co-movement between the interest rates and S\&P 500 surprises, a significant share of observations exhibit a positive co-movement. The authors' way to account for the positive co-movement is to attribute it to a shock that occurs systematically at the same time the FOMC announces its policy decisions, but that is different from a standard monetary policy shock. In particular, this additional shock is the disclosure of the FOMC's information about the present and future state of the US economy. Hence, by combining both the high-frequency surprises and imposing sign-restriction in their co-movements, the authors separately identify two structural FOMC shocks: a pure US monetary policy (MP) shock (which exhibits a negative co-movement between the interest rates and S\&P high-frequency surprises) and an information disclosure (ID) shock (which exhibits a positive co-movement between the interest rates and S\&P high-frequency surprises).

The high frequency surprise in the policy interest rate, $i^{Total}$, can be decomposed as
\begin{align}
    i^{\text{Total}} = i^{\text{MP}} + i^{\text{ID}}
\end{align}
where $i^{\text{MP}}$ is negatively correlated with the high frequency surprise of the $S\&P 500$ ``$s$'', and $i^{\text{ID}}$ is positively correlated with the ``$s$''. As shown by \cite{jarocinski2022central}, the sign restriction recovery of the structural shocks must satisfy the following decomposition
\begin{align}
    M = UC 
\end{align}
where $U'U$ is a diagonal matrix, $C$ takes the form of
\begin{align} \label{eq:matrix_rotation_appendix}
C =
\begin{pmatrix}
    1 & c^{MP}<0 \\
    1 & c^{ID}>0 \\
\end{pmatrix}
\end{align}
where $M=(i^{Total},s)$ is a $T \times 2$ matrix with $i^{Total}$ in the first column, $s$ in the second; $U = (i^{\text{MP}},i^{\text{ID}})$ is a $T \times 2$ matrix with $i^{\text{MP}}$ in the first column and $i^{\text{ID}}$ in the second column; and $T$ denoting the time length of the sample. By construction, $i^{\text{MP}}$ and $i^{\text{ID}}$ are mutually orthogonal. Matrix $C$ captures how $i^{\text{MP}}$ and $i^{\text{ID}}$ translates into financial market surprises.

The decomposition in \ref{eq:matrix_rotation_appendix} is not unique. In terms of \cite{jarocinski2022central} there is a range of rotations of matrices $U$ and $C$ that satisfy the sign restrictions $c^{MP}<0$ and $c^{ID}>0$.

The matrices $U$ and $C$ are computed as
\begin{align}
    U &= QPD \\
    C &= D^{-1} P' R
\end{align}
where the matrices $Q,P,D,R$ are obtained in three steps. 

\begin{itemize}
    \item[1.] Decompose matrix $M = UC$ into two orthogonal components using a QR decomposition such that
\end{itemize}
\begin{align}
    M &= QR \\
    Q'Q &=\begin{pmatrix}
        1 & 0 \\
        0 & 1 \\        
    \end{pmatrix} \\
    R&=\begin{pmatrix}
        r_{1,1}>0 & r_{1,2} \\
        0 & r_{2,2}>0 \\        
    \end{pmatrix}
\end{align}

\begin{itemize}
    \item [2.] Rotate these orthogonal components using the rotation matrix
    \begin{align}
    P &= \begin{pmatrix}
       \cos \left(\alpha\right) & \sin \left(\alpha\right) \\
       - \sin \left(\alpha\right) & \cos \left(\alpha\right)
    \end{pmatrix}
    \end{align}
    To satisfy the sign restrictions use any angle $\alpha$ in the following range
    \begin{align*}
        & \alpha \in \left( \left(1-w\right) \times \arctan \frac{r_{1,2}}{r_{2,2}} , \frac{w \times \pi}{2} \right) \qquad \text{if } r_{1,2} >0 \\
        & \alpha \in \left( 0 , w \times \arctan \frac{-r_{2,2}}{r_{1,2}} \right) \qquad \text{if } r_{1,2} \leq 0 \\
    \end{align*}
    where $w$ is weight, between 0 and 1, scaling the rotation angle. Setting $w = 0.5$ implies the median rotation angle, assumption used under the benchmark specification
\end{itemize}
\begin{itemize}
    \item[3.] Re-scale the resulting orthogonal components with a diagonal matrix $D$ to ensure that they add up to the interest rate surprises $i^{\text{Total}}$. It is straightforward to show that
    \begin{align}
        D = \begin{pmatrix}
            r_{1,1} \cos \left(\alpha\right) & 0 \\
            0 & r_{1,1} \sin \left(\alpha\right)
        \end{pmatrix}
    \end{align}
\end{itemize}

\begin{figure}[ht]
         \centering
         \includegraphics[width=10cm, height=6cm]{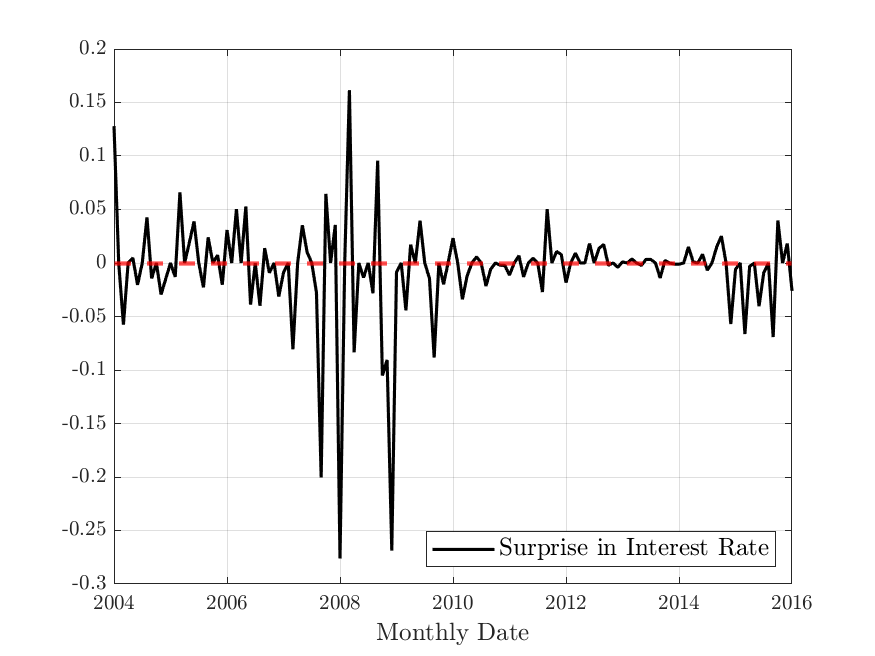}
         \caption{High-Frequency Financial Surprises around FOMC Meetings}
         \label{fig:FF4_shocks}
\end{figure}

\begin{figure}[ht]
         \centering
         \includegraphics[width=10cm, height=6cm]{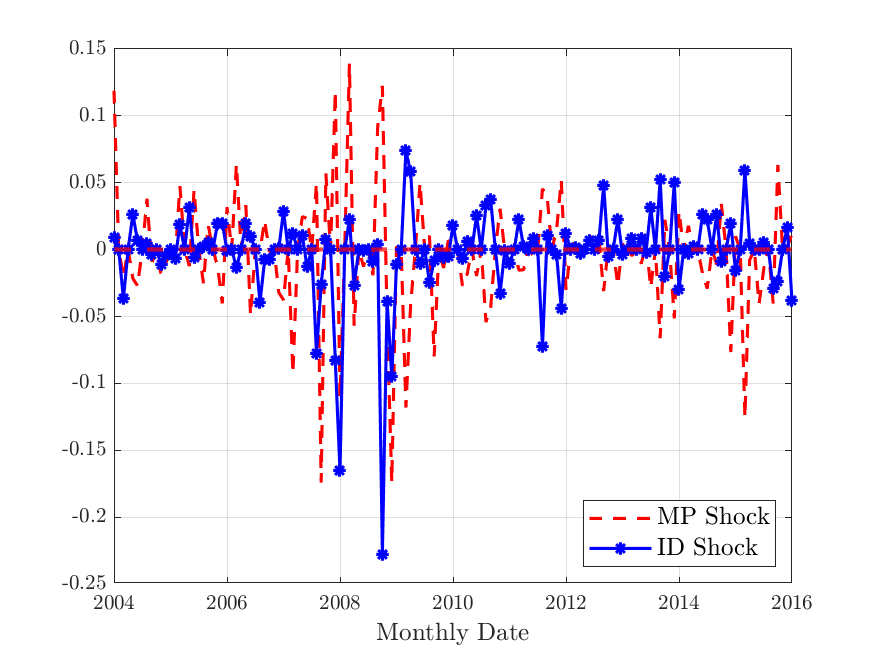}
         \caption{Structural FOMC Shocks}
         \label{fig:JK_Shocks}
\end{figure}

Figures \ref{fig:FF4_shocks} and \ref{fig:JK_Shocks} present the time series of the high frequency surprise of the interest rate  and of the two recovered structural for my benchmark specification, respectively.

The first robustness check in Section \ref{sec:robustness_checks_additional_results} which proposes an alternative approach to deal with the non-uniqueness problem of sign-restriction identification follows the approach of \cite{jarocinski2022central}. In \cite{jarocinski2022central}, the angle of rotation $\alpha$ of matrix $P$ is pinned down following these steps
\begin{itemize}
    \item[1.] Construct ``poor man's sign restrictions'' shocks such that
    
    \begin{align*}
        i^{\text{Total}} &= i^{MP} \quad \& \quad i^{\text{ID}} = 0 \qquad \text{if } i^{Total} \times s \leq 0 \\
        i^{\text{Total}} &= i^{ID} \quad \& \quad i^{\text{MP}} = 0 \qquad \text{if } i^{Total} \times s > 0 \\
    \end{align*}
\end{itemize}
In the data, \cite{jarocinski2022central} finds that the poor man's monetary policy shocks account for 88\% of the variance of the Federal Reserve's total interest rate surprises, i.e,
\begin{align*}
    \frac{var \left(i^{\text{MP}}\right)}{var \left(i^{\text{Total}}\right)} = 0.88
\end{align*}
To pin down the decomposition,  \cite{jarocinski2022central} impose that, as in the ``poor man's sign restrictions'' case, $var \left(i^{\text{MP}}\right)/ var \left(i^{\text{MP}}\right) = 0.88$. Additionally, \cite{jarocinski2022central} shows that the angle $\alpha$ can meets this condition can be recovered as 
\begin{align}
    \alpha = \arccos \sqrt{\frac{var \left(i^{\text{MP}}\right)}{var \left(i^{\text{Total}}\right)}}
\end{align}
In particular, the $\alpha$ that meets this condition is set equal to $\alpha = 0.8702$.

In Section \ref{sec:robustness_checks_additional_results} I carry out several robustness checks that considering different rotational angles.

\newpage
\section{Additional Results} \label{sec:appendix_additional_results}

\begin{figure}[ht]
         \centering
         \caption{Local Projection Impulse Response Functions \\ \footnotesize  Pooled Specification}
         \includegraphics[scale=0.325]{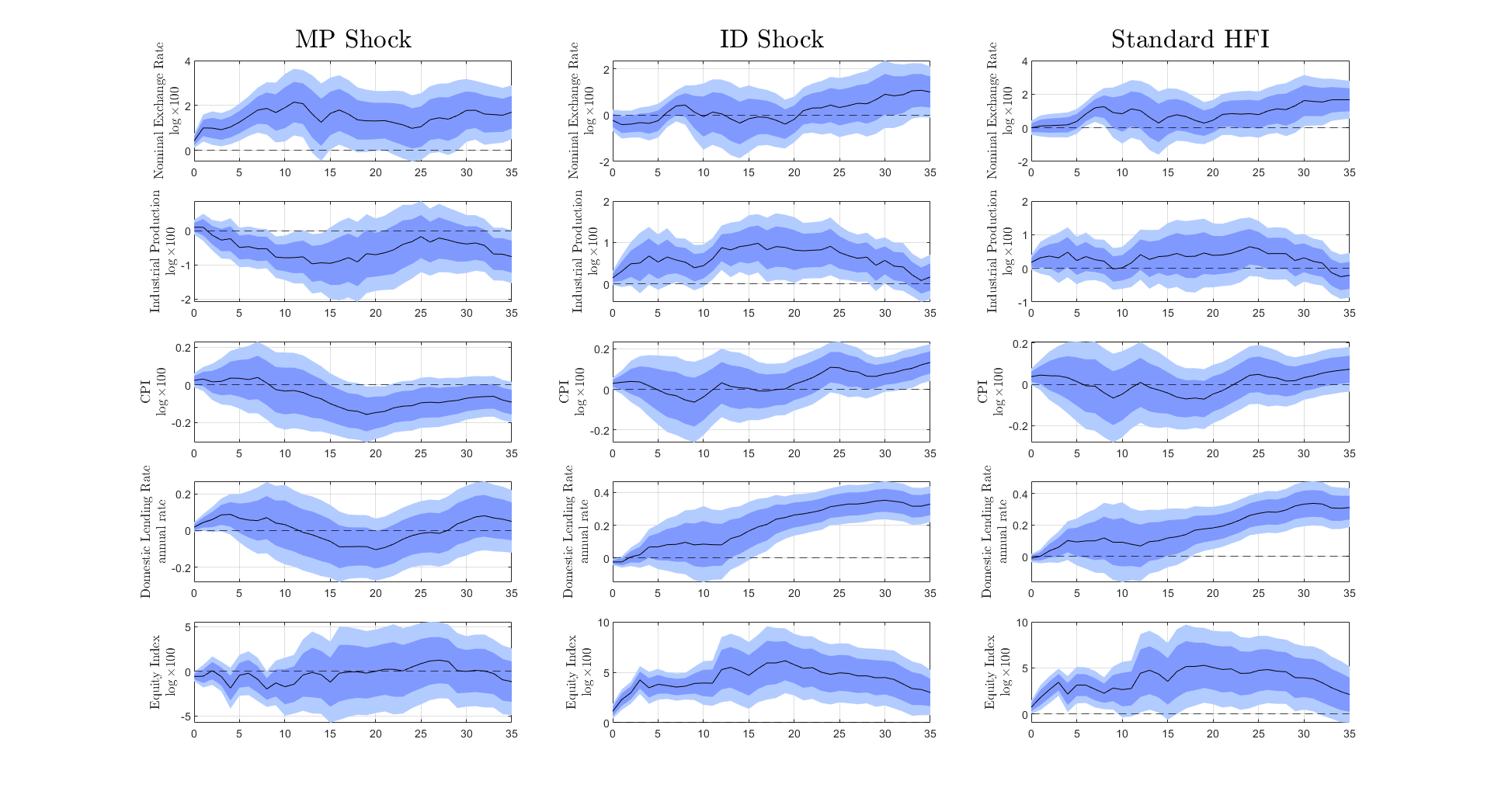}
         \label{fig:LP_Pooled}
         \floatfoot{\scriptsize \textbf{Note:} The black solid line represents the point estimate of coefficient for either one of the two FOMC shocks, or the high-frequency surprise of the policy interest rate composite. The dark shaded area represents within +/- one standard deviation, and the light shaded are represents the area within 1.65 standard deviations from the point estimate. The figure is comprised of 15 sub-figures ordered in five rows and three columns. Every row represents a different variable: (i) nominal exchange rate, (ii) industrial production index, (iii) consumer price index, (iv) lending rate, (v) equity index. The first column presents the results for the MP or ``Pure US Monetary Policy'' shock, the middle column presents the results for the ID or ``Information Disclosure'' shock, and the last column presents the results for the interest rate composite high frequency surprise or ``Standard HFI''. In the text, when referring to Panel $(i,j)$, $i$ refers to the row and $j$ to the column of the figure. For each shock, I standardize the coefficients such that they represent the dynamics of a one-standard deviation shock.}
\end{figure}

\begin{figure}[ht]
         \centering
         \caption{Local Projection Impulse Response Functions \\ \footnotesize  F.E. Specification}
         \includegraphics[scale=0.325]{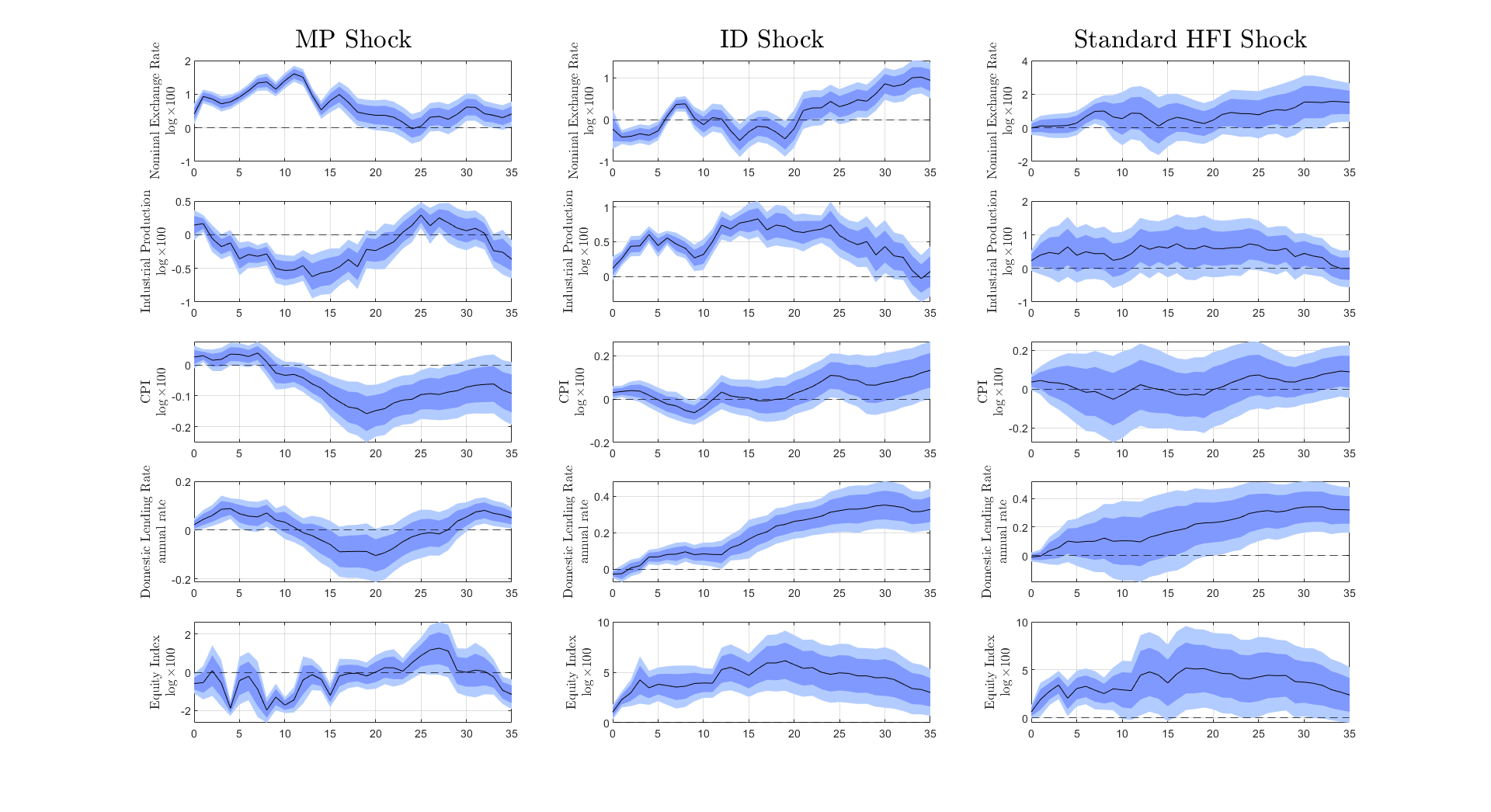}
         \label{fig:LP_Panel}
         \floatfoot{\scriptsize \textbf{Note:} The black solid line represents the point estimate of coefficient for either one of the two FOMC shocks, or the high-frequency surprise of the policy interest rate composite. The dark shaded area represents within +/- one standard deviation, and the light shaded are represents the area within 1.65 standard deviations from the point estimate. The figure is comprised of 15 sub-figures ordered in five rows and three columns. Every row represents a different variable: (i) nominal exchange rate, (ii) industrial production index, (iii) consumer price index, (iv) lending rate, (v) equity index. The first column presents the results for the MP or ``Pure US Monetary Policy'' shock, the middle column presents the results for the ID or ``Information Disclosure'' shock, and the last column presents the results for the interest rate composite high frequency surprise or ``Standard HFI''. In the text, when referring to Panel $(i,j)$, $i$ refers to the row and $j$ to the column of the figure. For each shock, I standardize the coefficients such that they represent the dynamics of a one-standard deviation shock.}
\end{figure}

\begin{figure}[ht]
         \centering
         \caption{Local Projection Impulse Response Functions \\ \footnotesize  F.E. + Time Trend Specification}
         \includegraphics[scale=0.325]{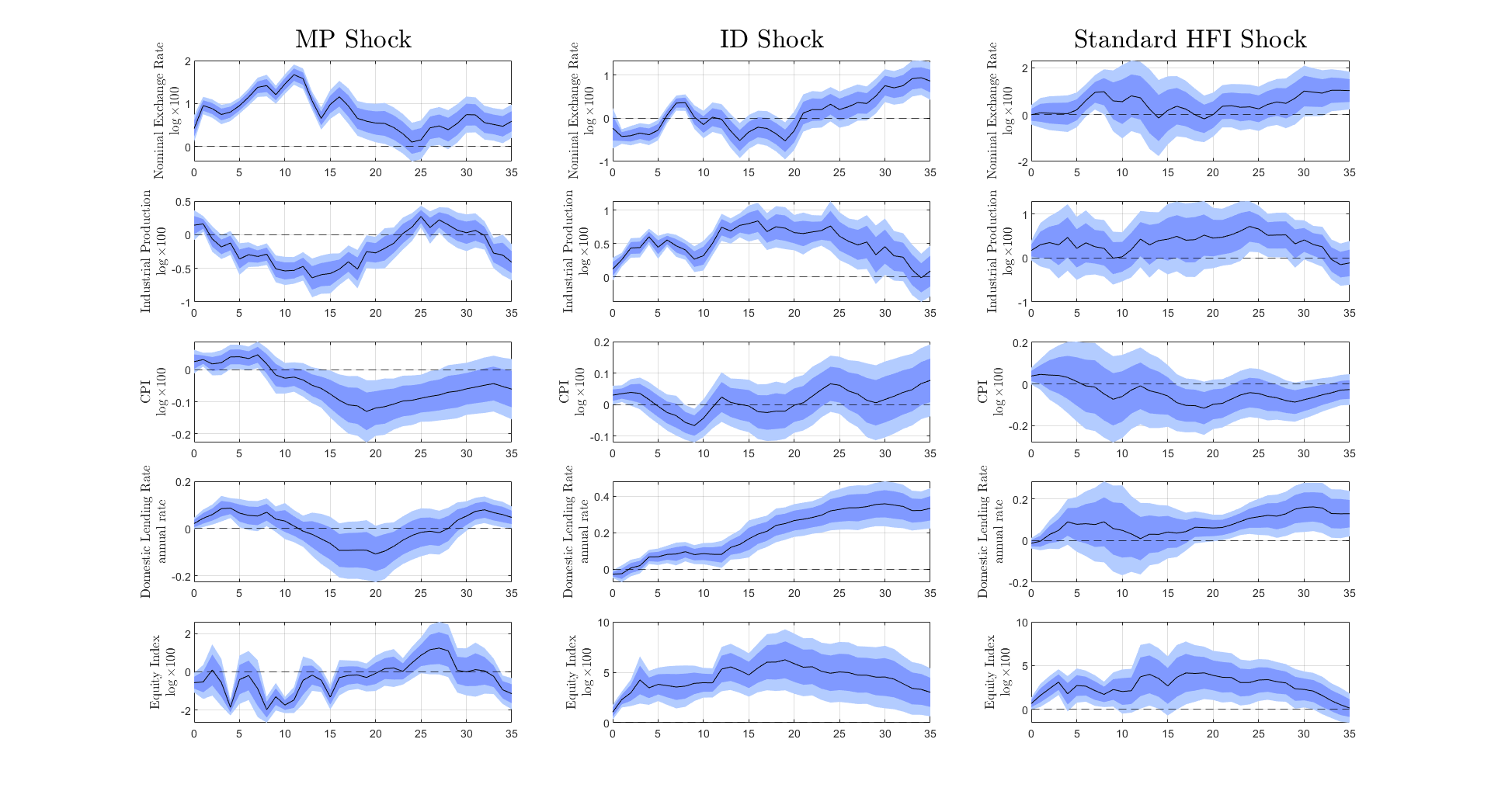}
         \label{fig:LP_Panel_Trend}
         \floatfoot{\scriptsize \textbf{Note:} The black solid line represents the point estimate of coefficient for either one of the two FOMC shocks, or the high-frequency surprise of the policy interest rate composite. The dark shaded area represents within +/- one standard deviation, and the light shaded are represents the area within 1.65 standard deviations from the point estimate. The figure is comprised of 15 sub-figures ordered in five rows and three columns. Every row represents a different variable: (i) nominal exchange rate, (ii) industrial production index, (iii) consumer price index, (iv) lending rate, (v) equity index. The first column presents the results for the MP or ``Pure US Monetary Policy'' shock, the middle column presents the results for the ID or ``Information Disclosure'' shock, and the last column presents the results for the interest rate composite high frequency surprise or ``Standard HFI''. In the text, when referring to Panel $(i,j)$, $i$ refers to the row and $j$ to the column of the figure. For each shock, I standardize the coefficients such that they represent the dynamics of a one-standard deviation shock.}
\end{figure}

\begin{figure}[ht]
         \centering
    \caption{Impulse Response to One-Standard-Deviation Shock \\ \footnotesize  Mean Group Estimator}
         \includegraphics[scale=0.325]{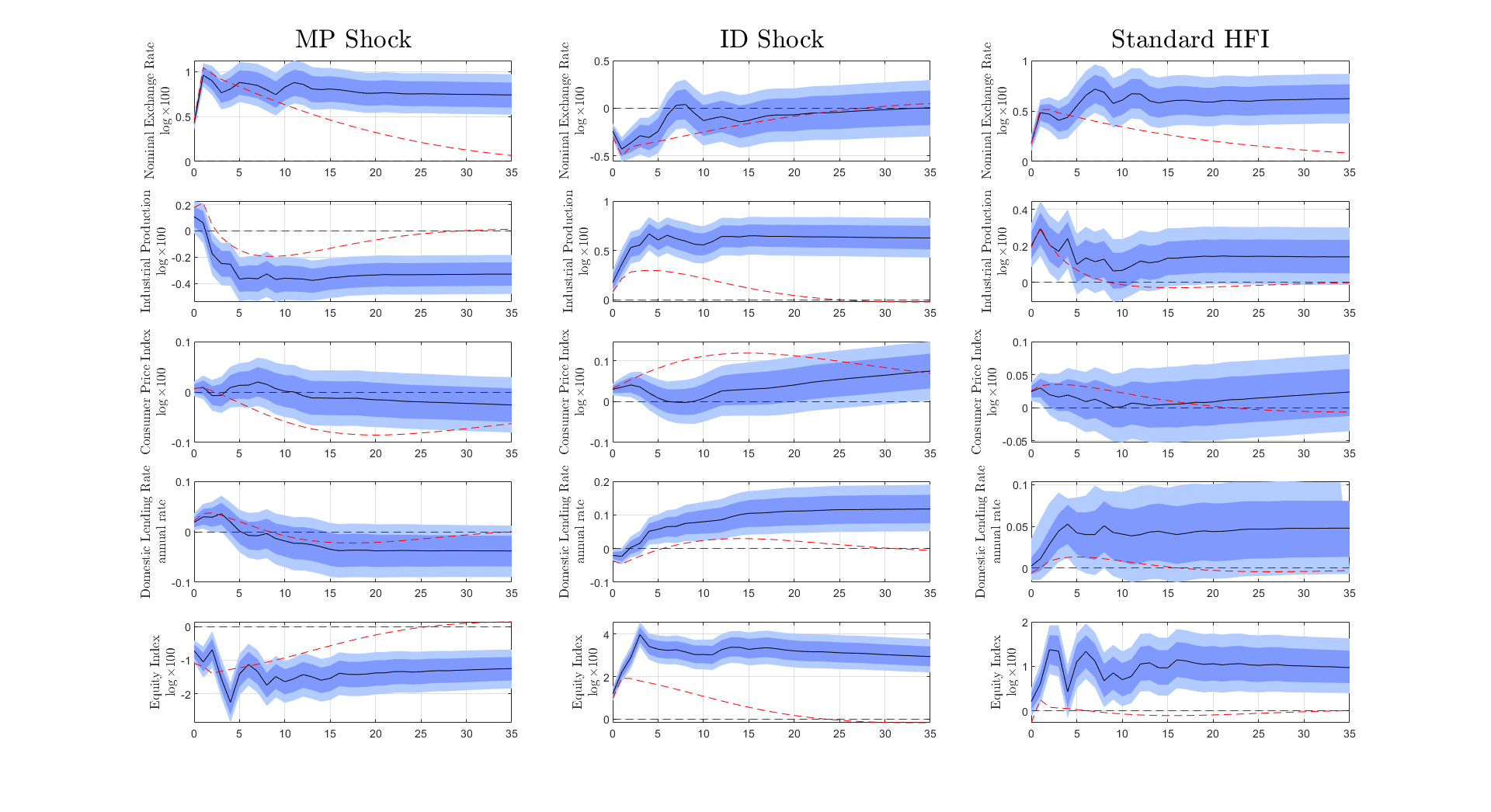}
         \label{fig:MG_Benchmark}
         \floatfoot{\scriptsize \textbf{Note:} The black solid line represents the median impulse response function. The dark shaded area represents the 16 and 84 percentiles. The light shaded are represents the 5 and 95 percentiles. The red dashed line is the mean group estimator which plots the average response across countries. The figure is comprised of 15 sub-figures ordered in five rows and three columns. Every row represents a different variable: (i) the MP structural FOMC shock, (ii) the ID structural FOMC shock, (iii) nominal exchange rate, (iv) industrial production index, (v) consumer price index, (vi) lending rate, (vii) equity index. The first column presents the results for the MP or ``Pure US Monetary Policy'' shock, the middle column the ID or ``Information Disclosure'' shock and the the last column presents the results following the ``Standard High Frequency'' identification shock. In the text, when referring to Panel $(i,j)$, $i$ refers to the row and $j$ to the column of the figure.}
\end{figure}

\begin{figure}[ht]
         \centering
    \caption{Impulse Response to One-Standard-Deviation Shock \\ \footnotesize  Mean Group Estimator}
         \includegraphics[scale=0.325]{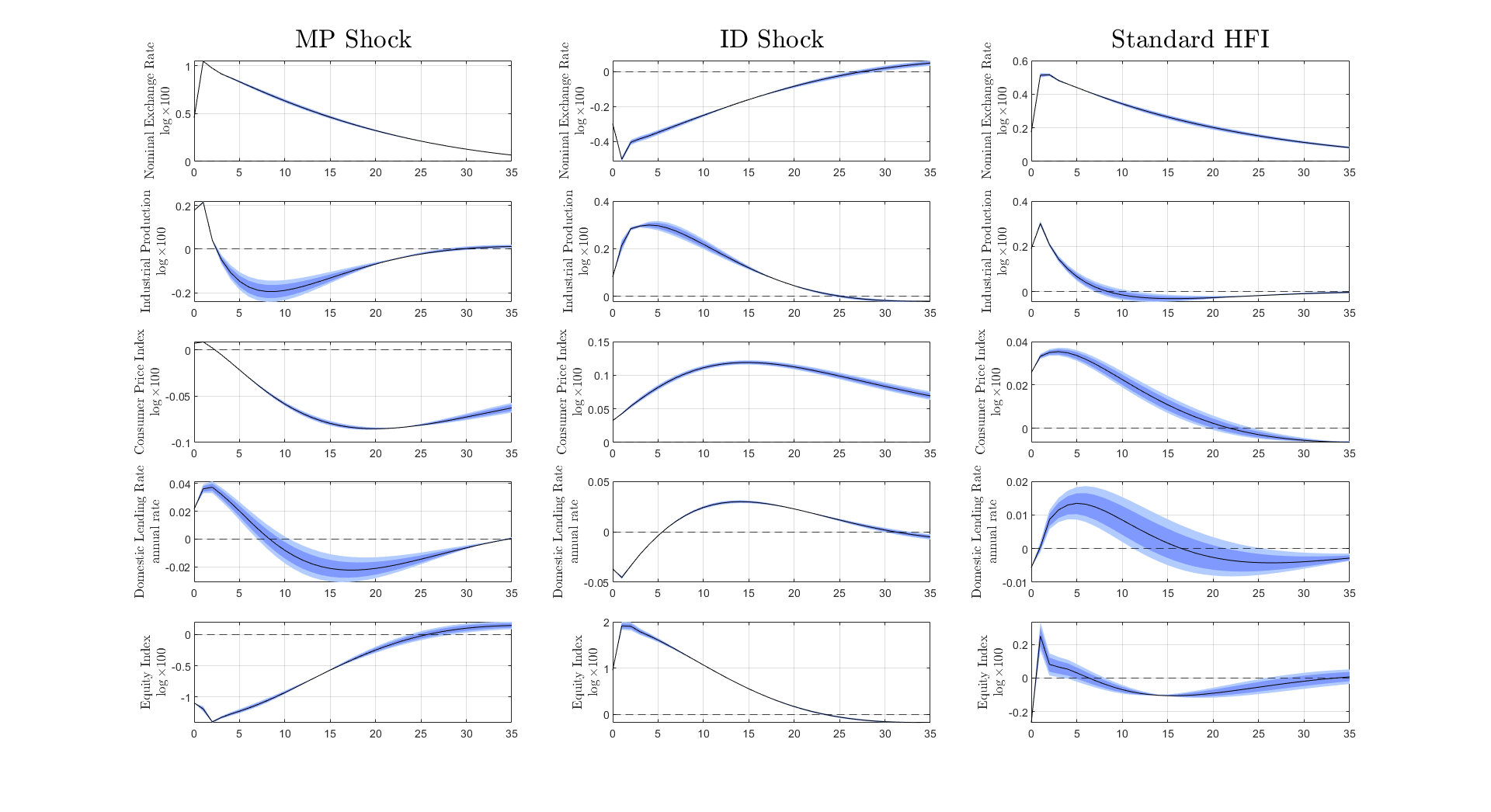}
         \label{fig:MG}
         \floatfoot{\scriptsize \textbf{Note:} The black solid line represents the median impulse response function. The dark shaded area represents the 16 and 84 percentiles. The light shaded are represents the 5 and 95 percentiles. The figure is comprised of 15 sub-figures ordered in five rows and three columns. Every row represents a different variable: (i) the MP structural FOMC shock, (ii) the ID structural FOMC shock, (iii) nominal exchange rate, (iv) industrial production index, (v) consumer price index, (vi) lending rate, (vii) equity index. The first column presents the results for the MP or ``Pure US Monetary Policy'' shock, the middle column the ID or ``Information Disclosure'' shock and the the last column presents the results following the ``Standard High Frequency'' identification shock. In the text, when referring to Panel $(i,j)$, $i$ refers to the row and $j$ to the column of the figure.}
\end{figure}

\begin{figure}[ht]
         \centering
         \caption{Impulse Response to One-Standard-Deviation Shock \\ \footnotesize  All Variables}
         \includegraphics[scale=0.325]{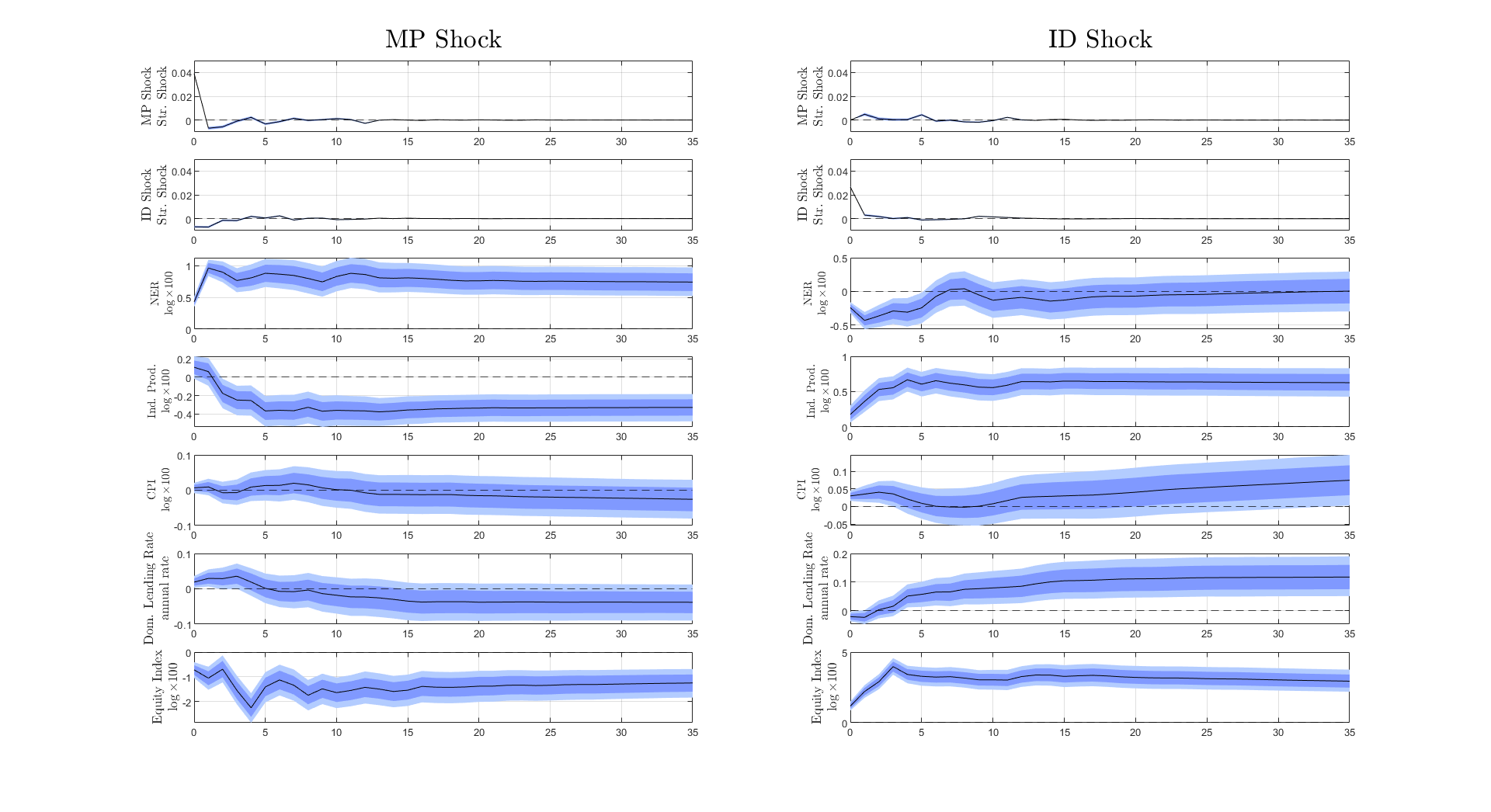}
         \label{fig:Benchmark_All_Variables}
         \floatfoot{\scriptsize \textbf{Note:} The black solid line represents the median impulse response function. The dark shaded area represents the 16 and 84 percentiles. The light shaded are represents the 5 and 95 percentiles. The figure is comprised of 14 sub-figures ordered in seven rows and two columns. Every row represents a different variable: (i) the MP structural FOMC shock, (ii) the ID structural FOMC shock, (iii) nominal exchange rate, (iv) industrial production index, (v) consumer price index, (vi) lending rate, (vii) equity index. The first column presents the results for the MP or ``Pure US Monetary Policy'' shock, the last or right column presents the results for the ID shock. In the text, when referring to Panel $(i,j)$, $i$ refers to the row and $j$ to the column of the figure.}
\end{figure}

\begin{figure}[ht]
         \centering
         \caption{Impulse Response to One-Standard-Deviation Shock \\ \footnotesize  MP Shock Reordering }
         \includegraphics[scale=0.325]{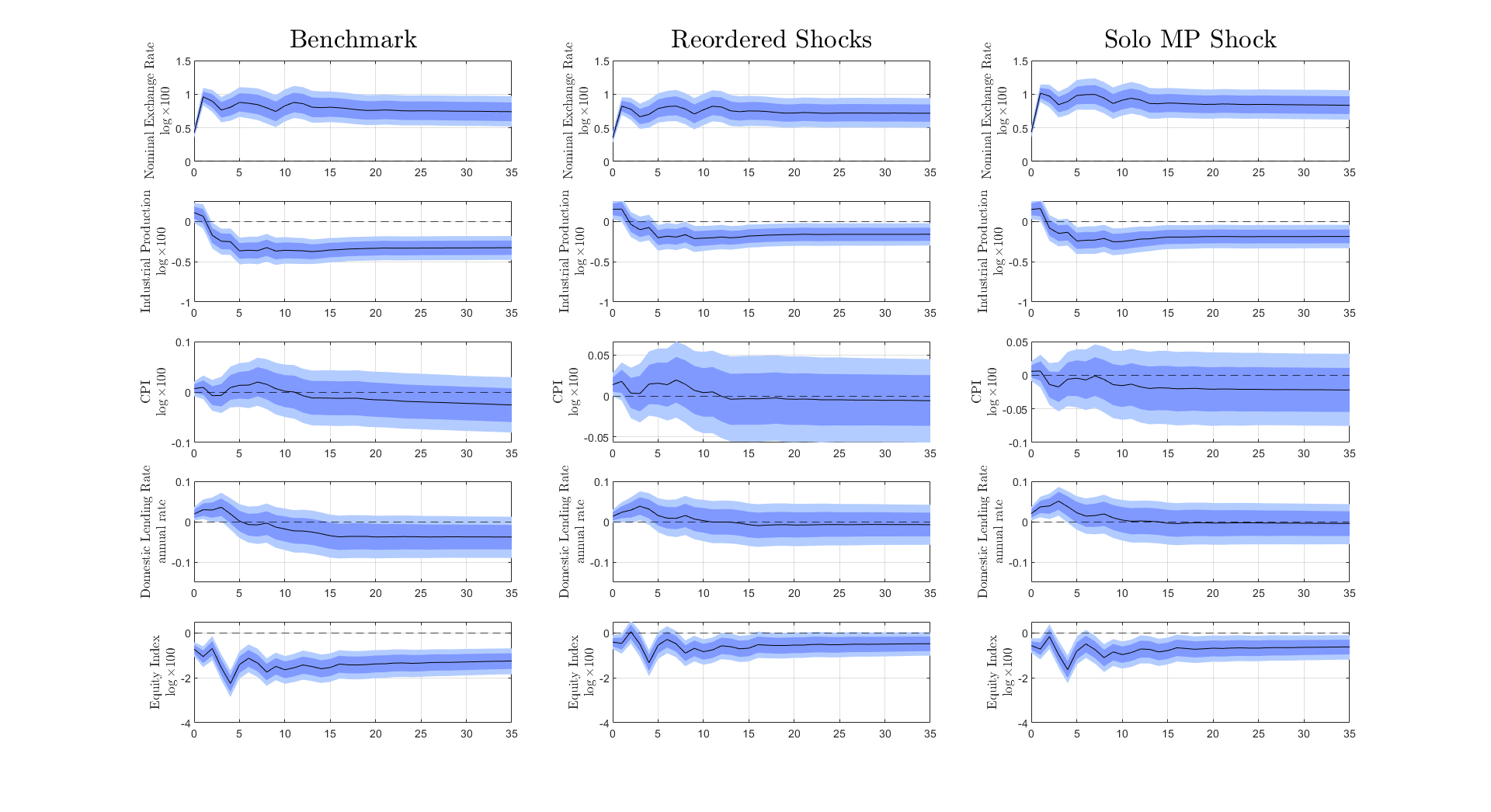}
         \label{fig:MP_Reordered}
         \floatfoot{\scriptsize \textbf{Note:} The black solid line represents the median impulse response function. The dark shaded area represents the 16 and 84 percentiles. The light shaded are represents the 5 and 95 percentiles. The figure is comprised of 15 sub-figures ordered in five rows and three columns. Every row represents a different variable: (i) the MP structural FOMC shock, (ii) the ID structural FOMC shock, (iii) nominal exchange rate, (iv) industrial production index, (v) consumer price index, (vi) lending rate, (vii) equity index. The first column presents the results for the MP shock using the benchmark specification. The middle column presents the results re-ordering the ID shock first and the MP shock second. The last column presents the results when only including the MP shock in the SVAR specification. In the text, when referring to Panel $(i,j)$, $i$ refers to the row and $j$ to the column of the figure.}
\end{figure}

\begin{figure}[ht]
         \centering
         \caption{Impulse Response to One-Standard-Deviation Shock \\ \footnotesize ID Shock Reordering }
         \includegraphics[scale=0.325]{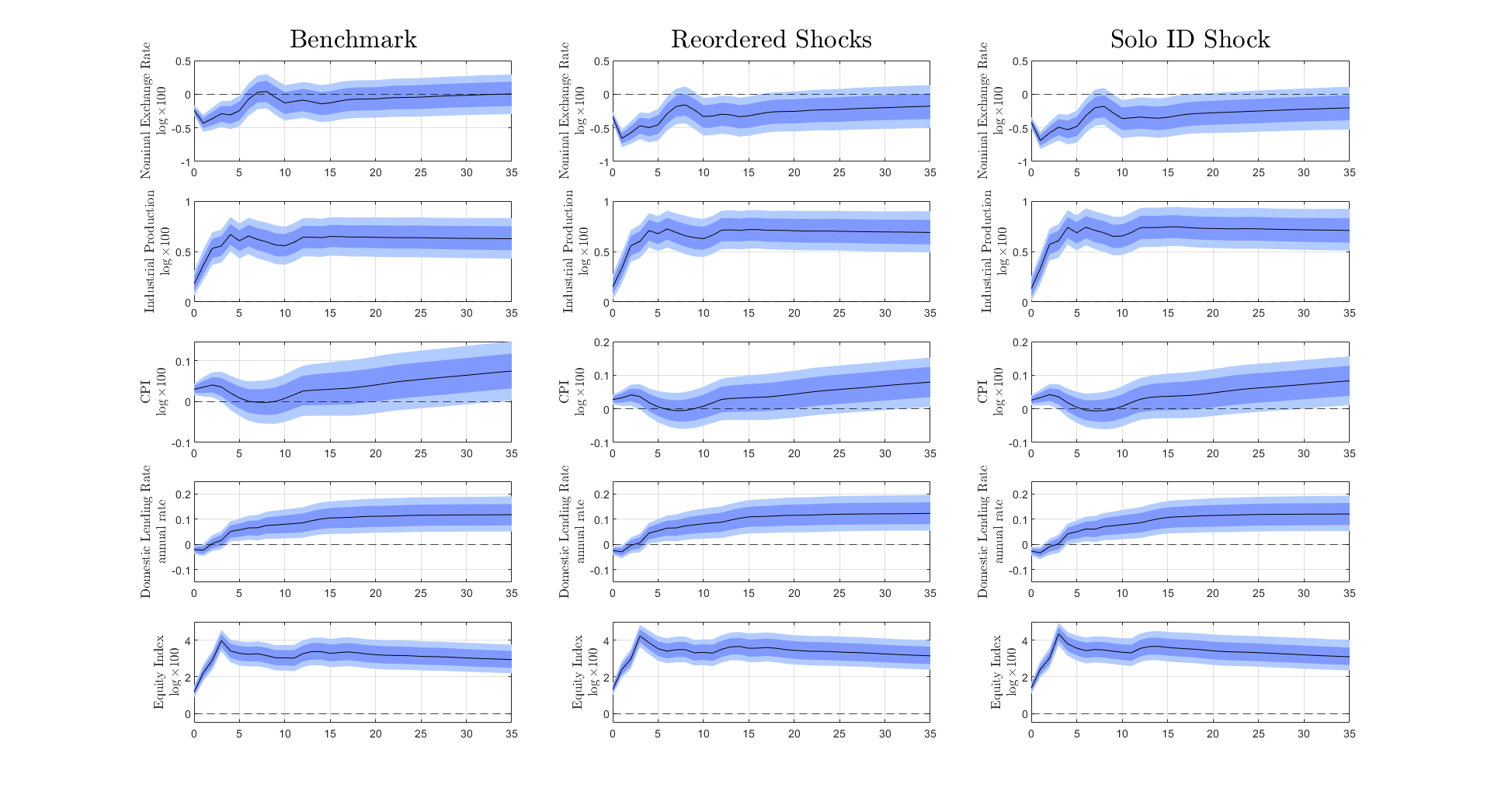}
         \label{fig:CBI_Reordered}
         \floatfoot{\scriptsize \textbf{Note:} The black solid line represents the median impulse response function. The dark shaded area represents the 16 and 84 percentiles. The light shaded are represents the 5 and 95 percentiles. The figure is comprised of 15 sub-figures ordered in five rows and three columns. Every row represents a different variable: (i) the MP structural FOMC shock, (ii) the ID structural FOMC shock, (iii) nominal exchange rate, (iv) industrial production index, (v) consumer price index, (vi) lending rate, (vii) equity index. The first column presents the results for the ID shock using the benchmark specification. The middle column presents the results re-ordering the ID shock first and the MP shock second. The last column presents the results when only including the ID shock in the SVAR specification. In the text, when referring to Panel $(i,j)$, $i$ refers to the row and $j$ to the column of the figure.}
\end{figure}

\begin{figure}[ht]
         \centering
         \caption{Impulse Response to One-Standard-Deviation Shock \\ \footnotesize  External Moment Condition }
         \includegraphics[scale=0.325]{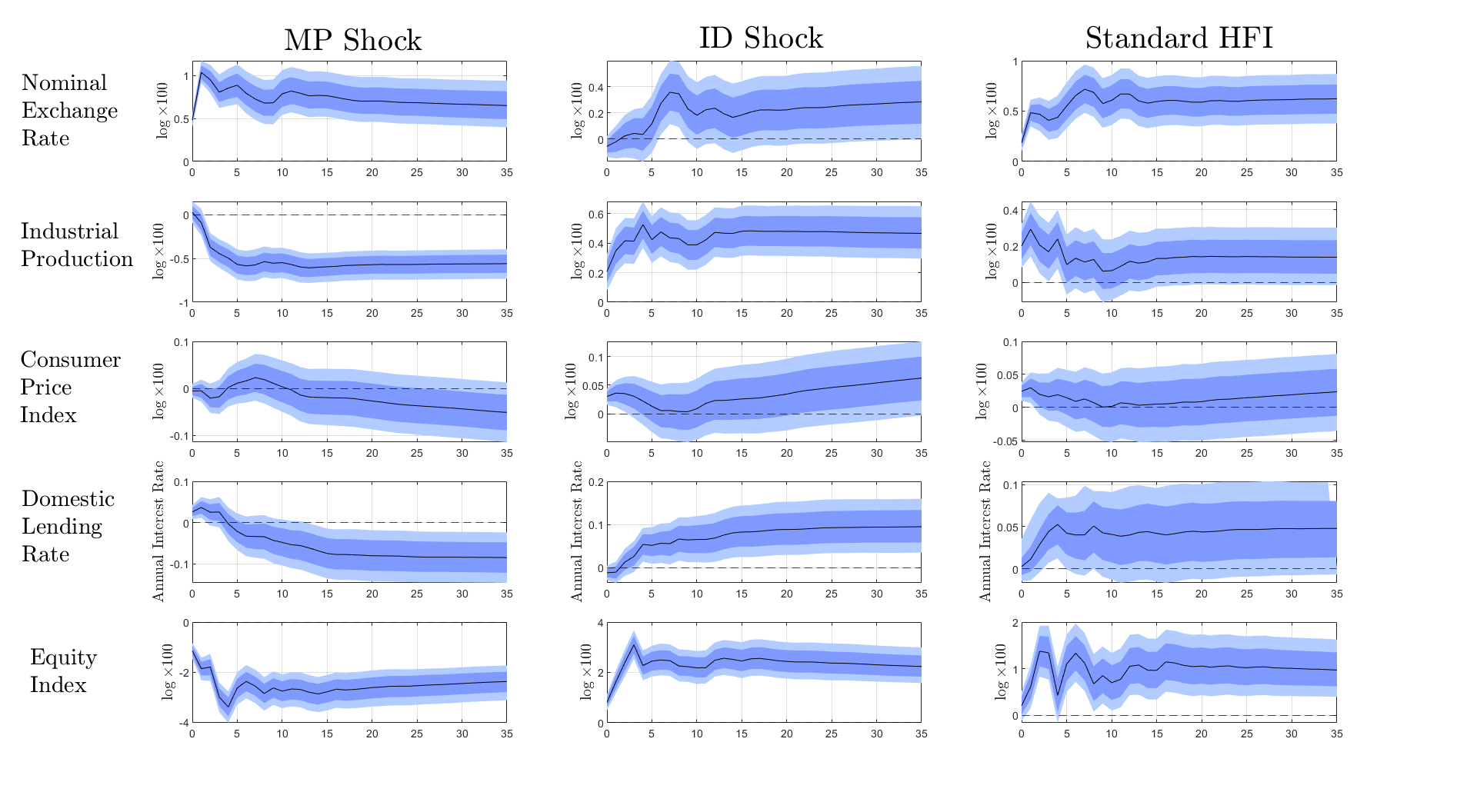}
         \label{fig:Benchmark_j}
         \floatfoot{\scriptsize \textbf{Note:} The black solid line represents the median impulse response function. The dark shaded area represents the 16 and 84 percentiles. The light shaded are represents the 5 and 95 percentiles. The figure is comprised of 15 sub-figures ordered in five rows and three columns. Every row represents a different variable: (i) the MP structural FOMC shock, (ii) the ID structural FOMC shock, (iii) nominal exchange rate, (iv) industrial production index, (v) consumer price index, (vi) lending rate, (vii) equity index. The first column presents the results for the MP or ``Pure US Monetary Policy'' shock, the middle column the ID or ``Information Disclosure'' shock and the the last column presents the results following the ``Standard High Frequency'' identification shock. In the text, when referring to Panel $(i,j)$, $i$ refers to the row and $j$ to the column of the figure.}
\end{figure}

\begin{figure}[ht]
         \centering
         \caption{Impulse Response to One-Standard-Deviation Shock \\ \footnotesize  Ext. Moment Condition - AE }
         \includegraphics[scale=0.325]{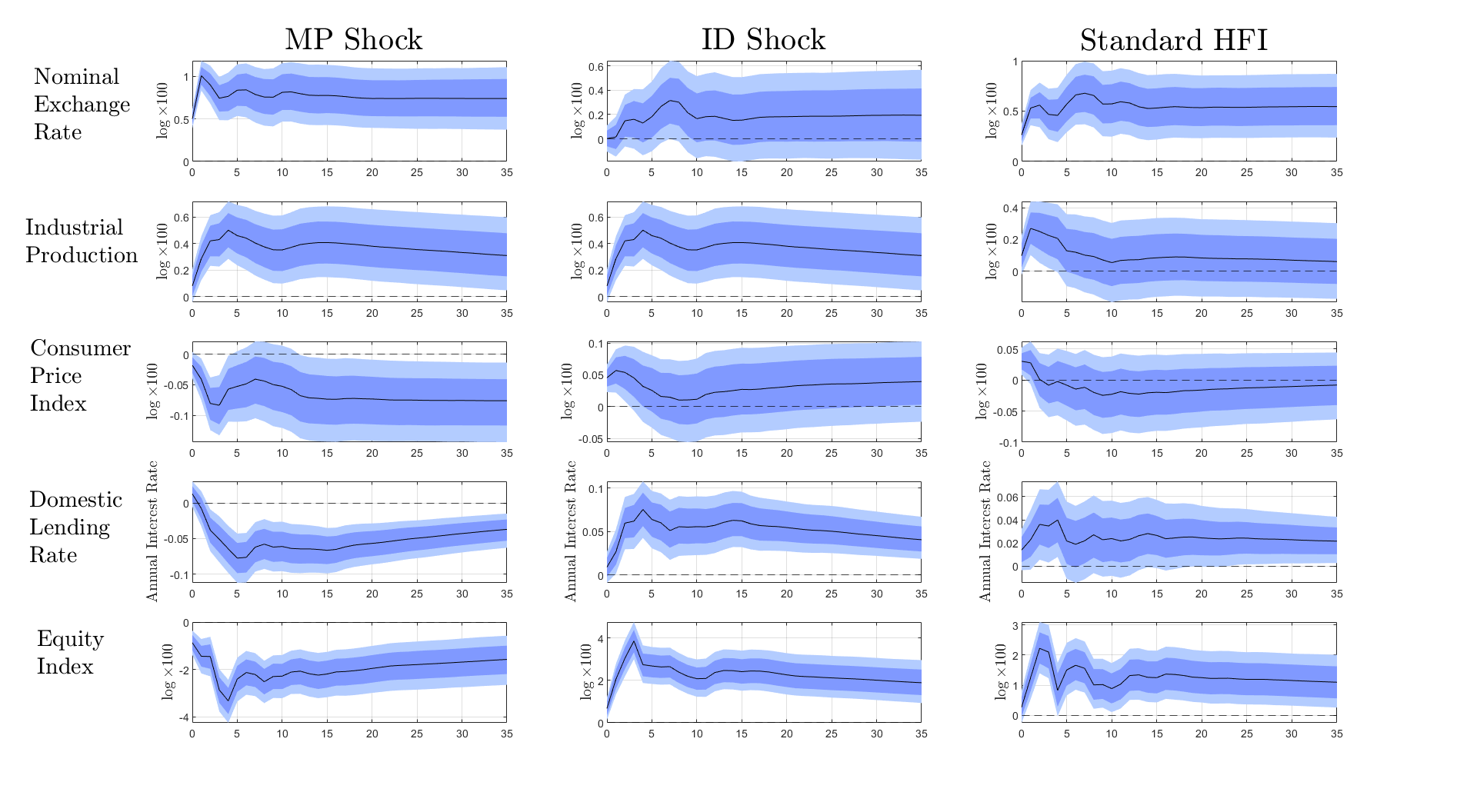}
         \label{fig:Benchmark_Adv_j}
         \floatfoot{\scriptsize \textbf{Note:} The black solid line represents the median impulse response function. The dark shaded area represents the 16 and 84 percentiles. The light shaded are represents the 5 and 95 percentiles. The figure is comprised of 15 sub-figures ordered in five rows and three columns. Every row represents a different variable: (i) the MP structural FOMC shock, (ii) the ID structural FOMC shock, (iii) nominal exchange rate, (iv) industrial production index, (v) consumer price index, (vi) lending rate, (vii) equity index. The first column presents the results for the MP or ``Pure US Monetary Policy'' shock, the middle column the ID or ``Information Disclosure'' shock and the the last column presents the results following the ``Standard High Frequency'' identification shock. In the text, when referring to Panel $(i,j)$, $i$ refers to the row and $j$ to the column of the figure.}
\end{figure}

\begin{figure}[ht]
         \centering
         \caption{Impulse Response to One-Standard-Deviation Shock \\ \footnotesize  Ext. Moment Condition - EM }
         \includegraphics[scale=0.325]{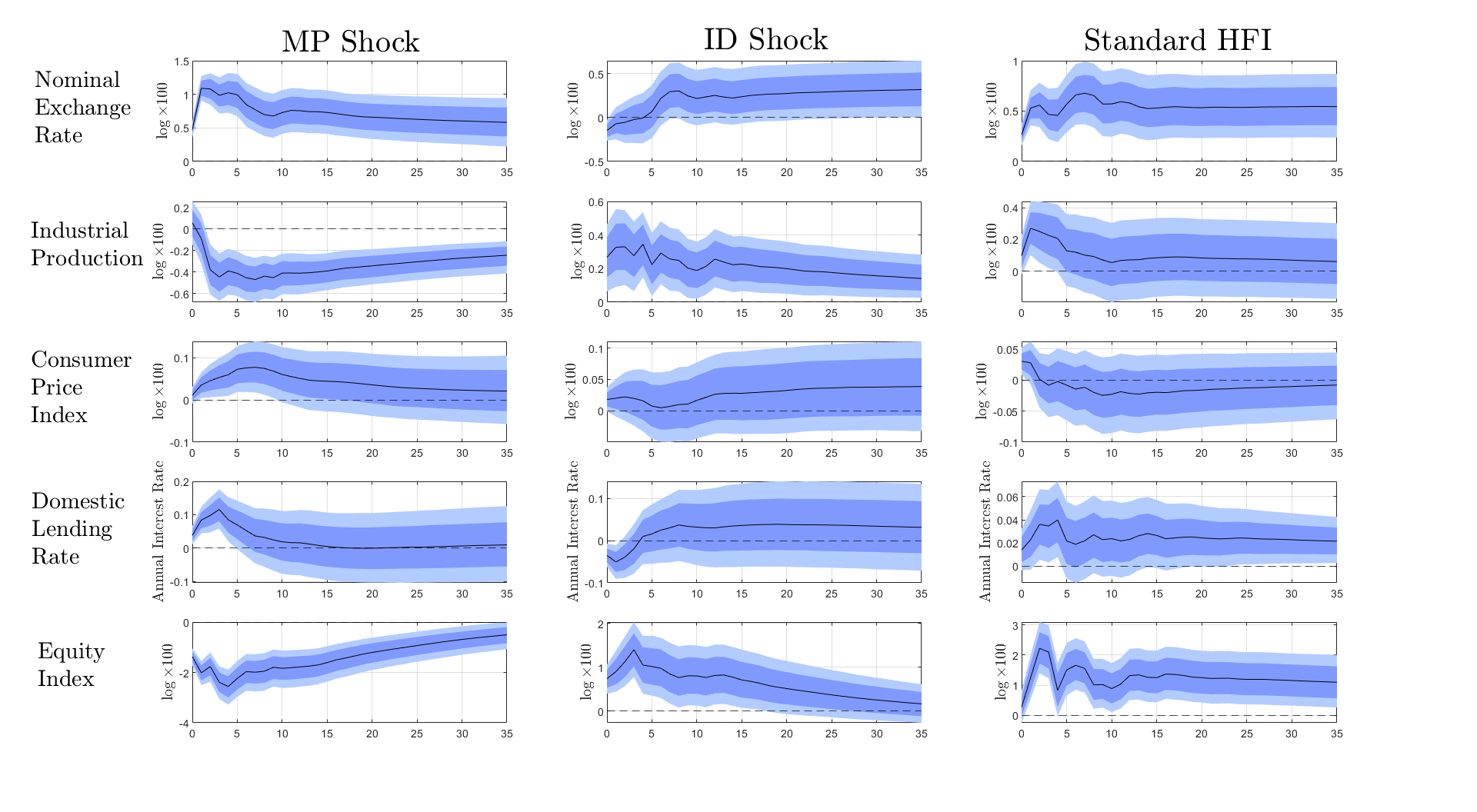}
         \label{fig:Benchmark_EM_j}
         \floatfoot{\scriptsize \textbf{Note:} The black solid line represents the median impulse response function. The dark shaded area represents the 16 and 84 percentiles. The light shaded are represents the 5 and 95 percentiles. The figure is comprised of 15 sub-figures ordered in five rows and three columns. Every row represents a different variable: (i) the MP structural FOMC shock, (ii) the ID structural FOMC shock, (iii) nominal exchange rate, (iv) industrial production index, (v) consumer price index, (vi) lending rate, (vii) equity index. The first column presents the results for the MP or ``Pure US Monetary Policy'' shock, the middle column the ID or ``Information Disclosure'' shock and the the last column presents the results following the ``Standard High Frequency'' identification shock. In the text, when referring to Panel $(i,j)$, $i$ refers to the row and $j$ to the column of the figure.}
\end{figure}

\begin{figure}[ht]
         \centering
         \caption{Impulse Response to One-Standard-Deviation Shock \\ \footnotesize  Uniform Prior over Rotations }
         \includegraphics[scale=0.325]{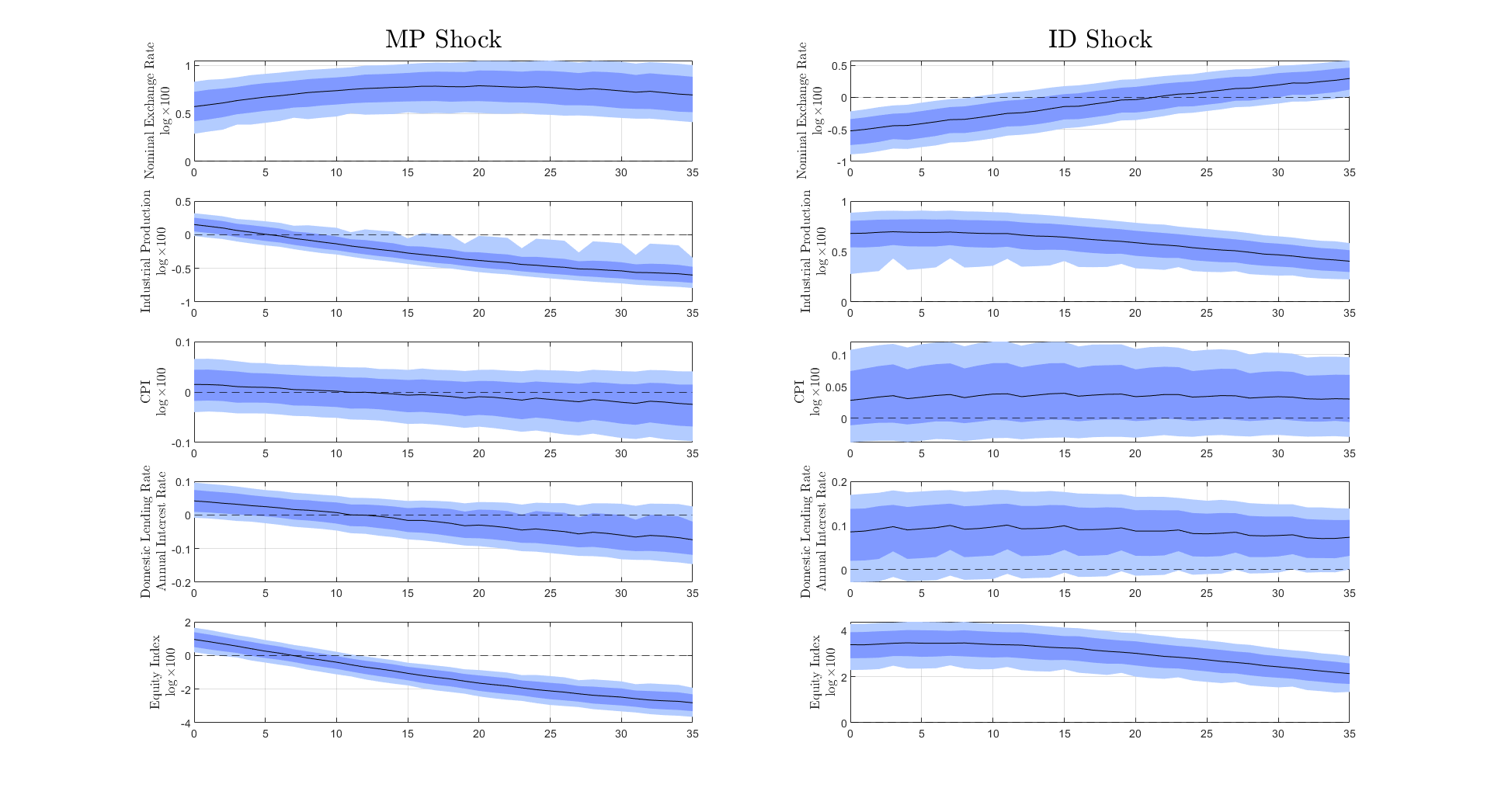}
         \label{fig:Sign_Restriction_90}
         \floatfoot{\scriptsize \textbf{Note:} The black solid line represents the median impulse response function. The dark shaded area represents the 16 and 84 percentiles. The light shaded are represents the 5 and 95 percentiles. The figure is comprised of 10 sub-figures ordered in five rows and two columns. Every row represents a different variable: (i) the MP structural FOMC shock, (ii) the ID structural FOMC shock, (iii) nominal exchange rate, (iv) industrial production index, (v) consumer price index, (vi) lending rate, (vii) equity index. The first column presents the results for the MP or ``Pure US Monetary Policy'' shock, the middle column the ID or ``Information Disclosure'' shock and the the last column presents the results following the ``Standard High Frequency'' identification shock. In the text, when referring to Panel $(i,j)$, $i$ refers to the row and $j$ to the column of the figure.}
\end{figure}

\begin{figure}[ht]
         \centering
         \caption{Impulse Response to One-Standard-Deviation Shock \\ \footnotesize  Medians across Rotations }
         \includegraphics[scale=0.325]{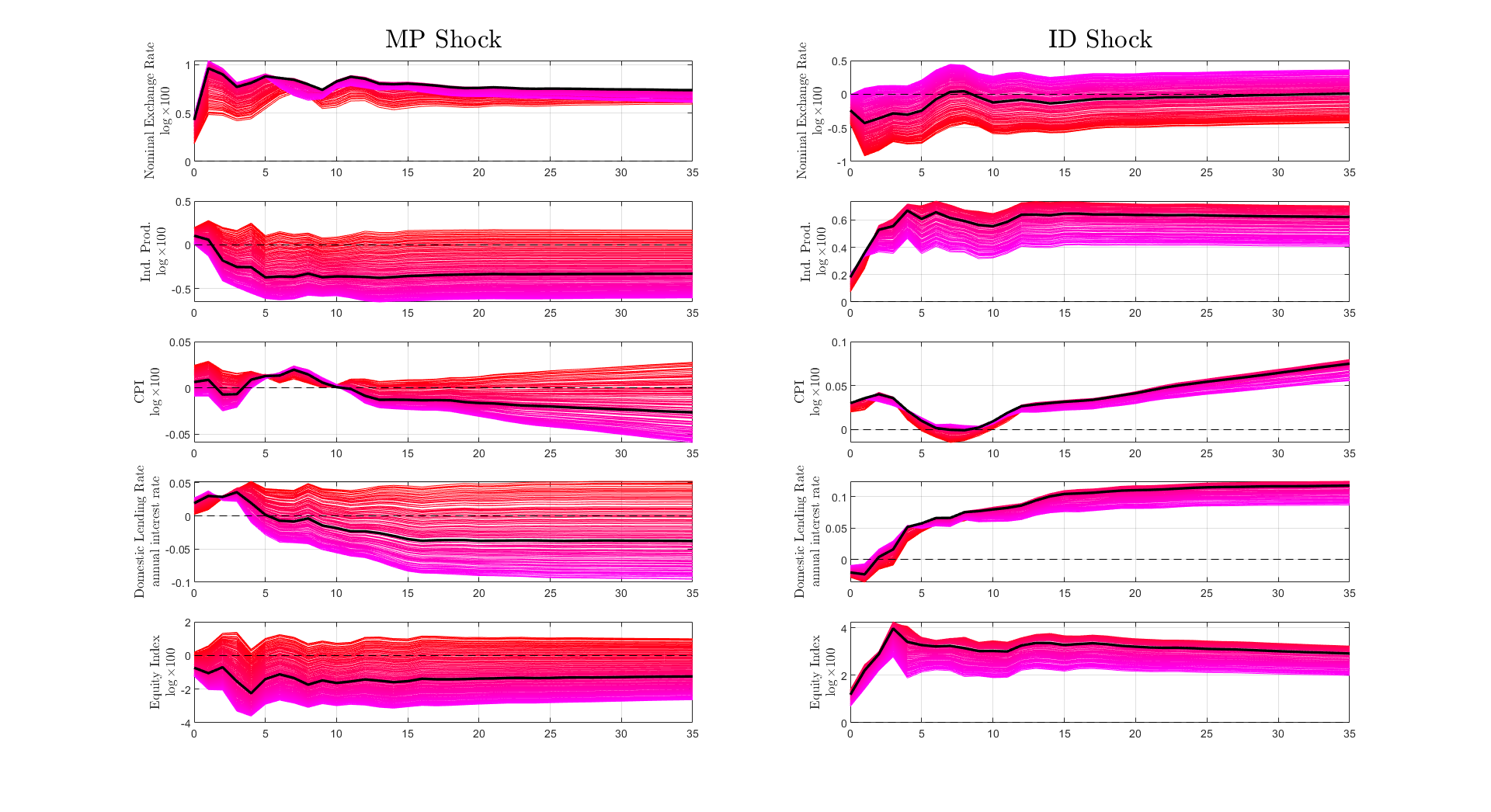}
         \label{fig:Sign_Restriction_Medians}
         \floatfoot{\scriptsize \textbf{Note:} Each line represents the median impulse response function of estimating the SVAR model using acceptable rotational angle $p$ where $p = 0.01,\ldots,0.99$. The color of the lines varies from red (closer to 1) to purple (closer to 99). The solid black line represents the median impulse response using the median acceptable rotational angle. The figure is comprised of 10 sub-figures ordered in five rows and two columns. Every row represents a different variable: (i) the MP structural FOMC shock, (ii) the ID structural FOMC shock, (iii) nominal exchange rate, (iv) industrial production index, (v) consumer price index, (vi) lending rate, (vii) equity index. The first column presents the results for the MP or ``Pure US Monetary Policy'' shock, the second column the ID or ``Information Disclosure'' shock. In the text, when referring to Panel $(i,j)$, $i$ refers to the row and $j$ to the column of the figure.}
\end{figure}

\newpage
\section{Alternative Strategies that Control for Information Effects} \label{sec:appendix_additional_results_strategies}

In this section, I show that the main results of this paper are also present when estimating the spillovers of a pure US monetary policy shock using alternative identification strategies which control for possible information effects around FOMC meetings. This result strengthens the argument that recent atypical international dynamics found when following the standard high-frequency identification strategy of US monetary policy shock, such as the expansion of industrial output, can be attributed to the systematic disclosure of information around FOMC meetings.

I consider two alternative identification strategies: the one proposed by \cite{miranda2021transmission} and the one presented by \cite{bauer2022reassessment}. These two alternative identification strategies control for information effects of the Federal Reserve's monetary policy through different approaches to that followed by \cite{jarocinski2020deconstructing}. First, \cite{miranda2021transmission} define US monetary policy shocks as exogenous shifts in the policy instrument that surprise market participants, are unforecastable, and are not due to the central bank’s systematic response to its own assessment of the macroeconomic outlook. To this end, the authors construct an instrument constructed by regressing the high-frequency market surprises in the fourth federal fund future onto a set of Greenbook forecasts for output, inflation and unemployment. Through this instrument, the authors can control for the signalling effect of monetary policy, i.e., the Federal Reserve responding to expected changes in macroeconomic and financial variables. Unlike \cite{jarocinski2020deconstructing}, this approach controls for the Federal Reserve's own information set which may or may not reflect or include the financial market's information.\footnote{The time series on the US monetary policy shocks under the two alternative identification strategies is sourced directly from the authors' websites. For the case of \cite{miranda2021transmission} the shocks can be recovered from \url{https://github.com/GRicco/info-policy-surprises}. This time series is only available up until December 2015. Thus, when estimating the model under this identification strategy the relevant sample is January 2004 to December 2015. For the case of \cite{bauer2022reassessment} the shocks can be recovered from \url{https://www.michaeldbauer.com/publication/mps/}.}

Second, \cite{bauer2022reassessment} define US monetary policy shocks as exogenous shifts in policy interest rates that are purged of any response of the Federal Reserve to macroeconomic news. To do so, the authors identify US monetary policy shocks by orthogonalizing the high-frequency surprises of the first component of four Eurodollar futures contracts of key macroeconomic or financial information available prior to the FOMC meeting.\footnote{Note that both the surprise in the Fed Funds Futures used by \cite{miranda2021transmission} and the Eurodollar future contracts used by \cite{bauer2022reassessment} are part of the policy interest rate composite used to construct the FOMC shocks in Section \ref{sec:main_results}.} The authors argue these approach purges the interest rate surprises of the "Fed Response to News" channel, in which incoming, publicly available economic news causes both the Fed to change monetary policy and the private sector to revise its forecasts. It is noteworthy to state that this identification strategy is based on a previous paper, \cite{bauer2020fed}, which argues that the  “Fed Information Effect”, as presented by \cite{jarocinski2022central}, is not quantitatively important. Furthermore, the authors argue that a plausible explanation to this effect is the “Fed Response to News” channel. However, as stressed by \cite{jarocinski2022central} the theoretical models presented in \cite{bauer2020fed,bauer2022reassessment} still predict a negative correlation between interest rate surprises and stock price surprises. Furthermore, even as the overall correlation between the high-frequency surprises between interest rates and the S\&P 500 is negative, to the extent that some variants of the “Fed Response to News” channel generates a positive correlation between these financial surprises, one can interpret the ID shock as a proxy for this effect too (see \cite{jarocinski2022central}).

\begin{figure}[ht]
         \centering
         \caption{Impulse Response to \cite{bauer2022reassessment} Shocks}
         \includegraphics[scale=0.325]{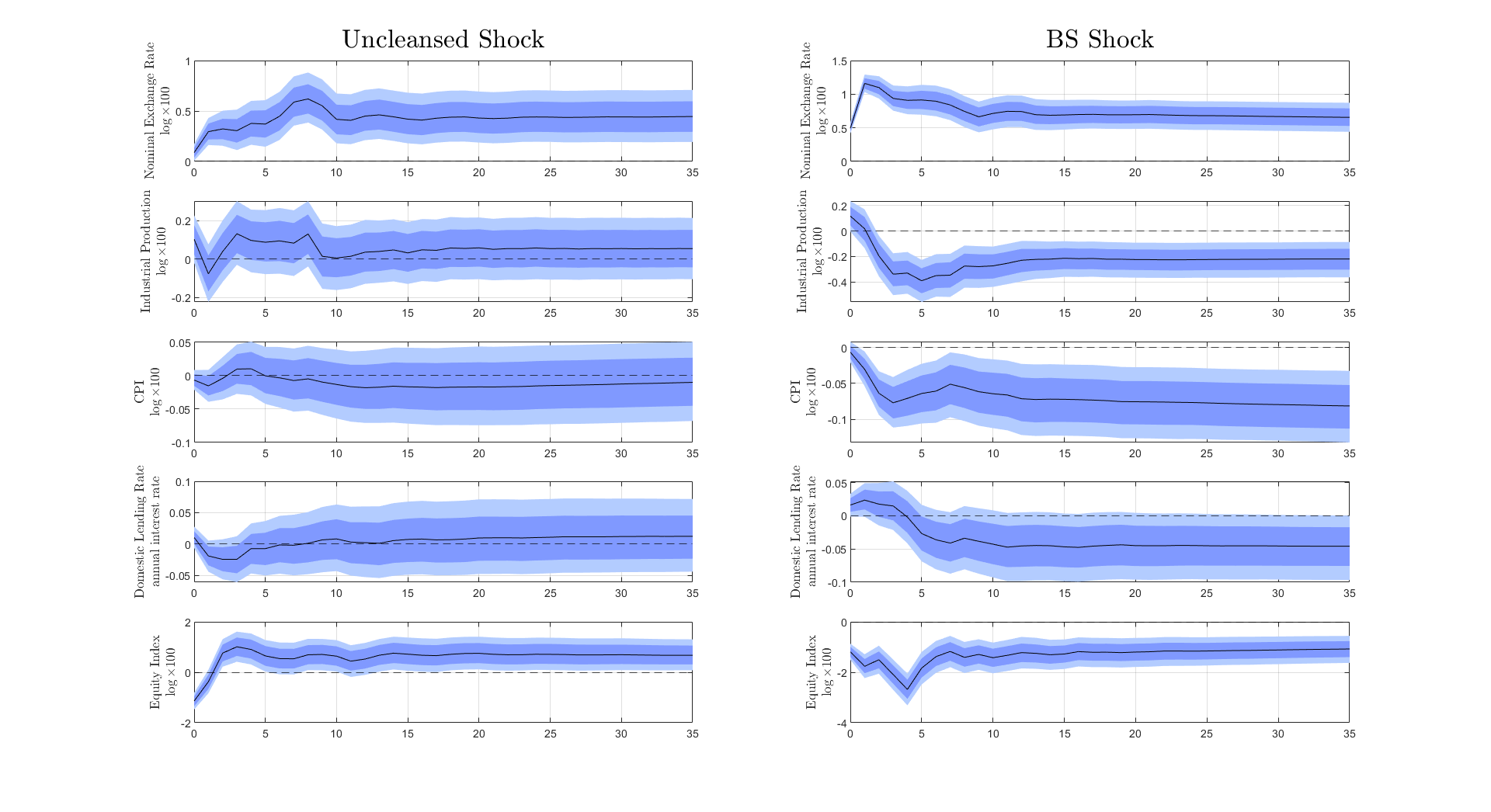}
         \label{fig:BS_vs_MPS}
         \floatfoot{\scriptsize \textbf{Note:} The black solid line represents the median impulse response function. The dark shaded area represents the 16 and 84 percentiles. The light shaded are represents the 5 and 95 percentiles. The figure is comprised of 10 sub-figures ordered in five rows and two columns. Every row represents a different variable: (i) nominal exchange rate, (ii) industrial production index, (iii) consumer price index, (iv) lending rate, (v) equity index. The first column presents the results for the un-cleansed or un-orthogonalized shock, the second column presents the results for the orthogonalized  shock. In the text, when referring to Panel $(i,j)$, $i$ refers to the row and $j$ to the column of the figure.}
\end{figure}

Figure \ref{fig:BS_vs_MPS} compares the impulse response functions using the two shocks constructed by \cite{bauer2022reassessment}: (i) the un-cleansed or un-orthogonalized shock, (ii) the cleansed or orthogonalized shock.
On the left column, the un-orthogonalized shock leads to a atypical dynamics, such as a non-depreciation on impact of the exchange rate and a mild expansion of the industrial production and equity indexes. On the right column, the orthogonalized shocks lead to a significant depreciation on impact which is persistent across time, and a significant drop in both the industrial production and equity index. Thus, the un-cleansed shocks lead to dynamics similar to those arising from following the ``Standard HFI'' strategy used in Section \ref{sec:main_results}. Additionally, using a monetary policy shock purged of information effects leads to results similar to those estimated for the MP shock in Section \ref{sec:main_results}.

\begin{figure}[ht]
         \centering
         \caption{Impulse Response to One-Standard-Deviation Shock \\ \footnotesize Benchmark Specification}
         \includegraphics[scale=0.325]{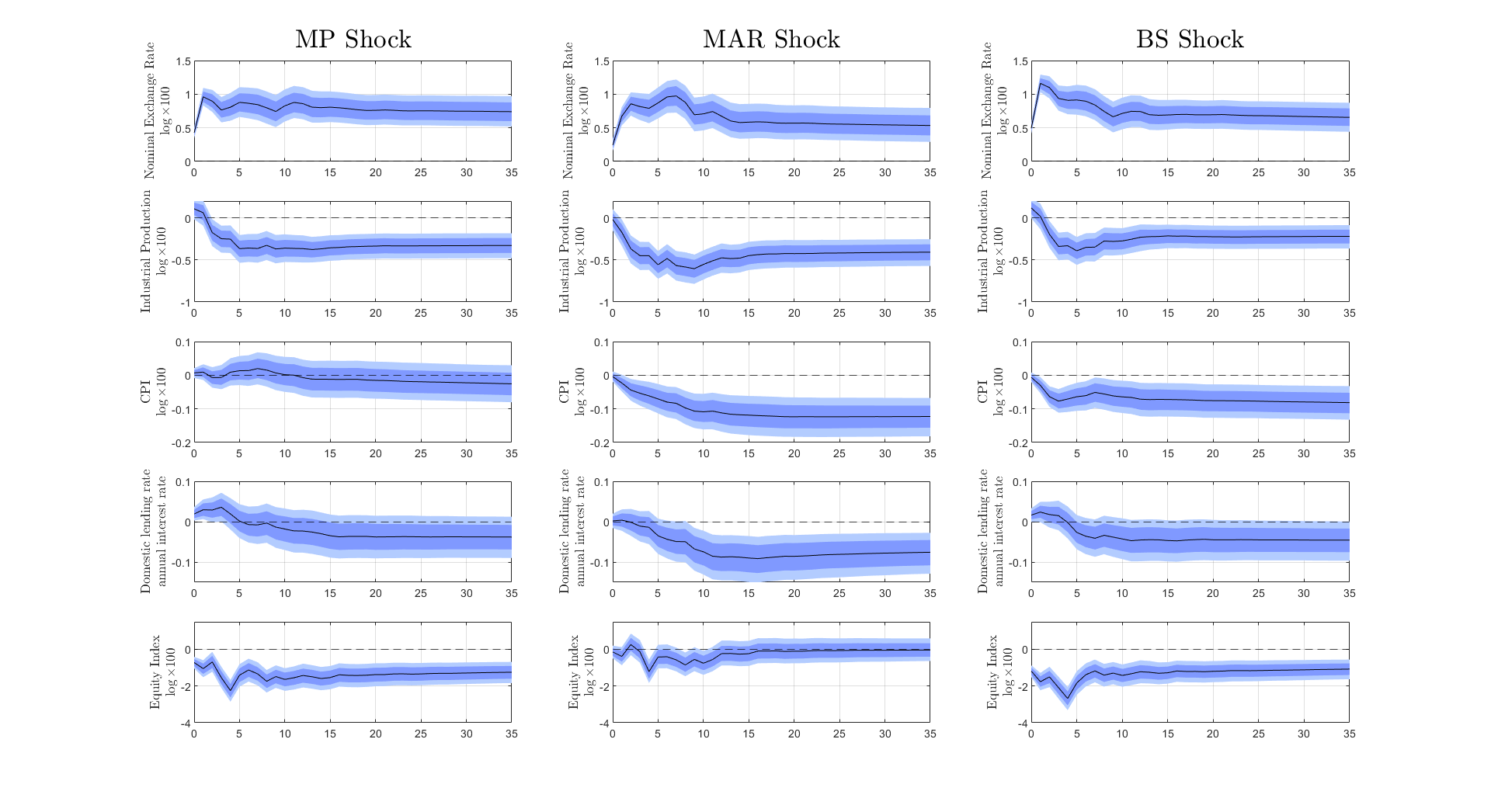}
         \label{fig:Other_Information}
         \floatfoot{\scriptsize \textbf{Note:} The black solid line represents the median impulse response function. The dark shaded area represents the 16 and 84 percentiles. The light shaded are represents the 5 and 95 percentiles. The figure is comprised of 15 sub-figures ordered in five rows and three columns. Every row represents a different variable: (i) nominal exchange rate, (ii) industrial production index, (iii) consumer price index, (iv) lending rate, (v) equity index. The first column presents the results for the MP or ``Pure US Monetary Policy'' shock under the benchmark specification, the middle column presents the results for MAR shock or \cite{miranda2020us}, and the last column presents the results for the BS shock or \cite{bauer2022reassessment}. In the text, when referring to Panel $(i,j)$, $i$ refers to the row and $j$ to the column of the figure.}
\end{figure}

Figure \ref{fig:Other_Information} compares the impulse response functions of a MP shock under the benchmark specifications, with the impulse response functions of a US monetary policy shock under the two alternative identification strategies.\footnote{Under the two alternative shocks I set the vector $m_t$ of structural shocks as solely containing the US monetary policy shock of either identification strategy. Note that the  methodology presented by \cite{bauer2022reassessment} expands the set of monetary policy announcements to include speeches by the Fed Chair, which essentially duplicates the number of announcements in their dataset. This leads to multiple announcements and high-frequency surprises within a month. In order to construct a monthly time series of shocks to introduce into vector $m_t$  I take the simple average of all shocks within the same month.}
The resulting impulse response functions are qualitatively in line with the main results presented in Section \ref{sec:main_results}. All three specifications of a US monetary policy shock leads to a nominal exchange rate depreciation, a hump-shaped fall in industrial production, a drop in the consumer price index, a mild increase in lending rates and a persistent drop in the equity index. Quantitatively the results are also significantly close across the three identification strategies. While the two alternative identification strategies predict a greater and more significant reduction in the consumer price index, the benchmark specification predicts a greater increase in domestic lending rates and a larger and more persistent drop in the equity index. 
In summary, the robustness check in which two alternative identification strategies which control for the disclosure of information by the Federal Reserve around FOMC meetings lead to similar qualitatively and quantitatively results to our benchmark results leads to two main conclusions. First, by controlling for the disclosure of information around FOMC meetings, a pure US monetary policy shock leads to negative spillovers over a panel of countries, consistent with the older and conventional view. The second conclusion is that the systematic disclosure of information about the state of the US economy during FOMC announcements leads to recent atypical dynamics, such as those presented by \cite{ilzetzki2021puzzling} and the ones presented in Section \ref{sec:main_results} under the ``Standard HFI'' strategy. Once the ID shocks are taking into account, whether interpreted as the “Fed Information Effect” or the “Fed Response to News” channel, any atypical dynamics disappear. 

\newpage
\section{Potential Sources of Quantitative Heterogeneity} \label{sec:appendix_heterogeneity}

Following, I describe two potential sources of quantitatively heterogeneity in the international impact of US interest rates: (i) countries' exchange rate regimes and (ii) countries' reliance on the export of commodity goods. I briefly comment how previous literature has theorized that these country characteristics may influence the quantitative impact of the spillovers of US interest rates and describe my results.

\noindent
\textbf{Country classification for heterogeneity identification.} Next, I describe the different classification/categorization of countries used in Section \ref{sec:robustness_checks_additional_results}. In particular, I classified countries according to their exchange rate regimes and Emerging Market economies according to their commodity good dependence.

First, I classified/categorized countries according to their exchange rate regimes. In particular, I considered two partitions: (i) Peru vs the rest of the sample, (ii) Peru and Indonesia vs the rest of the sample. This sample partition is constructed following the exchange rate classification built by \cite{ilzetzki2017country} and additional evidence coming from both policy makers and other papers in the literature. All countries part of the Euro Zone are classified as freely floating. Hungary, Iceland and Sweden are classified as de facto narrow band with the Euro, de-coupling their currencies from the US dollar, given that the Euro is classified as freely floating.
    \begin{table}[ht]
        \centering
        \caption{Exchange Rate Classification \\ \footnotesize Sourced from \cite{ilzetzki2017country}}
        \footnotesize
        \label{tab:data_details_exchange_rate}
        \begin{tabular}{l l}
\multicolumn{2}{c}{Emerging Markets} \\ \hline \hline
Brazil 	&	2003-2007: Freely Floating, 2008-2016: Managed Floating	\\
Chile	&	2000-2007: De facto moving band that is narrower than or equal to +/-5\% \\
& 2008-2016: Managed Floating 	\\
Colombia	&	Managed Floating	\\
Hungary & De facto crawling band +/- 2\% since 2009. Euro. Broader band prior. \\
Indonesia	&	Starting 2005 De facto crawling band +/-2 to 5\% range.	\\
Mexico	&	2000-2008: De facto moving band that is narrower than or equal to +/-5\% \\
& 2009-2016 Managed Floating. 	\\
Peru	&	Starting from 2002. De facto crawling band +/-2\%. US dollar	\\
Philippines	&	Starting from 2002. De facto crawling band +/-2\%. US dollar	\\
South Africa	&	2000-2016 Freely floating	\\
			\\
\multicolumn{2}{c}{Advanced Economies} \\ \hline \hline
Australia	&	2000-2016: Freely floating	\\
Canada	&	2003-Onward: Freely Floating	\\
France   & Freely Floating \\
Iceland & Manage floating \\
Italy   & Freely Floating \\
Japan	&	2000-2016: Freely Floating	\\
South Korea	&	2000-2003: De facto moving band that is narrower than or equal to +/-5\% \\
& 2004-2009 Managed Floating \\
& 2010-2016: De facto moving band that is narrower than or equal to +/-5\% 	\\
The Netherlands   & Freely Floating \\
Sweden   & De facto moving band +/-2\% since 2008. Euro. \\
        \end{tabular}
    \end{table}
Table \ref{tab:data_details_exchange_rate} presents the exchange rate classifications from \cite{ilzetzki2017country}. Evidence suggests that Peru has the tightest exchange rate regime (with a de facto moving band of less than 2\%), with Indonesia being a close second (de facto moving band with periods below 2\% and other periods below 5\%). Additionally, as commented in Section \ref{sec:robustness_checks_additional_results}, both the Central Banks of Peru and Indonesia, and the economic literature have highlighted or studied these countries as case studies of foreign exchange rate intervention policies, see \cite{rossini2013foreign,rossini2019international,castillo2021foreign} for the case of Peru and \cite{warjiyo2013indonesia} for the case of Indonesia. While South Korea also presents a de facto narrow band the Central Bank of Korea publicly denies interventions. As commented by \cite{ryoo2013foreign}, the Korean Central Bank does not carry out FX interventions directly and only does it sporadically through private banks. Consequently, in order to construct the sub-sample between countries which employ tight exchange rate regimes and the rest of the sample I only consider Peru and the ``Peru \& Indonesia'' group. 

Second, I carried out a sample partition exercise according to Emerging Market's commodity dependence. 
\begin{table}[ht]
    \centering
    \caption{Classification of Countries according to Commodity Dependence}
    \footnotesize
    \label{tab:data_details_commodity}
    \begin{tabular}{l c c c}
Emerging Markets & \multicolumn{2}{c}{Share of Commodity Goods in Total Exports}  & Classification  \\  
	&	2009-2010	&	2014-2015	&	\\ \hline \hline
Brazil	&	63	&	63	&	High Commodity Dependence	\\
Chile	&	88	&	86	&	High Commodity Dependence	\\
Colombia	&	76	&	81	&	High Commodity Dependence	\\
Indonesia	&	62	&	58	&	Low Commodity Dependence	\\
Mexico	&	25	&	19	&	Low Commodity Dependence	\\
Peru	&	89	&	88	&	High Commodity Dependence	\\
Philippines	&	15	&	19	&	Low Commodity Dependence	\\
South Africa	&	54	&	55	&	Low Commodity Dependence \\ \hline \hline
    \end{tabular}
\end{table}
The first two columns of Table \ref{tab:data_details_commodity} presents data on the share of commodity goods in countries' total export baskets as constructed by the UNCTAD's ``The State of Commodity Dependence'' report. All countries in the sample export commodity goods, with Peru showing the greatest share of commodity goods in total exports (89\% and 88\% for the 2009-2010 and 2014-2015 periods), and Mexico showing the lowest share of commodity goods in total exports (25\% and 19\% for the 2009-2010 and 2014-2015 periods). 

I classify the countries in the Emerging Market economies sample into ``High Commodity Dependence'' and ``Low Commodity Dependence'' according to whether they are on the top/bottom half of share of commodity goods in total exports. I follow this approach as I only seek to explore commodity goods as a potential source of heterogeneity. It is clear that there are other potential sample partitions. Additionally, there are other exercises to test commodity exports as potential sources of heterogeneity on the impact of spillovers of US interest rates. I leave these exercises for future research.

First, I study whether countries' exchange rate regimes may influence the quantitative impact of the spillovers of US interest rates. The advantages and disadvantages of different exchange rate regimes is a classic yet still open question in international economics (see \cite{mundell1963capital,levy2003float,edwards2005flexible}). Moreover, the evidence is also mixed in terms of exchange rate regimes shaping the impact of US interest rates. On the one hand, \cite{di2008impact} provides evidence that for a panel of advanced economies the impact of higher foreign interest rates lead to a recession for countries with a fixed exchange rate. On the other hand, \cite{rey2015dilemma} argues that economies can not insulate from the US interest rate, even with flexible exchange rate regimes.\footnote{\cite{dedola2017if} and \cite{degasperi2020global} also present evidence that the impact of US interest rates is similar across countries with different exchange rate regimes.}

I carry two different partitions of the countries in the sample according to their exchange rate regimes. The first compares the resulting dynamics between a sample of only Peru versus the rest of the countries in the sample. The Peruvian Central Bank actively and publicly intervenes in foreign exchange rate markets to stabilize the nominal exchange rate. Also, the Peruvian case has been considered as a case study for the benefits and costs of FX interventions (see \cite{rossini2019international}, \cite{castillo2021foreign}  and \cite{camara2021FXI}).\footnote{Furthermore, under the exchange rate regime classification by \cite{ilzetzki2017country}, the country falls under tight or narrow exchange rate management. } Figures \ref{fig:ERR_MP_Peru} and \ref{fig:ERR_CBI_Peru} in Appendix \ref{sec:appendix_additional_results} presents the impulse response functions to the MP and ID shocks for each sub-sample.  On the one hand, across the two FOMC shocks, Peru exhibits a lower impact on the nominal exchange rate and on domestic lending rates as expected given its Central Bank interventions in foreign markets. The more moderate impact on the nominal exchange rate and domestic lending is particularly significant when comparing with a sample of Emerging Markets which excludes Peru (right-most column of Figures \ref{fig:ERR_MP_Peru} and \ref{fig:ERR_CBI_Peru}). On the other hand, the median response on industrial production are quantitatively close, while the impact on the equity index is significantly greater for Peru. 

The second country partition groups Peru and Indonesia as countries with a ``dirty float'' exchange rate regime versus the rest of the countries in the sample. The grouping of Peru and Indonesia is motivated by both countries being classified as ``\textit{tight}'' exchange rate managers according to \cite{ilzetzki2017country}.  Figures \ref{fig:ERR_MP_PyI} and \ref{fig:ERR_CBI_PyI} in Appendix \ref{sec:appendix_additional_results} shows the impulse response functions of the MP and ID shock for each sub-sample, respectively. The insights of the first partition specification still hold. 

Second, I study whether a country's share of commodity goods in total exports matters for the transmission of US interest rate spillovers. In order to test the potential role of commodity goods as a source of heterogeneity in the spillovers of US interest rates I partition the sample of  Emerging Market economies into ``High Commodity Dependence'' and ``Low Commodity Dependence'' countries. This partition is based on countries' share of commodity goods in their export basket as constructed by the UNCTAD's ``The State of Commodity Dependence'' for the year 2016.\footnote{For access to the UNCTAD report see \url{https://unctad.org/webflyer/state-commodity-dependence-2016}.} The high commodity dependence sample is comprised of Brazil, Chile, Colombia and Peru; while the low commodity dependence sample is comprised of Indonesia, Mexico, Philippines and South Africa. Hungary is dropped of the sample as it is not reported in the UNCTAD state of commodity dependence.

\begin{figure}[ht]
         \centering
         \caption{Impulse Response to One-Standard-Deviation Shock \\ \footnotesize  EMs without Hungary}
         \includegraphics[scale=0.325]{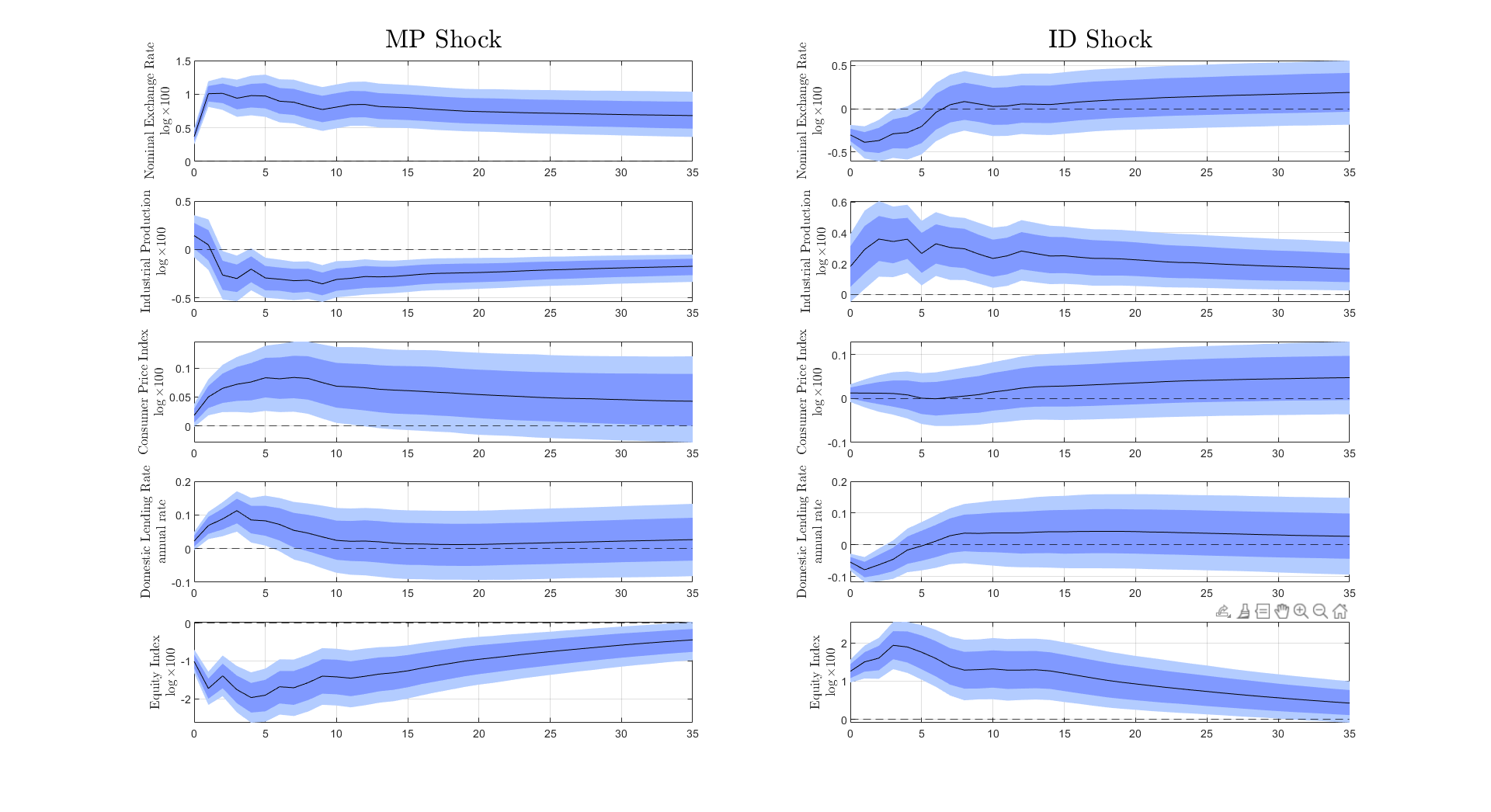}
         \label{fig:EMC}
         \floatfoot{\scriptsize \textbf{Note:} The black solid line represents the median impulse response function. The dark shaded area represents the 16 and 84 percentiles. The light shaded are represents the 5 and 95 percentiles. The figure is comprised of 10 sub-figures ordered in five rows and two columns. Every row represents a different variable: (i) the MP structural FOMC shock, (ii) the ID structural FOMC shock, (iii) nominal exchange rate, (iv) industrial production index, (v) consumer price index, (vi) lending rate, (vii) equity index. The first column presents the results for the MP or ``Pure US Monetary Policy'' shock, the second column the ID or ``Information Disclosure'' shock. In the text, when referring to Panel $(i,j)$, $i$ refers to the row and $j$ to the column of the figure.}
\end{figure}

\begin{figure}[ht]
         \centering
         \caption{Impulse Response to One-Standard-Deviation MP Shock \\ \footnotesize  Commodity Dependence}
         \includegraphics[scale=0.325]{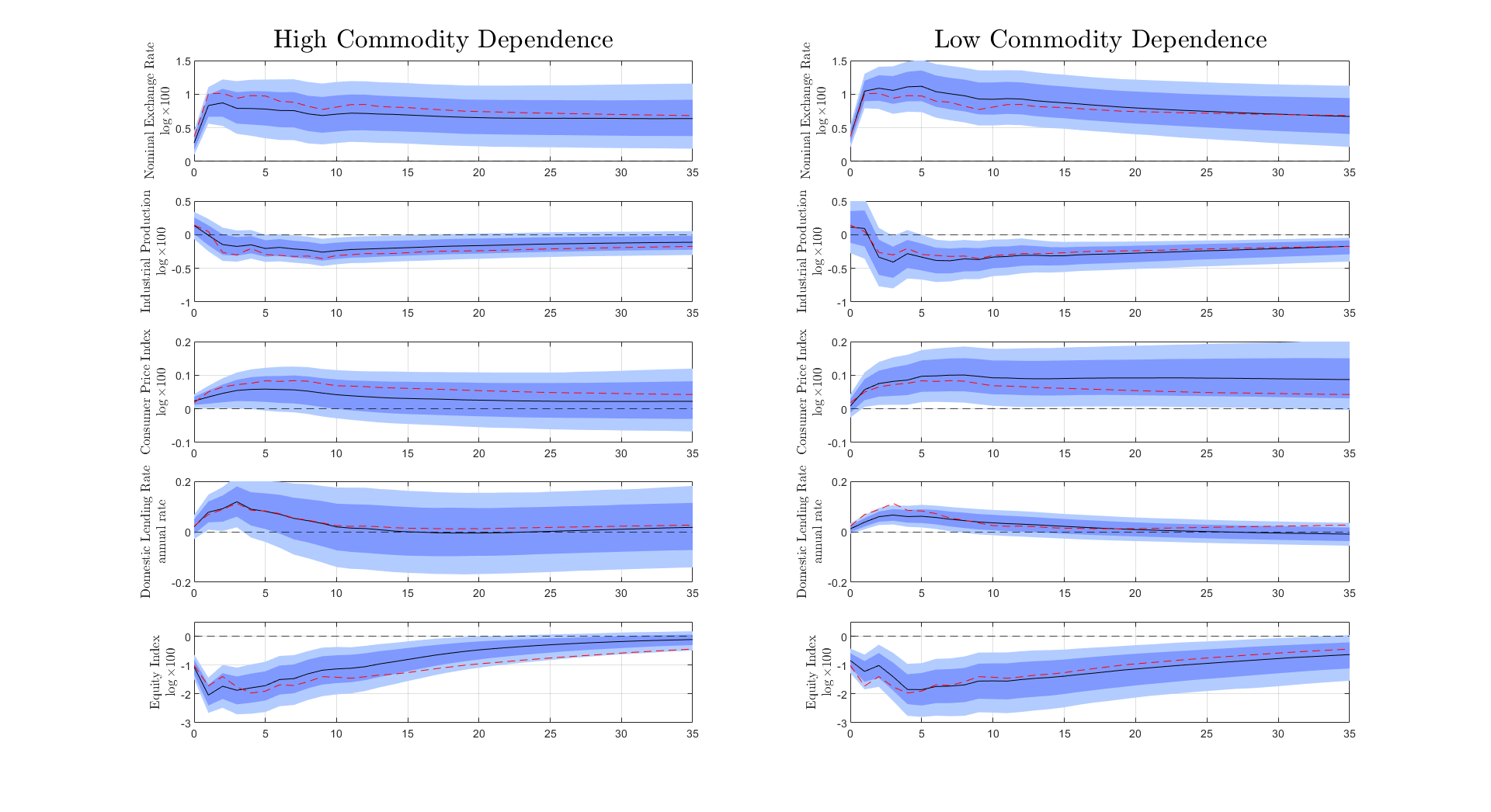}
         \label{fig:EMC_MP}
         \floatfoot{\scriptsize \textbf{Note:} The black solid line represents the median impulse response function. The dark shaded area represents the 16 and 84 percentiles. The light shaded are represents the 5 and 95 percentiles. The figure is comprised of 10 sub-figures ordered in five rows and two columns. Every row represents a different variable: (i) the MP structural FOMC shock, (ii) the ID structural FOMC shock, (iii) nominal exchange rate, (iv) industrial production index, (v) consumer price index, (vi) lending rate, (vii) equity index. The first column presents the results for the MP or ``Pure US Monetary Policy'' shock, the second column the ID or ``Information Disclosure'' shock. In the text, when referring to Panel $(i,j)$, $i$ refers to the row and $j$ to the column of the figure.}
\end{figure}

\begin{figure}[ht]
         \centering
         \caption{Impulse Response to One-Standard-Deviation ID Shock \\ \footnotesize  Commodity Dependence}
         \includegraphics[scale=0.325]{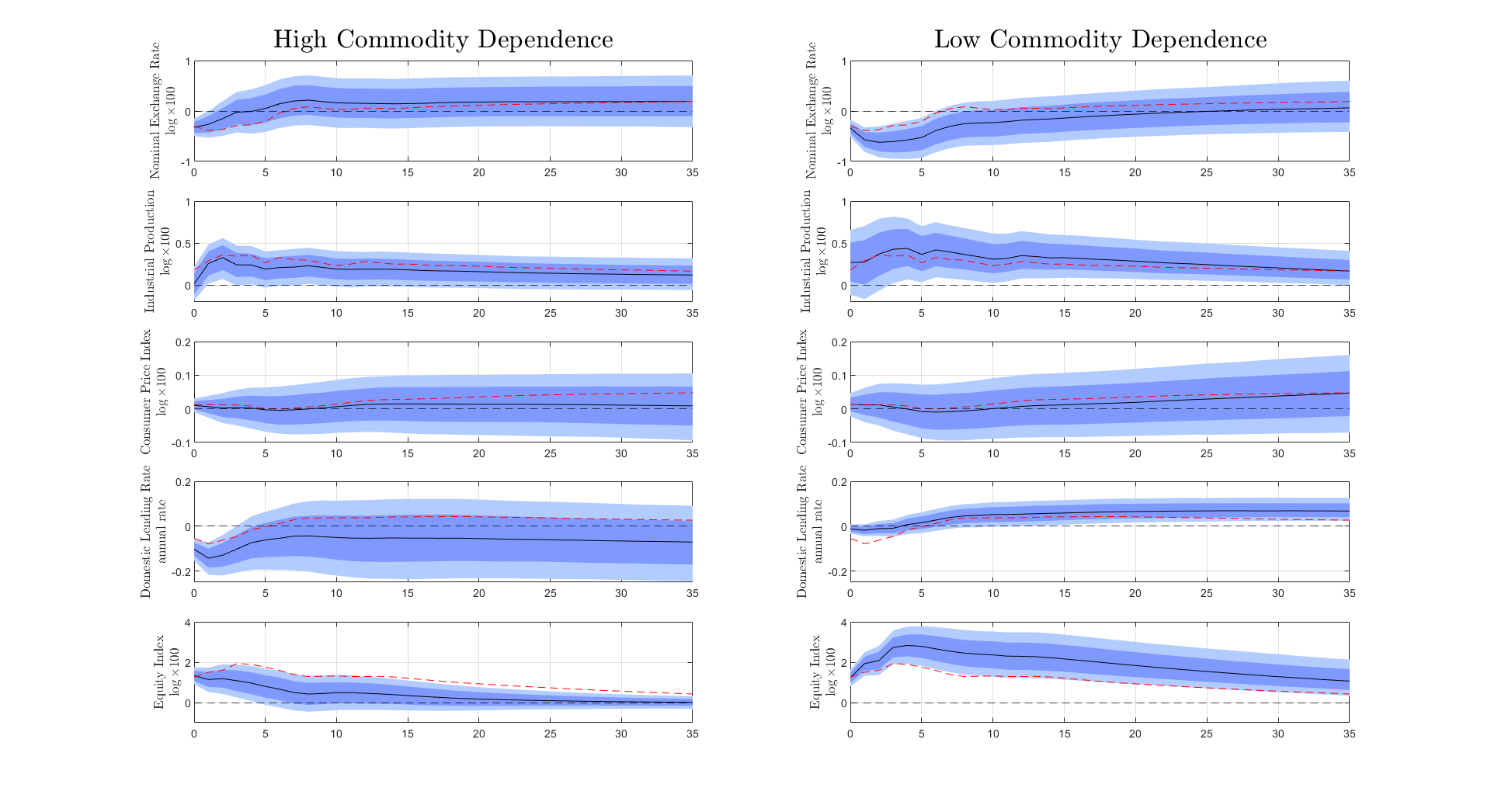}
         \label{fig:EMC_CBI}
         \floatfoot{\scriptsize \textbf{Note:} The black solid line represents the median impulse response function. The dark shaded area represents the 16 and 84 percentiles. The light shaded are represents the 5 and 95 percentiles. The figure is comprised of 10 sub-figures ordered in five rows and two columns. Every row represents a different variable: (i) the MP structural FOMC shock, (ii) the ID structural FOMC shock, (iii) nominal exchange rate, (iv) industrial production index, (v) consumer price index, (vi) lending rate, (vii) equity index. The first column presents the results for the MP or ``Pure US Monetary Policy'' shock, the second column the ID or ``Information Disclosure'' shock. In the text, when referring to Panel $(i,j)$, $i$ refers to the row and $j$ to the column of the figure.}
\end{figure}

Figures \ref{fig:EMC_MP} and \ref{fig:EMC_CBI} present the impulse response functions for the high and low commodity dependence samples respectively. I plot a dashed red line with the median response for the sample of Emerging Market economies (see Figure \ref{fig:EMC}). In response to a MP shock, the low commodity dependence sample exhibits a greater nominal exchange rate depreciation, a greater pass through to consumer prices, a greater increase in EMBI spreads and a larger drop in the equity index, than the high commodity dependence sample. Similar dynamics arise in response to a ID shock. While this may be driven by large commodity exporters being having a greater tradable sector than other countries which reduces the economic contraction necessary to adjust to a given cut-off capital outflow, as suggested by \cite{cavallo2008does}, I leave this question for future research.

\begin{figure}[ht]
         \centering
         \caption{IRF to One-Standard-Deviation MP Shock \\ \footnotesize Exchange Rate Regimes - Peru}
         \includegraphics[scale=0.325]{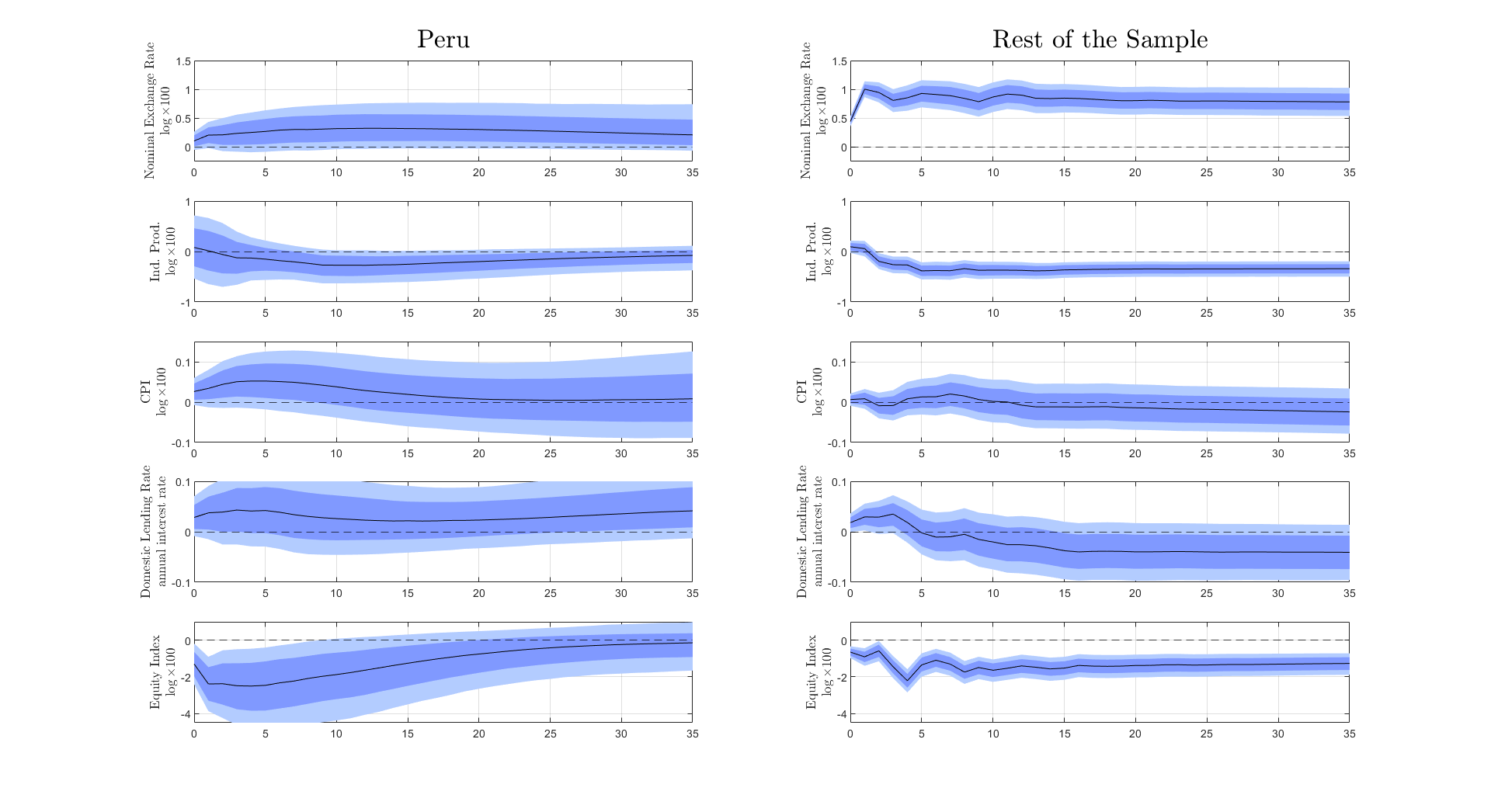}
         \label{fig:ERR_MP_Peru}
         \floatfoot{\scriptsize \textbf{Note:} The black solid line represents the median impulse response function. The dark shaded area represents the 16 and 84 percentiles. The light shaded are represents the 5 and 95 percentiles. The figure is comprised of 15 sub-figures ordered in five rows and three columns. Every row represents a different variable: (i) nominal exchange rate, (ii) industrial production index, (iii) consumer price index, (iv) lending rate, (v) equity index. The first column presents the results for Peru, the middle column presents the results for a sample of both Advanced and Emerging Market economies except Peru, the third or right column presents the results for a sample of Emerging Market economies except Peru. In the text, when referring to Panel $(i,j)$, $i$ refers to the row and $j$ to the column of the figure.}
\end{figure}

\begin{figure}[ht]
         \centering
         \caption{IRF to One-Standard-Deviation CBI Shock \\ \footnotesize Exchange Rate Regimes - Peru}
         \includegraphics[scale=0.325]{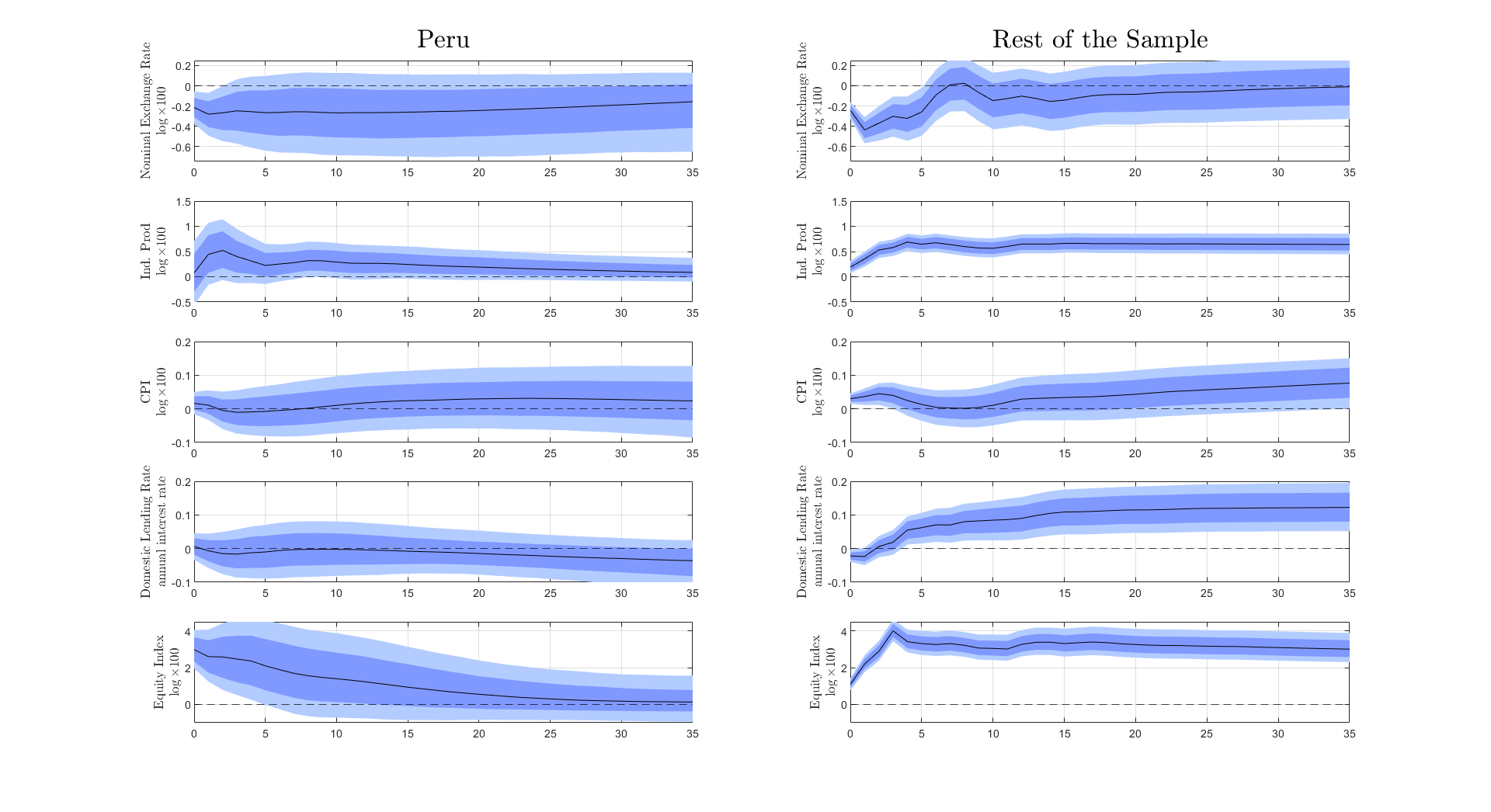}
         \label{fig:ERR_CBI_Peru}
         \floatfoot{\scriptsize \textbf{Note:} The black solid line represents the median impulse response function. The dark shaded area represents the 16 and 84 percentiles. The light shaded are represents the 5 and 95 percentiles. The figure is comprised of 15 sub-figures ordered in five rows and three columns. Every row represents a different variable: (i) nominal exchange rate, (ii) industrial production index, (iii) consumer price index, (iv) lending rate, (v) equity index. The first column presents the results for Peru, the middle column presents the results for a sample of both Advanced and Emerging Market economies except Peru, the third or right column presents the results for a sample of Emerging Market economies except Peru. In the text, when referring to Panel $(i,j)$, $i$ refers to the row and $j$ to the column of the figure.}
\end{figure}

\begin{figure}[ht]
         \centering
         \caption{IRF to One-Standard-Deviation MP Shock \\ \footnotesize Exchange Rate Regimes - P \& I}
         \includegraphics[scale=0.325]{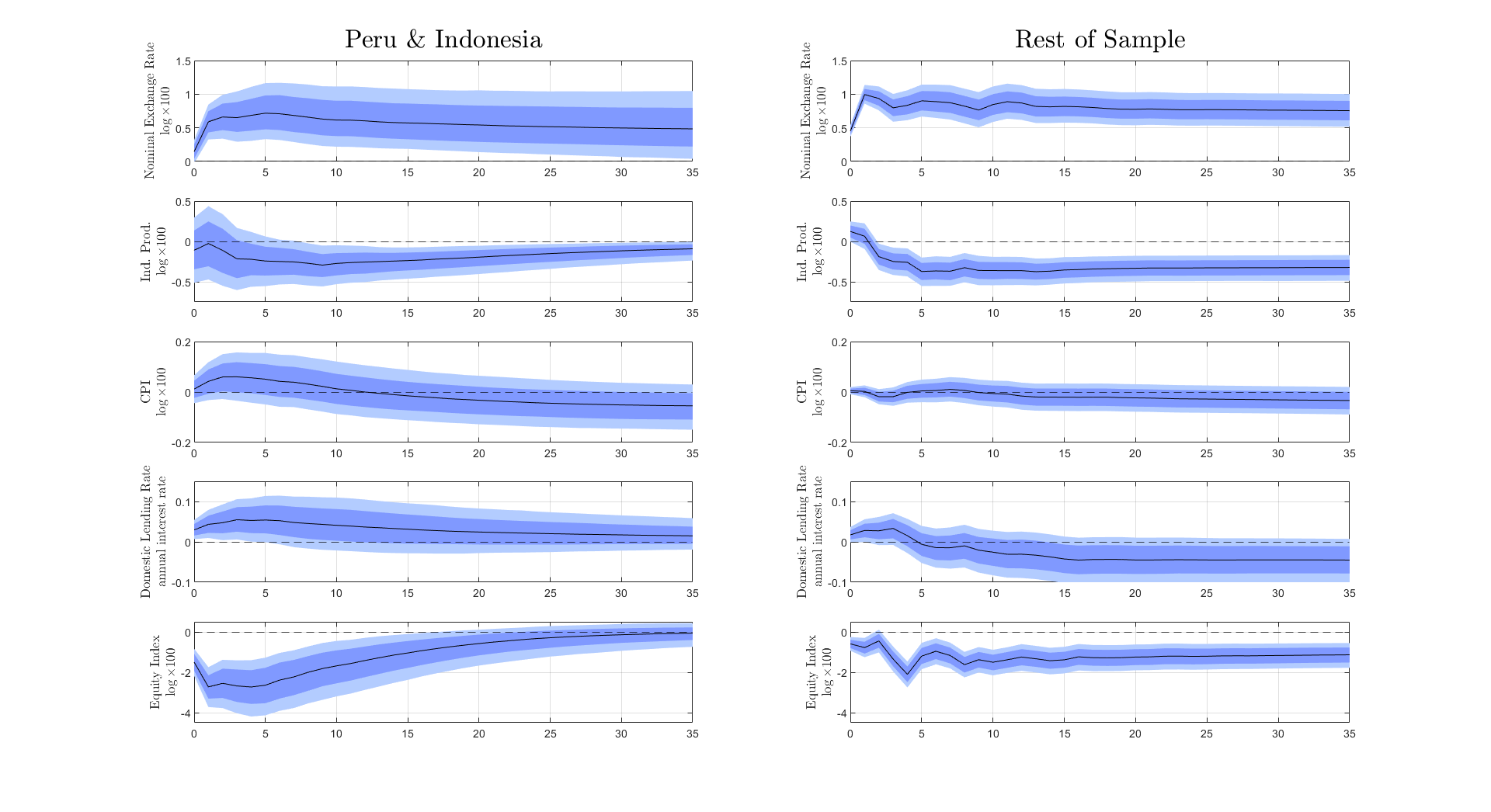}
         \label{fig:ERR_MP_PyI}
         \floatfoot{\scriptsize \textbf{Note:} The black solid line represents the median impulse response function. The dark shaded area represents the 16 and 84 percentiles. The light shaded are represents the 5 and 95 percentiles. The figure is comprised of 15 sub-figures ordered in five rows and three columns. Every row represents a different variable: (i) nominal exchange rate, (ii) industrial production index, (iii) consumer price index, (iv) lending rate, (v) equity index. The first column presents the results for a sample of Peru and Indonesia, the middle column presents the results for a sample of both Advanced and Emerging Market economies except Peru and Indonesia, the third or right column presents the results for a sample of Emerging Market economies except Peru and Indonesia. In the text, when referring to Panel $(i,j)$, $i$ refers to the row and $j$ to the column of the figure.}
\end{figure}

\begin{figure}[ht]
         \centering
         \caption{IRF to One-Standard-Deviation ID Shock \\ \footnotesize Exchange Rate Regimes - P \& I}
         \includegraphics[scale=0.325]{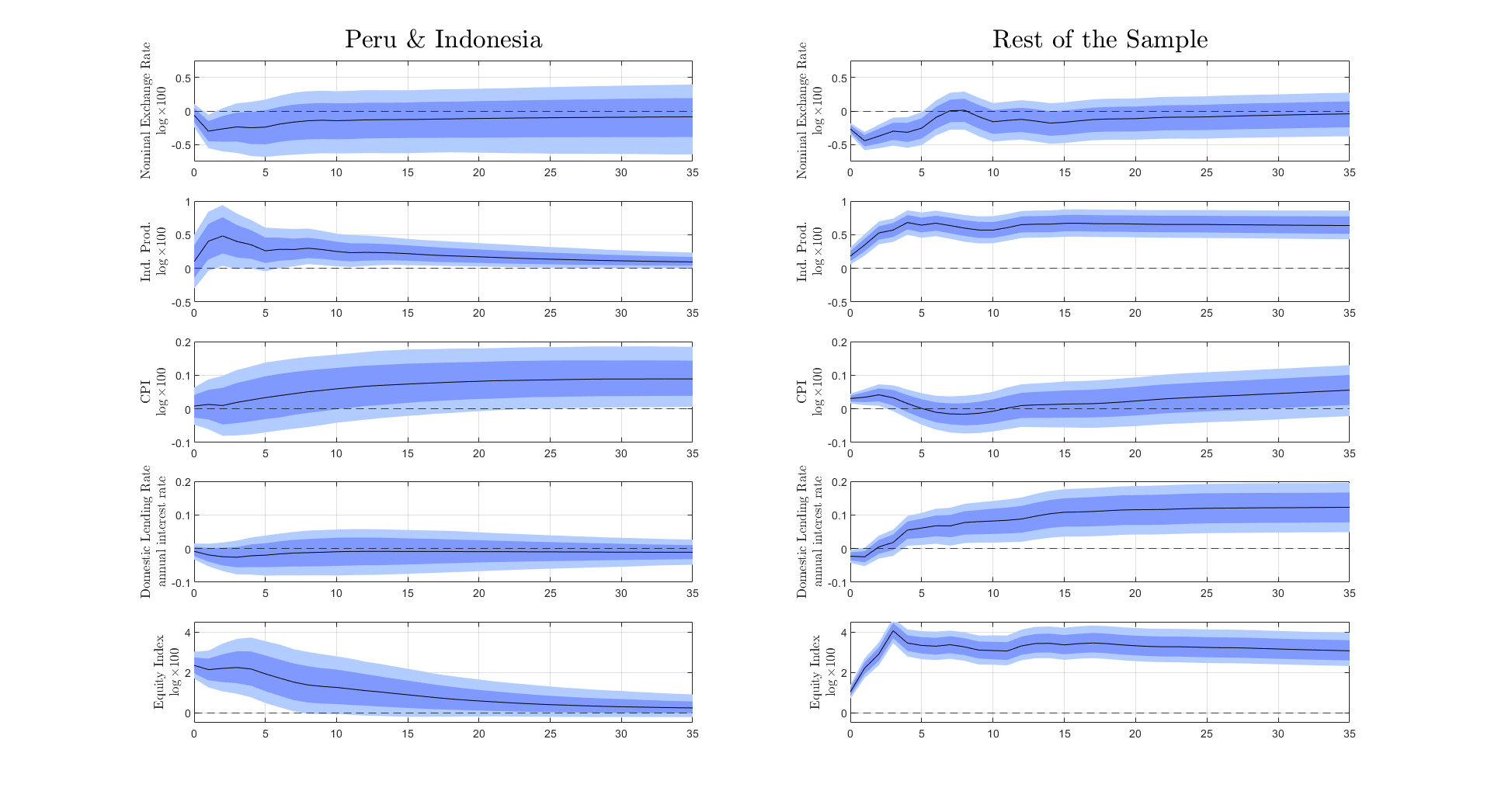}
         \label{fig:ERR_CBI_PyI}
         \floatfoot{\scriptsize \textbf{Note:} The black solid line represents the median impulse response function. The dark shaded area represents the 16 and 84 percentiles. The light shaded are represents the 5 and 95 percentiles. The figure is comprised of 15 sub-figures ordered in five rows and three columns. Every row represents a different variable: (i) nominal exchange rate, (ii) industrial production index, (iii) consumer price index, (iv) lending rate, (v) equity index. The first column presents the results for a sample of Peru and Indonesia, the middle column presents the results for a sample of both Advanced and Emerging Market economies except Peru and Indonesia, the third or right column presents the results for a sample of Emerging Market economies except Peru and Indonesia. In the text, when referring to Panel $(i,j)$, $i$ refers to the row and $j$ to the column of the figure.}
\end{figure}

\end{document}